# Variance partitioning in multilevel models for count data

George Leckie, William Browne, Harvey Goldstein,

Centre for Multilevel Modelling

School of Education

University of Bristol, UK

Juan Merlo

Unit for Social Epidemiology, Faculty of Medicine

Lund University

Malmö, Sweden

Peter Austin

Institute of Health Management, Policy and Evaluation,

University of Toronto

Ontario, Canada

**Address for correspondence**

Professor George Leckie

Centre for Multilevel Modelling

School of Education

University of Bristol

35 Berkeley Square

Bristol

BS8 1JA

United Kingdom

g.leckie@bristol.ac.uk

**Acknowledgements of support**

This research was funded by UK Economic and Social Research Council grant ES/R010285/1. This work was produced using statistical data from Office for National Statistics (ONS), UK. The use of the ONS statistical data in this work does not imply the endorsement of the ONS in relation to the interpretation or analysis of the statistical data. We are very grateful for the helpful comments made by the Associate Editor and the three reviewers.

## Author Note

# Variance partitioning in multilevel models for count data

## Abstract

A first step when fitting multilevel models to continuous responses is to explore the degree of clustering in the data. Researchers fit variance-component models and then report the proportion of variation in the response that is due to systematic differences between clusters. Equally they report the response correlation between units within a cluster. These statistics are popularly referred to as variance partition coefficients (VPCs) and intraclass correlation coefficients (ICCs). When fitting multilevel models to categorical (binary, ordinal, or nominal) and count responses, these statistics prove more challenging to calculate. For categorical response models, researchers appeal to their latent response formulations and report VPCs/ICCs in terms of latent continuous responses envisaged to underly the observed categorical responses. For standard count response models, however, there are no corresponding latent response formulations. More generally, there is a paucity of guidance on how to partition the variation. As a result, applied researchers are likely to avoid or inadequately report and discuss the substantive importance of clustering and cluster effects in their studies. A recent article drew attention to a little-known exact algebraic expression for the VPC/ICC for the special case of the two-level random-intercept Poisson model. In this article, we make a substantial new contribution. First, we derive exact VPC/ICC expressions for more flexible negative binomial models that allows for overdispersion, a phenomenon which often occurs in practice. Then we derive exact VPC/ICC expressions for three-level and random-coefficient extensions to these models. We illustrate our work with an application to student absenteeism.

## Introduction

Multilevel models (mixed-effects, random effects, or hierarchical linear models) are now a standard generalization of conventional regression models for analyzing clustered and longitudinal data in the social, psychological, behavioral and medical sciences. Examples include students within schools, respondents within neighborhoods, patients with hospitals, repeated measures within subjects, and panel survey waves on households. Multilevel models have been further generalized to handle a wide range of response types, including, continuous, categorical (binary or dichotomous, ordinal, and nominal or discrete choice), count, and survival responses. Standard introductions to these models can be found in textbooks by Goldstein (2011b), Hox et al. (2017), Raudenbush and Bryk (2002), and Snijders and Bosker (2012).

A natural first step in any multilevel analysis is to report the degree of clustering in the response since it is the assumed presence of clustering which is the fundamental motivation for fitting multilevel models. Confirming that there is a statistically significant degree of clustering is not enough. One must additionally communicate the practical importance of clustering and this is done by reporting variance partition coefficients (VPCs; Goldstein et al., 2002) and intraclass correlation coefficients (ICCs). Consider a study of the relationships between a continuous student math score $y_{ij}$ for student $i$ ($i = 1, \dots, n_j$) in school $j$ ($j = 1, \dots, J$) and a range of student and school characteristics $\mathbf{x}_{ij}$. Suppose we simply fit a linear regression model $y_{ij} = \mathbf{x}'_{ij}\boldsymbol{\beta} + e_{ij}$. Here the concern is student math scores will correlate within schools (within-cluster dependence), even after adjusting for the covariates $\text{Corr}\left(y_{ij}, y_{i'j} \middle| \mathbf{x}_{ij}, \mathbf{x}_{i'j}\right) > 0$, thereby violating the linear regression assumption of independent residuals $\text{Corr}\left(e_{ij}, e_{i'j} \middle| \mathbf{x}_{ij}, \mathbf{x}_{i'j}\right) = 0$. Such clustering would typically be due to unmodelled student and school influences on math scores that vary between schools (between-cluster heterogeneity). A two-level linear regression $y_{ij} = \mathbf{x}'_{ij}\boldsymbol{\beta} + u_j + e_{ij}$ attempts to account for these influences, and therefore clustering, by including a school random-intercept effect $u_j$. The total residual variance is then decomposed into separate within- and between-school components $\sigma_e^2$ and $\sigma_u^2$. The proportion of response variance which lies between schools (conditional on any covariates) can then be estimated and reported using the VPC. This statistic is calculated as the ratio of the estimated between-school variance to the total residual variance, $\sigma_u^2/(\sigma_u^2 + \sigma_e^2)$. The VPC therefore summarizes the "importance" of the clusters in influencing the response, in our case the importance of schools in influencing student outcomes above and beyond the modelled student and school characteristics. As such, the VPC is widely quoted in multilevel studies. In the case of the current random-intercept model, the expression for the VPC is also the expression for the expected correlation between two students from the same school (conditional on any covariates), and when this interpretation is favored the expression is referred to as the ICC. As a result of their common expression, the two terms and interpretations are used interchangeably. This is poor practice, especially as the VPC and ICC expressions diverge as soon as random coefficients are added to the model (Goldstein et al., 2002). It is therefore prudent to always apply the correct interpretation to each statistic. In this article, we shall focus primarily on expressions for the VPC and therefore the VPC interpretation. For readers seeking further details on the VPC and ICC for two-level variance-components, random-intercept and random-coefficient continuous response models, this can be found in Supplemental materials S1.

VPCs (and ICCs) can also be calculated in three- and higher-level models for continuous responses as well as in models with more complex cross-classified, multiple membership, spatial, and dyadic random effect structures (Leckie, 2013a, 2013b, 2013c). In these settings, VPCs are often used to ascertain the relative importance of different sources of clustering in influencing the response, for example, in establishing the relative importance of family, school and area effects on student attainment (Rasbash et al., 2010). VPCs can also be calculated after fitting models with random coefficients. In all these cases, the VPC expressions become more complex and don't necessarily continue to have parallel ICC interpretations, but these issues are well described in the literature (see textbook references).

For multilevel models for categorical responses (e.g., binary, ordinal, and nominal logistic regression), calculating VPCs (and ICCs) is less straightforward as these statistics typically no longer have closed-form expressions (they involve integrating out the random effects which can only be achieved via numerical integration). The standard approach to this problem is to appeal to the latent-response formulations of these models and to report VPCs in terms of unobserved continuous variables envisaged to underlie the observed categorical responses (see textbook references). For example, in a study of whether students pass an exam, we would appeal to the notion of a continuous exam score scale underlying the observed binary pass or fail status and we would report the VPC in terms of this latent variable, that is, in terms of the propensity for the student to pass the exam. An appealing feature of this approach is that it allows one to calculate VPCs for categorical responses using essentially the same expressions as those derived for continuous responses.

For multilevel models for count responses (e.g., Poisson and negative binomial models), however, there are no corresponding latent-response formulations (Skrondal and Rabe-Hesketh, 2004) and so it is less obvious how one should calculate the VPC (or ICC). More generally there is a paucity of guidance on how to partition response variation when modelling counts. As a result, applied researchers are likely to avoid or inadequately report and discuss the substantive importance of clustering and cluster effects in their studies.

A notable exception is the work by Stryhn et al. (2006) and later Austin et al. (2017) who show that the VPC (and ICC) for the special case of a two-level random-intercept Poisson model does have a closed-form and so can be calculated exactly with a simple algebraic expression. However, for many applications, researchers now routinely apply multilevel negative binomial models to account for overdispersion (the phenomenon whereby the variance of the observed counts is larger than that implied by the expectation) and so exact algebraic expressions for the VPC are also needed for these more flexible models. Likewise, many researchers now routinely fit count models allowing for three or more levels or random coefficients and so VPC expressions must also be extended to account for these modelling extensions.

Given the lack of VPC (and ICC) expressions for count response models, the only approach available to researchers is the simulation method proposed by Goldstein et al. (2002) and further illustrated by Browne et al. (2005), Leckie and Goldstein (2015), and Leckie et al. (2012). While these articles discuss this method in the context of two-, three- and four-level binary response models, the method readily extends to the count models discussed here. Indeed, Austin et al. (2017) used this simulation method to confirm that the exact algebraic expression introduced there for the VPC for the special case of the two-level random-intercept Poisson model gives the correct value. The simulation method involves using the fitted model to simulate new count responses and then calculating the within- and between-cluster variances for these simulated data. The VPC can then be calculated in the usual way. The principal disadvantage of using the simulation method to calculate the VPC is that it has not been implemented in software, forcing researchers to write their own code which is error prone, especially for models with complex random effect structures, and it is perhaps for these reasons why one sees so few applications of the method in practice.

In this article, we propose a simpler approach. We derive exact algebraic expressions for the VPC and ICC for the two-level random-intercept Poisson and negative binomial model (mean dispersion or NB2 version) and the three-level and random-coefficient extensions of these models. We focus on these two models as they are most widely discussed (see textbook references). However, we note that other multilevel models for count responses are increasingly being implemented in standard software (R, SAS, SPSS, Stata). We therefore additionally review two of these models and derive their VPC and ICC expressions (Supplemental materials S3): the Poisson model with an overdispersion random effect and the constant dispersion or NB1 version of the negative binomial model. We motivate and illustrate each methodological development with an application to student absenteeism. We confirm our derivations are correct by showing they give the same VPC and ICC estimates for our application as those based on the simulation method (Supplemental materials S8). We share the Stata and R code and output for our application in order to assist readers in applying our VPC and ICC expressions in their own work (Supplemental materials S9 and S10).

## Application: Student absenteeism from school

Student absenteeism and its detrimental effects on student learning are ongoing concerns in the US (EPI, 2018), UK (DfE, 2019a) and many other countries. In response, school accountability systems are increasingly monitoring student absenteeism rates alongside more traditional attainment and progress measures (Leckie and Goldstein, 2017). Student absenteeism is known to vary by student demographic and socioeconomic characteristics, factors which also vary across school intakes. It would therefore seem important to adjust student absenteeism rates for school differences in student composition before making any attempt to hold schools accountable for their performance (Prior et al., 2019). In this application, we explore these ideas using multilevel models for count data. These models and notion of adjusting student outcomes for student characteristics when comparing schools are analogous to those used to estimate school value-added effects on student attainment (Castellano and Ho, 2013; Goldstein, 1997; Leckie and Goldstein, 2009, 2019; OECD, 2008; Raudenbush and Willms, 1995).

Our study relates to students in London schools who completed their compulsory secondary schooling at the end of the 2016/17 academic year (students aged 15/16 in UK school year 11; equivalent to US 10th grade). The study was granted ethical approval by the Ethics Committee at the Faculty of Social Sciences and Law, University of Bristol. The data are drawn from the national pupil database, a census of all students in state-maintained schools in England (DfE, 2019b). The data are a three-level hierarchy consisting of 66,955 students (level-1) nested in 434 schools (level-2) nested in 32 school districts (level-3). The response is a count of the number of days students were absent from school during the academic year. Figure 1 presents the frequency distribution. The distribution is positively skewed, with students, on average, absent for 8.41 days over the academic year, but with an interquartile range spanning from 2 to 10 days (Variance 124.39; Min 0; Max 156). Figure 2 illustrates variation in the mean number of days by district (left; Mean 8.36; IQR = 7.56, 9.10) and by school (right; Mean = 8.46; IQR = 6.56, 9.76). We see meaningful differences in student absence rates, especially between schools.

We will start by fitting “empty” or “null” two-, and three-level random-intercept Poisson and negative binomial models with no covariates. In the continuous response case, these models are referred to as variance-component models and we use that terminology here. The aim of these initial analyses is to quantify the degree of clustering and overdispersion in the data and we do this using our VPC (and ICC) expressions. Thus, we will use the VPC expressions to summarize the importance of district and school differences in student absenteeism rates in explaining why some students miss more school days than others. We will then pick a preferred model and extend it by entering student characteristics as covariates. Here our aim is to not just study the predictors of student absenteeism, but to adjust for these factors so that the resulting predicted school random effects provide more meaningful estimates of school influences on students that are plausibly related to factors within schools' control. At this point we will recalculate the VPCs and explore how they now vary as a function of the covariates and we investigate the predicted school effects. Last, we shall fit a random-coefficients model, in which we allow the effect of one of the student covariates to vary across schools, and we will explore the implications this has for calculating and interpreting the VPC.

We fit all models using maximum likelihood estimation (via adaptive quadrature) as implemented in the "mepoisson" and "menbreg" commands in Stata (StataCorp, 2019). See Rabe-Hesketh and Skrondal (2012, Chapter 13) for an excellent introduction to multilevel models for count responses using Stata. These models can equally be fitted by maximum likelihood estimation in other standard software (R, SAS, SPSS) or by Markov chain Monte Carlo (MCMC) methods.

## Two-level variance-components Poisson model for count responses

In this section, we review the two-level variance-components Poisson model and then present expressions for the VPC and ICC. We delay the introduction of covariates to later. This model and VPC and ICC expressions were also the focus of Austin et al. (2017), but here we explore in more detail the assumptions the model makes regarding the observed counts in order to help readers better understand the expressions and their interpretations. These steps will then help the reader when we explore more complex count models which are the primary focus of this paper and which were not considered by Austin et al. For readers seeking a refresher on the conventional (single-level) Poisson model (and other count models), we recommend the textbook by Long (1997).

The two-level variance-components Poisson model, like all multilevel models, is a conditional (cluster-specific) model as opposed to a marginal (population-averaged) model: estimation and inference are based on conditioning on the values of the random effects as well as the covariates as opposed to conditioning only on the covariates and averaging over the population of clusters. The model is therefore described in terms of the conditional expectation and variance of the response. Importantly, conditional and marginal models for count responses make different inferences. This contrasts conditional and marginal models for continuous responses where inferences coincide. Our view is that in applied research it is usually informative to explore and contrast both approaches. Indeed, as we shall go on to discuss, the VPC and ICC are marginal statistics derived from the marginal expectation, variance, covariance and correlation of the responses. It turns out, one can recover these marginal statistics from the conditional model, but these statistics are rarely presented in textbook and other discussions of the model (Supplemental materials S2). We therefore review each marginal statistic below prior to presenting our expressions for the VPC and ICC. We hope that this treatment provides a further useful resource for readers. We provide full derivations in Supplemental materials S4.

**Model**

Let $y_{ij}$ denote the count for unit $i$ ($i = 1, \dots, n_j$) in cluster $j$ ($j = 1, \dots, J$). In terms of our application, the units will be students, the clusters schools, and the count will be the number of days each student is absent from school over the course of the school year. We can then write the two-level variance-component Poisson model for $y_{ij}$ as follows

$$y_{ij}|\mu_{ij} \sim \text{Poisson}(\mu_{ij})$$
$$\ln(\mu_{ij}) = \beta_0 + u_j \tag{1}$$
$$u_j \sim N(0, \sigma_u^2)$$

where $\mu_{ij}$ denotes the expected count, $\beta_0$ denotes the intercept, and $u_j$ is the cluster random intercept effect, assumed normally distributed with zero mean and variance $\sigma_u^2$.

**Conditional statistics**

The conditional expectation of $y_{ij}$ (given $u_j$) is given by

$$\mu_{ij}^C \equiv \text{E}(y_{ij}|u_j) = \exp(\beta_0 + u_j) \tag{2}$$

which in this model is simply equal to $\mu_{ij}$ in Equation 1, but this will not be the case in the next model, hence the introduction of the “C” superscript.

The conditional variance is given by

$$\omega_{ij}^C \equiv \text{Var}(y_{ij}|u_j) = \mu_{ij}^C \tag{3}$$

Thus, the conditional variance of the counts is assumed to equal the conditional expectation. In practice, this equi-dispersion assumption often fails, with the variance of the observed counts in many clusters being larger or smaller than that implied by the cluster mean, phenomena known as overdispersion (extra-Poisson variability) or underdispersion, respectively. Overdispersion is far more common than underdispersion and is typically attributed to unobserved unit-level covariates. In terms of our application, such covariates might include student prior attainment, demographics, or socioeconomic status and we will introduce these into the analysis later.

The conditional covariance and correlation between the response measurements on two units $i$ and $i'$ from the same cluster $j$, $y_{ij}$ and $y_{i'j}$ are zero and so the observed counts are assumed conditionally independent (sometimes referred to as the local independence assumption). The conditional covariance and correlation between response measurements on two units from two different clusters are also assumed equal to zero.

## Marginal statistics

The marginal expectation of $y_{ij}$ (now averaged over $u_j$) is given by

$$\mu_{ij}^{M} \equiv \mathrm{E}(y_{ij}) = \exp(\beta_0 + \sigma_u^2/2) \tag{4}$$

The marginal variance of $y_{ij}$ is given by

$$\omega_{ij}^{M} \equiv \mathrm{Var}(y_{ij}) = \mu_{ij}^{M} + \left(\mu_{ij}^{M}\right)^2\{\exp(\sigma_u^2) - 1\} \tag{5}$$

The marginal variance is therefore a quadratic function of the marginal expectation and is larger than the marginal expectation if there is clustering, $\sigma_u^2 > 0$.

The marginal covariance of $y_{ij}$ and $y_{i'j}$ is given by

$$\mathrm{Cov}(y_{ij}, y_{i'j}) = \left(\mu_{ij}^{M}\right)^{2}\{\exp(\sigma_u^2) - 1\} \tag{6}$$

and the associated correlation can then be calculated in the usual way to give

$$\mathrm{ICC}_{ij,i'j} \equiv \mathrm{Corr}(y_{ij}, y_{i'j}) = \frac{\left(\mu_{ij}^{M}\right)^{2}\{\exp(\sigma_u^2)-1\}}{\mu_{ij}^{M} + \left(\mu_{ij}^{M}\right)^{2}\{\exp(\sigma_u^2)-1\}} \tag{7}$$

This marginal correlation can be interpreted as the ICC as it is the response correlation between two units in the same cluster. In contrast, the marginal covariance and correlation between response measurements on two units from two different clusters is equal to zero as the units do not share a cluster random effect.

The VPC is defined as the proportion of the marginal response variance which lies between clusters. Thus, we can only calculate the VPC after we have partitioned the marginal variance into level-specific components. The expression for the marginal variance (Equation 5) does just this (Supplemental materials S4.2). Specifically, the first term $\mu_{ij}^{M}$ captures the average variance within clusters in the observed unit-level counts $y_{ij}$ around the expected counts $\mu_{ij}^{C}$ (Equation 3) (the unit-level or level-1 component), while the second term $\left(\mu_{ij}^{M}\right)^{2}\{\exp(\sigma_u^2) - 1\}$ captures the variance between clusters in their expected counts $\mu_{ij}^{C}$ attributable to the cluster random intercept effect $u_j$ (the cluster-level or level-2 component). The expression for the VPC can then be derived in the usual way: as the ratio of the level-2 component of the marginal variance divided by the summation of the level-2 and -1 components to give

$$\mathrm{VPC}_{ij} = \frac{\overbrace{\left(\mu_{ij}^{M}\right)^{2}\{\exp(\sigma_u^2)-1\}}^{\text{level}-2\text{ variance}}}{\underbrace{\left(\mu_{ij}^{M}\right)^{2}\{\exp(\sigma_u^2)-1\}}_{\text{level}-2\text{ variance}} + \underbrace{\mu_{ij}^{M}}_{\text{level}-1\text{ variance}}} \tag{8}$$

and this expression is identical to that for the marginal correlation or ICC given in Equation 7. This expression, but where $\exp(\beta_0 + \sigma_u^2/2)$ is substituted in for $\mu_{ij}^M$, was published in Austin et al. (2017). Studying Equation 8, we see that the VPC is an increasing function of both the marginal expectation $\mu_{ij}^M$ and the cluster variance $\sigma_u^2$. This VPC is strictly the level-2 VPC. The level-1 VPC – the proportion of the marginal response variance which lies within clusters – is then simply equal to one minus the level-2 VPC.

**Application continued**

Model 1 is a two-level variance-components Poisson model (Equation 1). The model includes a school random intercept to investigate and account for potential school clustering. The estimated intercept is 2.085. The model estimates the school variance to be 0.100 and a likelihood-ratio test confirms that this between school variation in student absence rates is statistically significant (Model 1 vs. a conventional single-level Poisson model, but with no covariates; results not shown; $\chi_1^2 = 53194, p < 0.001$). What is less clear is the practical or substantive importance of these estimates. In particular, is a value of 0.100 for the school variance big? Does school clustering matter? Do we care? We can answer these questions using the VPC, ICC and other estimated marginal statistics.

The parameter estimates imply a marginal expectation of 8.46 (application of equation 4) and a marginal variance of 15.98 (application of Equation 5). The marginal expectation approximately equals the sample mean number of days (8.41), but the marginal variance is far below the sample variance of 124.39 and so this model proves inadequate for these data. To understand this last point, it proves revealing to decompose the estimated marginal variance into level-specific components. The school component equals 7.52 while the student component equals 8.46. The resulting school VPC equals 0.47 and so a very high 47% of the marginal variance is due to systematic differences between schools (application of Equation 8). This suggests that school level factors, or at least school-level variation in student characteristics, account for almost half the modelled variation in student absenteeism between students. The ICC in this model (application of Equation 7) equals the VPC: the response correlation between two students in the same school is very high at 0.47. Finally, note that the student component of the marginal variance appears identical to the marginal expectation, both are estimated as 8.46. This is indeed the case; in this model these two terms are equal by definition (see the last term in the denominator of Equation 8). In the next section, we will explore the negative binomial model which relaxes this constraint and therefore allows the data to be more dispersed than that implied by the mean and so will likely provide a better fit to the data and a more reasonable estimate of the marginal variance.

We note that when desired, interval estimates can also be calculated for the VPCs (and ICCs) (Goldstein et al., 2002) and for that matter each of the other marginal statistics. When models are fitted by maximum likelihood estimation, a 95% confidence interval for each statistic can be constructed via the delta method (the VPC is a non-linear combination of the model parameters) or via multilevel bootstrapping (e.g., fitting the model to 1000 bootstrapped samples to obtain a sampling distribution for the VPC) (Goldstein, 2011a). When the model is fitted by MCMC methods, a 95% credible interval can be calculated using the MCMC chain for the posterior distribution of the VPC.

## Two-level variance-component Negative binomial model for count responses

In this section, we shift our focus to the more flexible two-level variance-component negative binomial model (mean dispersion or NB2 version) which allows for overdispersion. This work now moves beyond that presented in Austin et al. (2017). In particular, the expression we present for the VPC and ICC is an important new result. See Supplemental materials S3.2 for a parallel presentation of the constant dispersion or NB1 version of this model and associated VPC and ICC expressions. We provide full derivations of all marginal statistics in Supplemental materials S4.

### Model

The negative binomial model (mean dispersion or NB2 version) is an extension of the Poisson model that adds a normally distributed unit-level overdispersion random effect to represent omitted unit-level variables that are envisaged to be driving any overdispersion. In contrast to the conventional cluster random intercept effect, this does not induce any dependence among the units. The model can be written as

$$\begin{gathered} y_{ij}|\mu_{ij} \sim \text{Poisson}(\mu_{ij}) \\ \ln(\mu_{ij}) = \beta_0 + u_j + e_{ij} \\ u_j \sim N(0, \sigma_u^2) \\ \exp(e_{ij}) \sim \text{Gamma}\left(\frac{1}{\alpha}, \alpha\right) \end{gathered} \tag{9}$$

where $e_{ij}$ denotes the overdispersion random effect. Thus, in this model, two units with the same random intercept effect value may nonetheless differ in their expected counts $\mu_{ij}$, with such differences attributed to the two units differing in terms of their values on the omitted unit-level variables. The exponentiated overdispersion random effect $\exp(e_{ij})$ is assumed gamma distributed with shape and scale parameters $1/\alpha$ and $\alpha$ and is therefore distributed with mean 1 and variance or overdispersion parameter $\alpha$. The larger $\alpha$ is, the greater the overdispersion. When $\alpha = 0$, the model simplifies to the Poisson model (Equation 1) and so we can conduct a likelihood-ratio test to compare the two models to see whether the estimated overdispersion is statistically significant.

**Conditional statistics**

In this model, we can again calculate the conditional expectation and variance of the response. However, here we must first integrate out the overdispersion random effect since this is not typically of substantive interest. The conditional expectation of $y_{ij}$ (given $u_j$ but averaged over $e_{ij}$) has the same form as in the Poisson model (Equation 2), with

$$\mu_{ij}^{C} \equiv \mathrm{E}(y_{ij}|u_j) = \exp(\beta_0 + u_j) \tag{10}$$

and we see that, in contrast to the Poisson model, $\mu_{ij}^{C} \neq \mu_{ij}$. The conditional variance is then given by

$$\omega_{ij}^{C} \equiv \mathrm{Var}(y_{ij}|u_j) = \mu_{ij}^{C} + (\mu_{ij}^{C})^2\alpha \tag{11}$$

Thus, the conditional variance is now a quadratic function of the conditional expectation and is larger than the conditional expectation if $\alpha > 0$. Therefore, the usual variance-mean relationship for the Poisson model (Equation 3) is relaxed, allowing overdispersion with respect to the conditional expectation ($\omega_{ij}^C > \mu_{ij}^C$).

**Marginal statistics**

The marginal expectation of $y_{ij}$ (now averaged over $u_j$ as well as $e_{ij}$) is given by

$$\mu_{ij}^M \equiv \mathrm{E}(y_{ij}) = \exp(\beta_0 + \sigma_u^2/2) \qquad (12)$$

which is the same as that for the Poisson model (Equation 4).

The marginal variance of $y_{ij}$ is given by

$$\omega_{ij}^M \equiv \mathrm{Var}(y_{ij}) = \mu_{ij}^M + (\mu_{ij}^M)^2\{\exp(\sigma_u^2)(1+\alpha) - 1\} \qquad (13)$$

which differs from that for the Poisson model (Equation 5) via the inclusion of the additional multiplicative term $(1+\alpha)$. Thus, in this model, the marginal variance is larger than the marginal expectation if there is clustering $\sigma_u^2 > 0$ *or* overdispersion $\alpha > 0$.

The marginal covariance of $y_{ij}$ and $y_{i'j}$ (averaged over $u_j$, $e_{ij}$ and $e_{i'j}$) is given by

$$\mathrm{Cov}(y_{ij}, y_{i'j}) = (\mu_{ij}^M)^2\{\exp(\sigma_u^2) - 1\} \qquad (14)$$

and is the same as that for the Poisson model (Equation 6). The marginal correlation of $y_{ij}$ and $y_{i'j}$ is then given by

$$\text{ICC}_{ij,i'j} \equiv \text{Corr}(y_{ij}, y_{i'j}) = \frac{\left(\mu_{ij}^M\right)^2 \{\exp(\sigma_u^2) - 1\}}{\mu_{ij}^M + \left(\mu_{ij}^M\right)^2 \{\exp(\sigma_u^2)(1+\alpha) - 1\}} \tag{15}$$

and as with the Poisson model can be interpreted as the ICC. The expression differs from that for the Poisson model (Equation 7) only in the inclusion of the additional multiplicative term $(1 + \alpha)$ in the denominator.

As with the Poisson model, we can partition the marginal variance (Equation 13) into level-specific components which capture the within- and between-cluster variance in $y_{ij}$ (Supplemental materials S4.2). The resulting level-2 VPC is given by

$$\text{VPC}_{ij} = \frac{\overbrace{\left(\mu_{ij}^M\right)^2 \{\exp(\sigma_u^2) - 1\}}^{\text{level-2 variance}}}{\underbrace{\left(\mu_{ij}^M\right)^2 \{\exp(\sigma_u^2) - 1\}}_{\text{level-2 variance}} + \underbrace{\mu_{ij}^M + \left(\mu_{ij}^M\right)^2 \exp(\sigma_u^2)\alpha}_{\text{level-1 variance}}} \tag{16}$$

and this expression is identical (after rearranging terms) to that for the marginal correlation or ICC given in Equation 15.

Studying Equation 16, we see that, as in the Poisson case (Equation 8), the VPC is an increasing function of both the marginal expectation $\mu_{ij}^M$ and the cluster variance $\sigma_u^2$. However, the VPC is now also a decreasing function of the overdispersion parameter $\alpha$. This makes sense. As the overdispersion increases, all else equal, the more unmodelled variation there is at level-1 and so the VPC decreases.

Comparing the two VPC expressions (Equations 8 and 16), we see that the expression for the level-2 component of the marginal variance is the same and so it is only the expression for the level-1 component which varies across models. This makes sense as the models differ only in their treatment of overdispersion, which is viewed as a level-1 phenomenon. The overdispersion parameter in the negative binomial model leads the expression for the level-1 component of the marginal variance to exceed that of the Poisson model. The marginal variance is simply the summation of the level-2 and -1 variances and so is also expected to be higher in the negative binomial model compared to that of the Poisson model.

**Application continued**

Model 2 is a two-level variance-components negative binomial model (Equation 9). The model includes a student overdispersion random effect to account for any within-school variation due to omitted student influences. The overdispersion parameter estimated to be 0.877 and a likelihood-ratio test confirms that there is significant overdispersion (Model 2 vs. Model 1: $\chi^2_1 = 363096, p < 0.001$). What is less clear is the practical or substantive importance of this additional variation. Put simply, is a value of 0.877 for the overdispersion parameter big? Does overdispersion matter? Do we care? Here too, we can answer these questions using the estimated VPC, ICC and other marginal statistics.

Most importantly, the estimated marginal variance now increases from 15.98 to 84.10 (application of Equation 13). As expected, the school component remains approximately stable and so the increase in the marginal variance is brought about by the student component which increases nine-fold from 8.46 to 77.15 (it is no longer constrained to equal the marginal variance). This increase indicates that, even within schools, student absenteeism is far from a random Poisson process, rather the model suggests that there is substantial within school variability driven by omitted student characteristics. This increase in the student component in turn has a dramatic impact on the estimated VPC. The model now estimates the school VPC to be 0.08 (application of Equation 16), suggesting that it is in fact omitted student-specific factors rather than omitted school-specific factors that are likely the dominant cause of the variation in absenteeism. This estimated VPC is far lower than the estimate of 0.47 reported for Model 1. Thus, an important finding is that by ignoring overdispersion, the Poisson model grossly overestimates the relative importance of schools in these data. More generally, the Poisson VPC is biased upwards in the presence of overdispersion.

## Three-level variance-components models for count data and calculation of VPCs

In this section, we focus on three-level variance components negative binomial model and restrict our presentation of the marginal statistics to only the VPC and ICC. Recall, however, that the Poisson model is simply the special case of the negative binomial model with no overdispersion (see previous sections). The parallel VPC and ICC expressions for the Poisson model can therefore be obtained by setting $\alpha = 0$ in the expressions below. We provide full derivations of all marginal statistics in Supplemental materials S5.

### Model

Let $y_{ijk}$ denote the count for unit $i$ ($i = 1, \dots, n_j$) in cluster $j$ ($j = 1, \dots, J_k$) in supercluster $k$ ($k = 1, \dots, K$). In terms of our application, the units are students, the clusters schools, and the superclusters school districts. The three-level variance-components negative binomial model can then be written as

$$y_{ijk}|\mu_{ijk} \sim \text{Poisson}(\mu_{ijk})$$

$$\ln(\mu_{ijk}) = \beta_0 + v_k + u_{jk} + e_{ijk} \tag{17}$$

$$v_k \sim N(0, \sigma_v^2)$$

$$u_{jk} \sim N(0, \sigma_u^2)$$

$$\exp(e_{ijk}) \sim \text{Gamma}\left(\frac{1}{\alpha}, \alpha\right)$$

where $v_k$ is the new supercluster random-intercept effect assumed normally distributed with zero mean and variance $\sigma_v^2$ and all other terms are defined as before.

**Marginal statistics**

With two higher levels there is now interest in reporting the relative importance of both superclusters and clusters as separate sources of response variation. As in the simpler two-level setting we can decompose the marginal variance into level-specific components and we use these to construct different VPC statistics (Supplemental materials S5.2).

The level-3 VPC can then be written as

$$\text{VPC}(3)_{ijk} = \frac{\overbrace{\left(\mu_{ijk}^M\right)^2\{\exp(\sigma_v^2)-1\}}^{\text{level}-3\text{ variance}}}{\underbrace{\left(\mu_{ijk}^M\right)^2\{\exp(\sigma_v^2)-1\}}_{\text{level}-3\text{ variance}}+\underbrace{\left(\mu_{ijk}^M\right)^2\exp(\sigma_v^2)\{\exp(\sigma_u^2)-1\}}_{\text{level}-2\text{ variance}}+\underbrace{\mu_{ijk}^M+\left(\mu_{ijk}^M\right)^2\exp(\sigma_v^2+\sigma_u^2)\alpha}_{\text{level}-1\text{ variance}}} \tag{18}$$

and is interpreted as the proportion of response variance which lies between superclusters. This expression can also be interpreted as an ICC as it also gives the response correlation between two units in the same supercluster, but different clusters.

The level-2 VPC can be written as

$$\mathrm{VPC}(2)_{ijk} = \frac{\overbrace{\left(\mu_{ijk}^{M}\right)^{2}\exp(\sigma_v^2)\{\exp(\sigma_u^2)-1\}}^{\text{level}-2\text{ variance}}}{\underbrace{\left(\mu_{ijk}^{M}\right)^{2}\{\exp(\sigma_v^2)-1\}}_{\text{level}-3\text{ variance}}+\underbrace{\left(\mu_{ijk}^{M}\right)^{2}\exp(\sigma_v^2)\{\exp(\sigma_u^2)-1\}}_{\text{level}-2\text{ variance}}+\underbrace{\mu_{ijk}^{M}+\left(\mu_{ijk}^{M}\right)^{2}\exp(\sigma_v^2+\sigma_u^2)\alpha}_{\text{level}-1\text{ variance}}} \tag{19}$$

and is interpreted as the proportion of response variance which lies within superclusters, between clusters. This VPC does not have a corresponding ICC interpretation. This can be seen by realizing that the implied correlation would be between two units in different superclusters, but the same cluster and this is not a possibility in hierarchical data.

We can also calculate the proportion of response variance collectively attributable to superclusters and clusters. This VPC is calculated by replacing the numerator in the previous equations with the sum of the level-3 and level-2 variance components of the marginal variance.

$$\mathrm{VPC}(2,3)_{ijk} = \frac{\overbrace{\left(\mu_{ijk}^{M}\right)^{2}\{\exp(\sigma_v^2)-1\}}^{\text{level}-3\text{ variance}}+\overbrace{\left(\mu_{ijk}^{M}\right)^{2}\exp(\sigma_v^2)\{\exp(\sigma_u^2)-1\}}^{\text{level}-2\text{ variance}}}{\underbrace{\left(\mu_{ijk}^{M}\right)^{2}\{\exp(\sigma_v^2)-1\}}_{\text{level}-3\text{ variance}}+\underbrace{\left(\mu_{ijk}^{M}\right)^{2}\exp(\sigma_v^2)\{\exp(\sigma_u^2)-1\}}_{\text{level}-2\text{ variance}}+\underbrace{\mu_{ijk}^{M}+\left(\mu_{ijk}^{M}\right)^{2}\exp(\sigma_v^2+\sigma_u^2)\alpha}_{\text{level}-1\text{ variance}}} \tag{20}$$

This expression once again has an ICC interpretation, namely, the response correlation between two units in the same supercluster and the same cluster. This correlation will be stronger than that between two units in the same supercluster but different clusters (Equation 18) as here the units share not only unobserved supercluster influences (captured by the level-3 component in the numerator), but now additionally share unobserved cluster influences (capture by the level-2 component in the numerator). The level-1 VPC – the proportion of the marginal response variance which lies within clusters – is simply equal to one minus this joint level-3 and level-2 VPC.

**Application continued**

Model 3 is a three-level variance-components negative binomial model (Equation 17). The model includes a district random intercept effect to investigate and account for potential superclustering by district. The estimated intercept is effectively unchanged from that of Model 2 the two-level negative binomial model. The model estimates the district variance to be just 0.006, but a likelihood-ratio test confirms that this district variation in student absence rates is nonetheless statistically significant (Model 3 vs. Model 2: $\chi_1^2 = 7.29, p < 0.001$). The school variance in turn decreases by 0.006 from 0.093 to 0.087. The overdispersion parameter is also unchanged. This is expected as overdispersion is treated as a level-1 phenomenon and so is unaffected by whether the school variation is decomposed into separate within and between district components, as we have done here, or not. Here, we immediately see that districts are of little practical or substantive importance when studying student absenteeism as the estimated variance is only around a tenth of the magnitude of the school variance. We can confirm this using the estimated VPC statistics for three-level models (Equations 18, 19 and 20). Districts, schools, and students account for 0.5%, 8% and 92% of the variance in days absent. We will therefore not consider three-level models further in this application. Thus, our preferred model for these data is the two-level negative binomial model which allows for both school clustering and overdispersion due to omitted student characteristics.

## Random-intercept models with covariates for count responses

So far we have explored two- and three-level variance components Poisson and negative binomial models and we have shown how to calculate their VPC expressions. We now explore extending these models to include covariates.

## Model

Thus, for the two-level Poisson and negative binormal models (Equations 1 and 9) we now replace $\beta_0$ with $\mathbf{x}'_{ij}\boldsymbol{\beta}$ (in the three-level versions of these models replace $\beta_0$ with $\mathbf{x}'_{ijk}\boldsymbol{\beta}$) where $\mathbf{x}_{ij}$ denotes the vector of unit- and cluster-level covariates (including the intercept and any cross-level interactions), and $\boldsymbol{\beta}$ is the associated vector of regression coefficients. We note that where units have different exposures (e.g., students observed for different lengths of time) an offset (a covariate with regression coefficient set to 1) measuring the log exposure may be added to $\mathbf{x}_{ij}$ to account for the expected variation that this would produce in the observed counts. The exponentiated regression coefficients can be interpreted as incidence-rate ratios (IRRs) or ratios of expected counts (e.g., $\exp(\beta_1)$ gives the ratio of incidence rates or expected counts when $x_{1ij}$ increases by 1 unit, holding all other covariates constant).

## Conditional and marginal statistics

We must also update the expressions for the conditional and marginal statistics so that they now condition on $\mathbf{x}_{ij}$. Thus, we also replace $\beta_0$ with $\mathbf{x}'_{ij}\boldsymbol{\beta}$ in all these expressions. It should be clear that all these expressions, including the VPC and ICC, now depend on the covariates via the marginal expectation. When an offset is also included, these expressions will additionally be a function of this variable. It is therefore important to inspect how the estimated statistics vary as a function of the marginal expectation and potentially specific covariates. A simple approach is to compute each statistic for each unit in the data based on the covariate values for that unit and we shall follow this approach in the application (some readers may find it helpful to look ahead to Figure 3). Typically, one will also want to summarize these distributions. A natural choice is to report the means of these distributions (or perhaps the medians accompanied by the interquartile ranges to communicate their variability). We shall also follow this approach in the application. Alternatively, one might calculate each statistic at specific meaningful values of the covariates. For example, at the covariate values associated with prototypical units and clusters.

**Application continued**

Recall that the motivation for our application is that school accountability systems are increasingly using student absenteeism rates to hold schools to account and that a particular concern is that differences between schools may in part reflect school intake differences in student demographic and socioeconomic characteristics rather than school influences on students' absenteeism. Thus, the next step in our analysis is to adjust for student demographic and socioeconomic characteristics so that the resulting predicted school random effects provide more meaningful estimates of school influences on students. To improve our interpretation of the model results, we will also recalculate the VPCs and show how they now vary as a function of the covariates.

Table 2 presents the results. Model 4 is a two-level negative binomial model where we include seven student covariates: prior attainment (test score quintile, based on tests taken five years earlier just before the start of secondary schooling), age (season of birth: Autumn, Winter, Spring, Summer; note that grade retention and acceleration are not features of the UK education system so children vary only in their month of birth, not their academic year of birth), gender, ethnicity (white, mixed, Asian, black, other), language (English or not), special educational needs (SEN), and free school meals (FSM). Table S7.1 in the Supplemental materials presents summary statistics.

A likelihood-ratio test confirms that the current model is preferred to its empty counterpart (Model 4 vs. Model 2; $\chi_1^2 = 6609$, $p < 0.001$) and so adding the covariates improves the fit of the model. The current model also continues to be preferred to its single-level counterpart (Model 4 vs. a single-level negative binomial model with covariates; results not shown; $\chi_1^2 = 5653$, $p < 0.001$) indicating that significant school clustering remains even after adjusting for the covariates. Similarly, the current model continues to be preferred to its Poisson counterpart (a two-level random-intercept Poisson model with covariates; results not shown; $\chi_1^2 = 314336$, $p < 0.001$) and so overdispersion also remains in the residual variation, even after adjusting for the covariates.

Examining the parameter estimates, we see that all predictors are statistically significant. Student absenteeism in London is, on average, higher among lower prior attainers, older students, girls, white students, those who speak English as a first language, those with SEN, and those on FSM. These findings agree with recent applied work in this area (Prior et al., 2019). The exponentiated regression coefficients can be interpreted as incidence-rate ratios (IRRs) or ratios of expected counts. Consider, for example, the FSM estimate of 0.377. The estimated IRR is 1.46 (= exp(0.377)). Thus, FSM students are, on average, predicted to miss almost one and a half days for every day missed by otherwise equivalent non-FSM students. This differential is substantial.

In contrary to our expectation, introducing the student characteristics does not lead the school variance to decrease as we move from Model 2 to Model 4. This suggests that school differences in student absenteeism are not predicted by school differences in student characteristics. Indeed, the school variance increases from 0.093 to 0.103. Such increases can also occur when fitting multilevel models to continuous responses and various explanations have been given which are well summarized by Snijders and Bosker (2012, Chapter 7; e.g., cluster-level confounding and covariates with no clustering). We do not explore this issue further here. The overdispersion parameter decreases from 0.877 to 0.782.

Next consider the marginal statistics. The reported values are the sample means of the statistics where we have calculated each statistic for every student in the sample, respecting their covariate values. The VPC depends on the covariates via the marginal expectation (Equation 16). In Figure 3 we therefore plot the predicted VPC against the predicted marginal expectation for all students in the sample (top panel). The figure also plots the sample distribution of predicted marginal expectation values (bottom panel). The VPC increases from approximately 0.085 to 0.105 as we increase the marginal expectation from its minimum to maximum predicted values. In models where the covariates have greater explanatory power, the marginal expectation and therefore VPC would be expected to vary more. Rather than plot the VPC distribution, researchers will often prefer to report a single summary statistic. A natural choice is to report the mean, in our case 0.10. This VPC is slightly higher than that reported in the variance-components model, 0.08, suggesting that the covariates have explained a higher proportion of student variation in the data compared to the school variation which perhaps might be expected given the covariates are defined at the student level.

Table 2 presents sample mean values for the various other marginal statistics. These estimates are broadly similar to those reported for the two-level variance-components negative binomial model (Table 1, Model 2). This again suggests the covariates have low explanatory power. The most important predictors for student absenteeism would appear to lie beyond those available to us here. Again, this finding agrees with recent applied work in this area (Prior et al., 2019). Nonetheless, as detecting outlying schools was one of the motivations for this illustrative application, Figure 4 presents a scatterplot of the predicted school random effects from the current model (Model 4) against those based on the variance-components model (i.e., the null model; Model 2) (left panel). The figure also presents the scatterplot in terms of the ranks of these two sets of predicted school effects (right panel). The corresponding correlations are 0.91-0.92, which are relatively high and show that adjusting for school differences in student composition, at least with respect to the factors examined here, leads to only a modest reordering of schools.

An important question is: What would be the consequences of ignoring the significant and substantial residual overdispersion seen in these data? We can answer this question by fitting a Poisson version of the current two-level random-intercept model (not shown). We have already noted that this model provides a statistically worse fit, but what are the implications for the model results? The regression coefficients are almost identical. The standard errors, however, are approximately one third those in the negative binomial model. The Poisson estimates are therefore spuriously precise because they ignore the overdispersion. Accordingly, they should not be trusted. Note, however, that were the models to additionally include school-level covariates, then the standard errors on these regression coefficient would be expected to be far more similar across the Poisson and negative binomial versions of the model, as the precision with which the coefficients of cluster-level covariates are estimated is determined primarily by the cluster-level variance and the estimate of this parameter is similar in both models.

## Random coefficient models for count responses

In this section, we focus on the two-level random-coefficient negative binomial model (i.e., regression coefficients, not just the intercept, are now allowed to vary across clusters) and the associated VPC and ICC statistics. The parallel VPC and ICC expressions for the Poisson model can again be obtained by setting $\alpha = 0$ in the expressions below. The three-level versions of both models can be easily extended to include random coefficients in a parallel fashion (Supplemental materials S6).

**Model**

The two-level random-coefficient negative binomial model for $y_{ij}$ can be written as follows

$$y_{ij}|\mu_{ij} \sim \text{Poisson}(\mu_{ij})$$
$$\ln(\mu_{ij}) = \mathbf{x}'_{ij}\boldsymbol{\beta} + \mathbf{z}'_{ij}\mathbf{u}_j \quad (21)$$
$$\mathbf{u}_j \sim N(0, \boldsymbol{\Omega}_\mathbf{u})$$
$$\exp(e_{ij}) \sim \text{Gamma}\left(\frac{1}{\alpha}, \alpha\right)$$

where $\mathbf{z}_{ij}$ denotes a vector of unit- and cluster-level covariates (typically an intercept and a subset of the unit-level covariates in $\mathbf{x}_{ij}$) and $\mathbf{u}_j$ is the associated vector of cluster random coefficient effects, assumed multivariate normally distributed with zero mean vector and covariance matrix $\boldsymbol{\Omega}_\mathbf{u}$.

**Marginal statistics**

The expression for the VPC is as in Equation 16, but where $\sigma_u^2$ is replaced by the cluster variance function $\mathbf{z}'_{ij}\boldsymbol{\Omega}_\mathbf{u}\mathbf{z}_{ij}$. For simplicity, consider the special case of a model with a random intercept and one random coefficient associated with the first predictor $x_{1ij}$. In this case we have $\mathbf{z}'_{ij}\mathbf{u}_j = u_{0j} + u_{1j}x_{1ij}$ and $\mathbf{z}'_{ij}\mathbf{\Omega}_{\mathbf{u}}\mathbf{z}_{ij} = \sigma^2_{u0} + 2\sigma_{u01}x_{1ij} + \sigma^2_{u1}x^2_{1ij}$. The expression for the VPC is then given by

$$\text{VPC}_{ij} = \frac{\overbrace{\left(\mu^M_{ij}\right)^2\left\{\exp\left(\sigma^2_{u0}+2\sigma_{u01}x_{1ij}+\sigma^2_{u1}x^2_{1ij}\right)-1\right\}}^{\text{level-2 variance}}}{\underbrace{\left(\mu^M_{ij}\right)^2\left\{\exp\left(\sigma^2_{u0}+2\sigma_{u01}x_{1ij}+\sigma^2_{u1}x^2_{1ij}\right)-1\right\}}_{\text{level-2 variance}}+\underbrace{\mu^M_{ij}+\left(\mu^M_{ij}\right)^2\exp\left(\sigma^2_{u0}+2\sigma_{u01}x_{1ij}+\sigma^2_{u1}x^2_{1ij}\right)\alpha}_{\text{level-1 variance}}} \tag{22}$$

where the marginal expectation is defined as follows

$$\mu^M_{ij} = \exp\left\{\mathbf{x}'_{ij}\boldsymbol{\beta} + \left(\sigma^2_u + 2\sigma_{u01}x_{1ij} + \sigma^2_{u1}x^2_{1ij}\right)/2\right\} \tag{23}$$

The VPC is now a function of both the marginal expectation and the cluster variance function. This expression can also be interpreted as an ICC as it is also the response correlation between two units in the same cluster with the same covariate values. The expression becomes more complex if we wish to consider two units with different covariate values.

## Application continued

Model 4, the previous two-level random-intercept negative binomial model, predicted FSM students in London miss, on average, almost one and a half days for every day missed by otherwise equivalent non-FSM students. In Model 5 we now allow this average effect to vary across schools by introducing a random coefficient for FSM. The estimates are presented in Table 2. The model has two extra parameters, a random slope variance $\sigma^2_{u15}$ and a random intercept-slope covariance $\sigma_{u015}$. A likelihood-ratio test confirms that this model is statistically preferred to the simpler random-intercept model (Model 5 vs. Model 4: $\chi_2^2 = 170, p < 0.001$). The random FSM effects are assumed normally distributed with an estimated mean of 0.372 (IRR = 1.45) and an estimated variance of 0.035. The 95% limits of this distribution are 0.005 and 0.739 (IRR = 1.00, 2.09). Thus, the FSM gap in student absenteeism varies substantially across London schools with FSM students in some schools missing no more days, on average, than otherwise equivalent non-FSM students in these schools, but with FSM students in other schools missing, on average, over two days for every day missed by otherwise equivalent non-FSM students in those schools. This variation is substantial and would seem worthy of further investigation.

Figure 5 explores the relationship between the predicted VPC and the marginal expectation in the same way that we did for simpler random-intercept model (Figure 3). The predicted VPC now varies not only as a function of the marginal expectation, but also as a function of the estimated school variance function $\sigma_{u0}^2 + 2\sigma_{u015}x_{15ij} + \sigma_{u15}^2 x_{15ij}^2$, where $x_{15ij}$ denotes the dummy variable for FSM (Equation 22). The school variance function simplifies to $\sigma_{u0}^2$ for non-FSM students and $\sigma_{u0}^2 + 2\sigma_{u015} + \sigma_{u15}^2$ for FSM students and thus results in two predicted values: 0.116 for non-FSM students and 0.097 for FSM students. This in turn leads to two distinct relationships between the predicted VPC and the marginal expectation and these are plotted in the figure. The relationship for non-FSM students lies above that for FSM students. Thus, for any given predicted number of days absent, the estimated VPC is higher for non-FSM students than for FSM students. This suggests that the influence of school attended on student absenteeism is more pronounced for non-FSM students than FSM students. That is, non-FSM students appear more sensitive and susceptible to their environments with respect to being absent from school than FSM students. This is another interesting finding worthy of further exploration and again highlights the additional insights provided when one calculates VPCs in count models.

## Discussion

In this article, we have derived exact algebraic expressions for variance partition coefficients (VPCs) and intraclass correlation coefficients (ICCs) for multilevel Poisson and negative binomial models (mean dispersion or NB2 version) for count data. We have done this for two- and three-level variance-components and random-intercept versions of these models and we have shown how these can be extended to accommodate models with random coefficients. Parallel derivations are provided in Supplemental materials S3 for two further count models: the Poisson model with an overdispersion random effect and the constant dispersion or NB1 negative binomial model. This work significantly extends that of Stryhn (2006) and Austin et al. (2017) who focus only on the special case of a two-level random-intercept Poisson model. We are not aware of any other publications presenting the VPC (or ICC) expressions for the three more general count models that we consider in this article and that allow for overdispersion and so all of these results are new. We view these extensions as important as overdispersion is a phenomenon which typically occurs in practice.

While the presented VPC expressions have the same general form as those for continuous and categorical responses (when expressed in terms of the latent responses underlying the observed binary, ordinal, or nominal responses), they are nonetheless more complex, as the VPCs are functions of the covariates. One must therefore choose the covariate values at which to evaluate the VPCs. This is also the case in random-coefficient models for continuous and categorical responses, so this is not a new idea for readers familiar with those models (Goldstein et al., 2002). A natural choice is to simply calculate the VPC for each observation in the data and then report the mean (or perhaps the median accompanied by the interquartile range to communicate the variability). There is no reason why such calculations cannot be automated in software as standard postestimation commands and we encourage software developers to do this.

While we have focused on deriving exact algebraic expressions for VPCs in different count models, our research also suggests some more general recommendations for applied researchers analyzing multilevel count data.

First, when count data exhibit overdispersion, as is frequently the case, the standard multilevel Poisson model will prove inadequate. The regression coefficients and cluster variance are not expected to be systematically affected by the overdispersion. The VPC and ICC, however, will be biased upwards in the standard Poisson model and so the degree of clustering and the cluster effects will appear more important than they truly are. Furthermore, the standard errors of the regression coefficient for unit-level (level-1) covariates, but not cluster-level (level-2) covariates, will be biased downwards, potentially leading to Type I errors of inference and incorrect research conclusions.

Second, we have found the Poisson model to sometimes run into computational difficulties when fitted to data with substantial overdispersion. We therefore recommend negative binomial models since these account for overdispersion and so avoid the problems associated with the standard Poisson model. Of these, the mean dispersion is the more widely discussed and is the version we focused on in this article. However, in software where negative binomial models are not implemented, or where the researcher is more familiar with the Poisson model, one alternative would be to simply add a unit-level overdispersion random effect to the Poisson model (Supplemental materials S3). The resulting model is very similar in form to the mean dispersion negative binomial model and would be expected to lead to similar results. However, in contrast to the negative binomial model, this overdispersed Poisson model proves computationally burdensome and may prove prohibitive in large data settings as it requires integrating out the unit-level overdispersion random effect.

Our most important recommendation, however, is that researchers should always explore competing models on their data, just as we have done so here. One learns more from the data doing this than by restricting attention to any one model.

An additional benefit of our work is that we derive and present expressions for not just the VPC, but the marginal expectation, variance, covariance, correlation, and ICC for all four models considered in the article (Supplemental materials S3, S4, S5, S6). These expressions are rarely found in one place and so we hope that our treatment provides a useful resource for readers. For example, researchers can use these expressions when simulating count data or when designing simulation studies to choose true parameter values such that they imply a certain marginal expectation and variance and degree of clustering or overdispersion in the population.

Our research also suggests some areas for further work. First, we have explained how our exact algebraic expressions for the VPC and ICC for two-level models extend to three-level models and the same steps can be followed to further extend these expressions to four and higher-level settings. However, it is less obvious how they extend to cross-classified and multiple membership models and more work is required here.

Second, we have restricted our attention to the standard versions of the Poisson and negative binomial models which use the canonical log link. In some applications, researchers may wish instead to use the identity or power link functions and these will lead to different VPC and ICC expressions. More generally, there are a range of more complex multilevel count models which we have not explored, but for which it should be possible to extend the exact algebraic VPC expressions presented here to incorporate those modelling extensions. These include generalized negative binomial models, with-zeros or zero-inflated models, truncated and censored models, hurdle or two-part models, and mixture models (Cameron and Trivedi, 2013). We leave these extensions for future research.

Third, count data can be modelled as ordinal data, possibly after some grouping of counts to limit the number of observed categories. An advantage of this approach is that we can then appeal to the latent response formulation of ordinal models and their associated VPC and ICCexpressions. We can then model the counts as being due to a latent continuous process that, on crossing progressively higher thresholds, leads to progressively higher values of the observed count. This approach would seem most useful when there are just a few low observed counts, say 0, 1, and 2 or more, as when there are many categories, as there would be in our application, this would lead to many additional threshold parameters to be estimated. One solution would be to merge adjacent categories, but the resulting coarsening of the data will often prove unappealing. Goldstein and Kounali (2009) present an alternative solution which is to apply a smoothing function to the threshold parameters. Relevant to the current work, they show how with sufficient structure imposed on these threshold parameters the ordered probit model reduces to the standard Poisson model. They refer to this formulation of the Poisson model as the “Poison latent normal transformation”. Thus, in future work we shall explore whether this alternative formulation of the Poisson model in terms of a latent continuous response variable also leads to a simple and easy to interpret VPC and ICC expression which can complement those shown here for the observed count data.

## References

Austin, P. C., Stryhn, H., Leckie, G., & Merlo, J. (2017). Measures of clustering and heterogeneity in multilevel Poisson regression analyses of rates/count data. *Statistics in Medicine*, *37,* 572-589.

Browne, W. J., Subramanian, S. V., Jones, K., & Goldstein, H. (2005). Variance partitioning in multilevel logistic models that exhibit overdispersion. *Journal of the Royal Statistical Society: Series A (Statistics in Society)*, *168*, 599-613.

Cameron, A. C., & Trivedi, P. K. (2013). *Regression analysis of count data (2nd ed.)*. Cambridge university press.

Castellano, K. E., & Ho, A. D. (2013). *A Practitioner's Guide to Growth Models*. Council of Chief State School Officers.

DfE (2019a). A guide to absence statistics: March 2019. Department for Education, London.

DfE (2019b). National pupil database. Department for Education, London. URL: https://www.gov.uk/government/collections/national-pupil-database.

EPI (2018). Student absenteeism. Who misses school and how missing school matters for performance. Economic Policy Institute, Washington DC.

Goldstein, H. (1997). Methods in school effectiveness research. *School Effectiveness and School Improvement*, *8*, 369–395.

Goldstein, H. (2011a). Bootstrapping in Multilevel Models. In: Hox J. J., Roberts JK (eds). *Handbook of Advanced Multilevel Analysis*. Routledge: New York, NY,163-171.

Goldstein, H. (2011b). *Multilevel statistical models (4th ed.)*. Chichester, UK: Wiley.

Goldstein, H., Browne, W., & Rasbash, J. (2002). Partitioning variation in multilevel models. *Understanding Statistics: Statistical Issues in Psychology, Education, and the Social Sciences*, *1*, 223-231.

Goldstein, H., & Kounali, D. (2009). Multilevel multivariate modelling of childhood growth, numbers of growth measurements and adult characteristics. *Journal of the Royal Statistical Society: Series A (Statistics in Society)*, *172*, 599-613.

Hox, J. J., Moerbeek, M., & van de Schoot, R. (2017). *Multilevel analysis: Techniques and applications (3rd ed.)*. Routledge.

Leckie, G. (2013a). Cross-Classified Multilevel Models - Concepts. LEMMA VLE Module 12, 1-60. URL: http://www.bristol.ac.uk/cmm/learning/online-course/.

Leckie, G. (2013b). Multiple Membership Multilevel Models - Concepts. LEMMA VLE Module 13, 1-61. URL: http://www.bristol.ac.uk/cmm/learning/online-course/.

Leckie, G. (2013c). Three-Level Multilevel Models - Concepts. LEMMA VLE Module 11, 1-47. URL: http://www.bristol.ac.uk/cmm/learning/online-course/.

Leckie, G., & Goldstein, H. (2009). The limitations of using school league tables to inform school choice. *Journal of the Royal Statistical Society: Series A (Statistics in Society)*, *172*, 835-851.

Leckie, G., & Goldstein, H. (2015). A multilevel modelling approach to measuring changing patterns of ethnic composition and segregation among London secondary schools, 2001-2010. *Journal of the Royal Statistical Society: Series A (Statistics in Society)*, *178*, 405-422.

Leckie, G., & Goldstein, H. (2017). The evolution of school league tables in England 1992-2016: ‘contextual value-added’, ‘expected progress’ and ‘progress 8’. *British Educational Research Journal*, *43,* 193-212.

Leckie, G., & Goldstein, H. (2019). The importance of adjusting for pupil background in school value-added models: A study of Progress 8 and school accountability in England. *British Educational Research Journal*, *45,* 518-537.

Leckie, G., Pillinger, R., Jones, K., & Goldstein, H. (2012). Multilevel modelling of social segregation. *Journal of Educational and Behavioral Statistics*, *37,* 3-30.

Long, J. S. (1997). *Regression models for categorical and limited dependent variables*. Advanced quantitative techniques in the social sciences, 7. Sage.

OECD (2008). Measuring improvements in learning outcomes: Best practices to assess the value-added of schools. Paris: Organisation for Economic Co-operation and Development Publishing & Centre for Educational Research and Innovation.

Prior, L., Goldstein, H., & Leckie, G. (2019). School value-added models for multivariate academic and non-academic student outcomes: A more rounded approach to using student data to inform school accountability arXiv:2001.01996 [stat AP].

Rabe-Hesketh, S., & Skrondal, A. (2012). *Multilevel and longitudinal modeling using Stata (3rd ed., Vol. 2: Categorical, count and survival responses)*. College Station, TX: Stata Press.

Rasbash, J., Leckie, G., Pillinger, R., & Jenkins, J. (2010). Children's educational progress: partitioning family, school and area effects. *Journal of the Royal Statistical Society: Series A (Statistics in Society)*, *173,* 657-682.

Raudenbush, S. W., & Bryk, A. S. (2002). *Hierarchical linear models: Applications and data analysis methods (2nd ed.)*. Thousand Oaks, CA: Sage.

Raudenbush, S. W., & Willms, J. (1995). The estimation of school effects, *Journal of Educational and Behavioral Statistics*, *20*, 307-335.

Skrondal, A., & Rabe-Hesketh, S. (2004). *Generalized latent variable modeling: Multilevel, longitudinal, and structural equation models*. Boca Raton, FL: Chapman & Hall/CRC.

Snijders, T. A. B., & Bosker, R. J. (2012). *Multilevel analysis: An introduction to basic and advanced multilevel modelling (2nd ed.)*. London: Sage.

StataCorp. (2019). Stata statistical software: Release 16 [Computer software]. College Station, TX: StataCorp LLC. URL: http://www.stata.com.

Stryhn, H., Sanchez, J., Morley, P., Booker, C., & Dohoo, I. R. (2006). Interpretation of variance parameters in multilevel Poisson models. In *Proceedings of the 11th Symposium of the International Society for Veterinary Epidemiology an Economics*, 702–704. Cairns, Australia.

*Table* 1. Estimates for variance components models fitted to the student absenteeism data.

| | Model 1: Two-level variance-components Poisson model | Model 2: Two-level variance-components negative binomial model | Model 3: Three-level variance-components negative binomial model |
|---|---|---|---|
| | | Parameter estimates | |
| $\beta_0$ – Intercept | 2.085 (0.015) | 2.088 (0.015) | 2.086 (0.020) |
| $\sigma_v^2$ – District variance | | | 0.006 (0.003) |
| $\sigma_u^2$ – School variance | 0.100 (0.007) | 0.093 (0.007) | 0.087 (0.007) |
| $\alpha$ – Overdispersion | | 0.877 (0.005) | 0.877 (0.005) |
| | | Marginal statistics | |
| Marginal expectation | 8.46 | 8.45 | 8.44 |
| Marginal variance | 15.98 | 84.10 | 83.79 |
| District (level-3) component | | | 0.42 |
| School (level-2) component | 7.52 | 6.95 | 6.50 |
| Student (level-1) component | 8.46 | 77.15 | 76.87 |
| District (level-3) VPC | | | 0.005 |
| School (level-2) VPC | 0.47 | 0.08 | 0.08 |
| Student (level-1) VPC | 0.53 | 0.92 | 0.92 |
| | | Fit statistics | |
| Deviance | 785142 | 422046 | 422039 |

*Note*. Number of districts: $K = 32$; number of schools: $J = 434$; number of students: $N = 66{,}955$. Standard errors in parentheses.

*Table* 2. Estimates for two-level random-intercept and -coefficient models fitted to the student absenteeism data.

| | Model 4: Two-level random-intercept negative binomial model | Model 5: Two-level random-coefficient negative binomial model |
|---|---|---|
| | Parameter estimates | |
| $\beta_0$ – Intercept | 2.126 (0.021) | 2.126 (0.021) |
| $\beta_1$ – Prior attainment: Quintile 2 | -0.051 (0.012) | -0.048 (0.012) |
| $\beta_2$ – Prior attainment: Quintile 3 | -0.118 (0.012) | -0.116 (0.012) |
| $\beta_3$ – Prior attainment: Quintile 4 | -0.222 (0.012) | -0.219 (0.012) |
| $\beta_4$ – Prior attainment: Quintile 5 | -0.330 (0.014) | -0.326 (0.014) |
| $\beta_5$ – Age: Spring born | 0.026 (0.011) | 0.026 (0.011) |
| $\beta_6$ – Age: Winter born | 0.077 (0.011) | 0.078 (0.011) |
| $\beta_7$ – Age: Autumn born | 0.112 (0.011) | 0.112 (0.011) |
| $\beta_8$ – Female | 0.122 (0.009) | 0.122 (0.009) |
| $\beta_9$ – Ethnicity: Mixed | -0.073 (0.014) | -0.074 (0.014) |
| $\beta_{10}$ – Ethnicity: Asian | -0.194 (0.013) | -0.198 (0.013) |
| $\beta_{11}$ – Ethnicity: Black | -0.422 (0.011) | -0.421 (0.011) |
| $\beta_{12}$ – Ethnicity: Other | -0.194 (0.017) | -0.195 (0.017) |
| $\beta_{13}$ – Language not English | -0.244 (0.009) | -0.242 (0.009) |
| $\beta_{14}$ – Special Educational Needs (SEN) | 0.267 (0.011) | 0.267 (0.011) |
| $\beta_{15}$ – Free school meal (FSM) | 0.377 (0.008) | 0.372 (0.013) |
| $\sigma^2_{u0}$ – School intercept variance | 0.103 (0.007) | 0.116 (0.009) |
| $\sigma^2_{u15}$ – School FSM variance | | 0.035 (0.005) |
| $\sigma^2_{u0,15}$ – School intercept-FSM covariance | | -0.027 (0.005) |
| $\alpha$ – Overdispersion | 0.782 (0.005) | 0.775 (0.005) |
| | Marginal statistics | |
| Marginal expectation | 8.50 | 8.52 |
| Marginal variance | 87.05 | 87.20 |
| School (level 2) variance | 8.71 | 9.04 |
| Student (level 1) variance | 78.34 | 78.17 |
| School (level 2) VPC | 0.10 | 0.10 |
| Student (level 1) VPC | 0.90 | 0.90 |
| | Fit statistics | |
| Deviance | 415438 | 415268 |

*Note*. Number of schools: $J = 434$; number of students: $N = 66{,}955$. Reference categories. Prior attainment: Quintile 1 (lowest prior attainment); Age: Summer born (youngest in year); Ethnicity: White. Standard errors in parentheses. Sample average values are reported for the marginal statistics as each statistic is a function of the covariates.

*Figure* 1. Distribution of number of days absent over the school year.

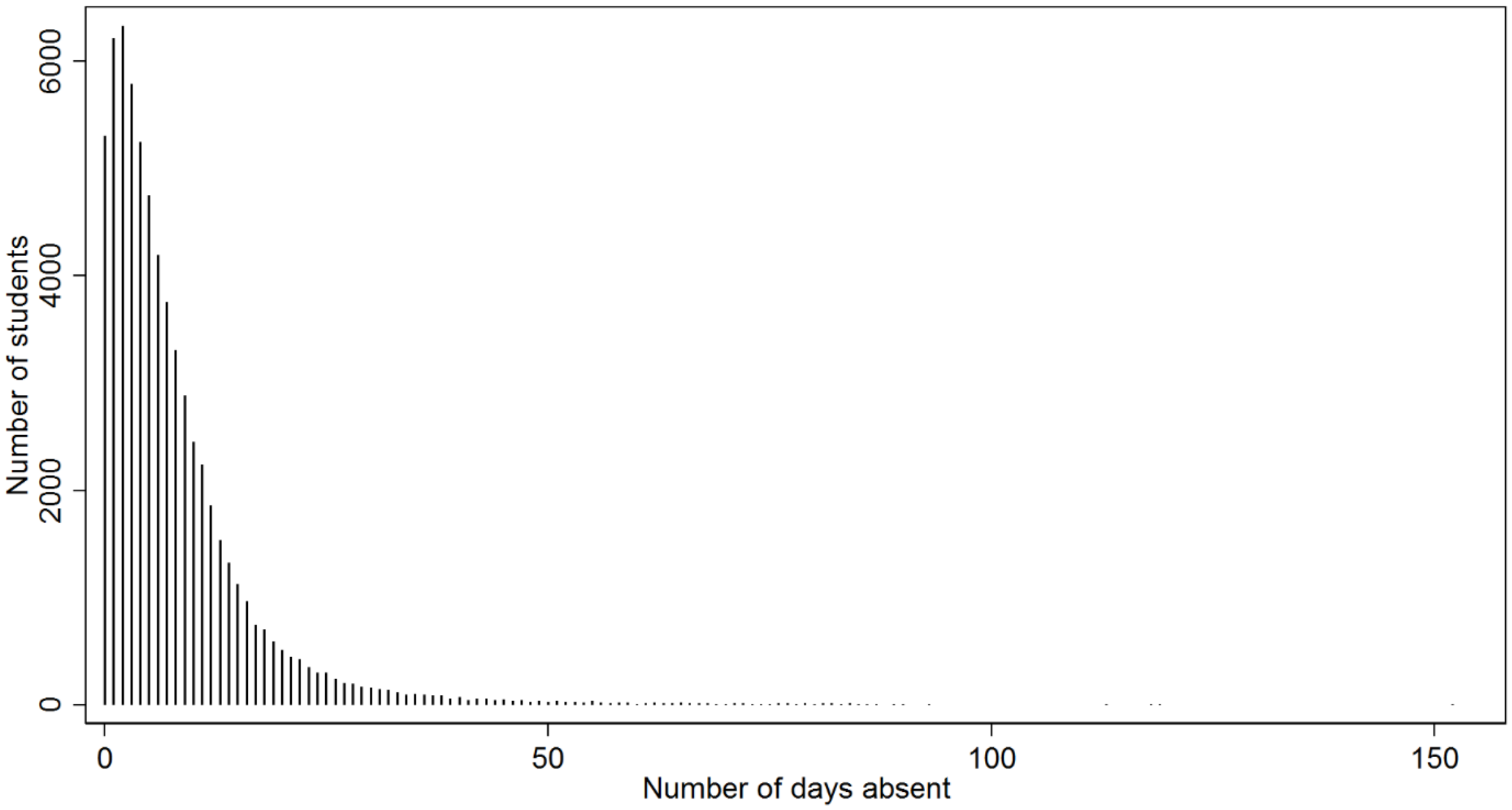

*Figure* 2. Mean number of days absent by district (left panel) and school (right panel). The horizontal line depicts the student sample mean.

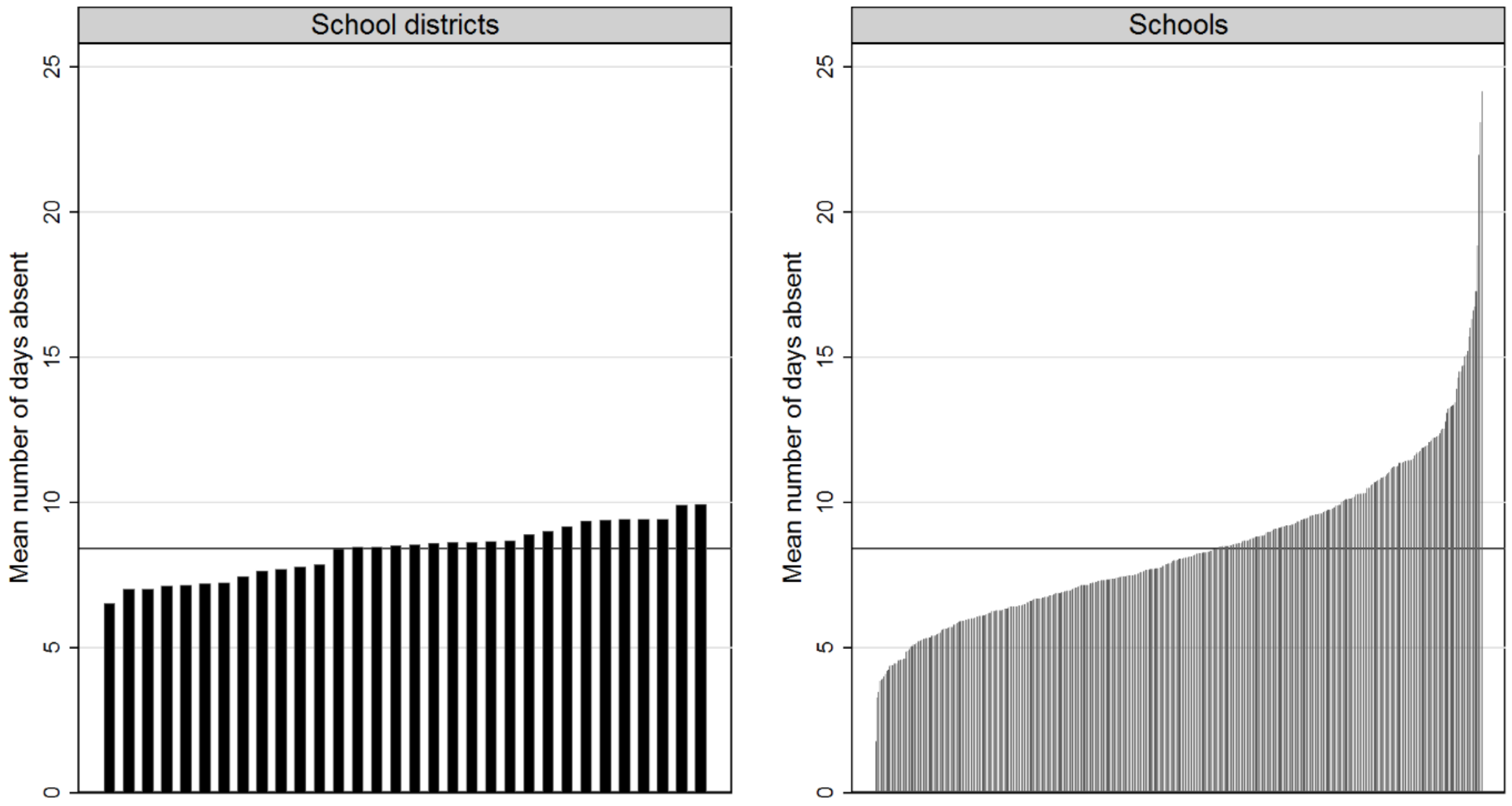

*Figure* 3. Relationship between the predicted school VPC and marginal expectation (top panel). Distribution of predicted student-level marginal expectation values (bottom panel). Plots are based on Model 2.

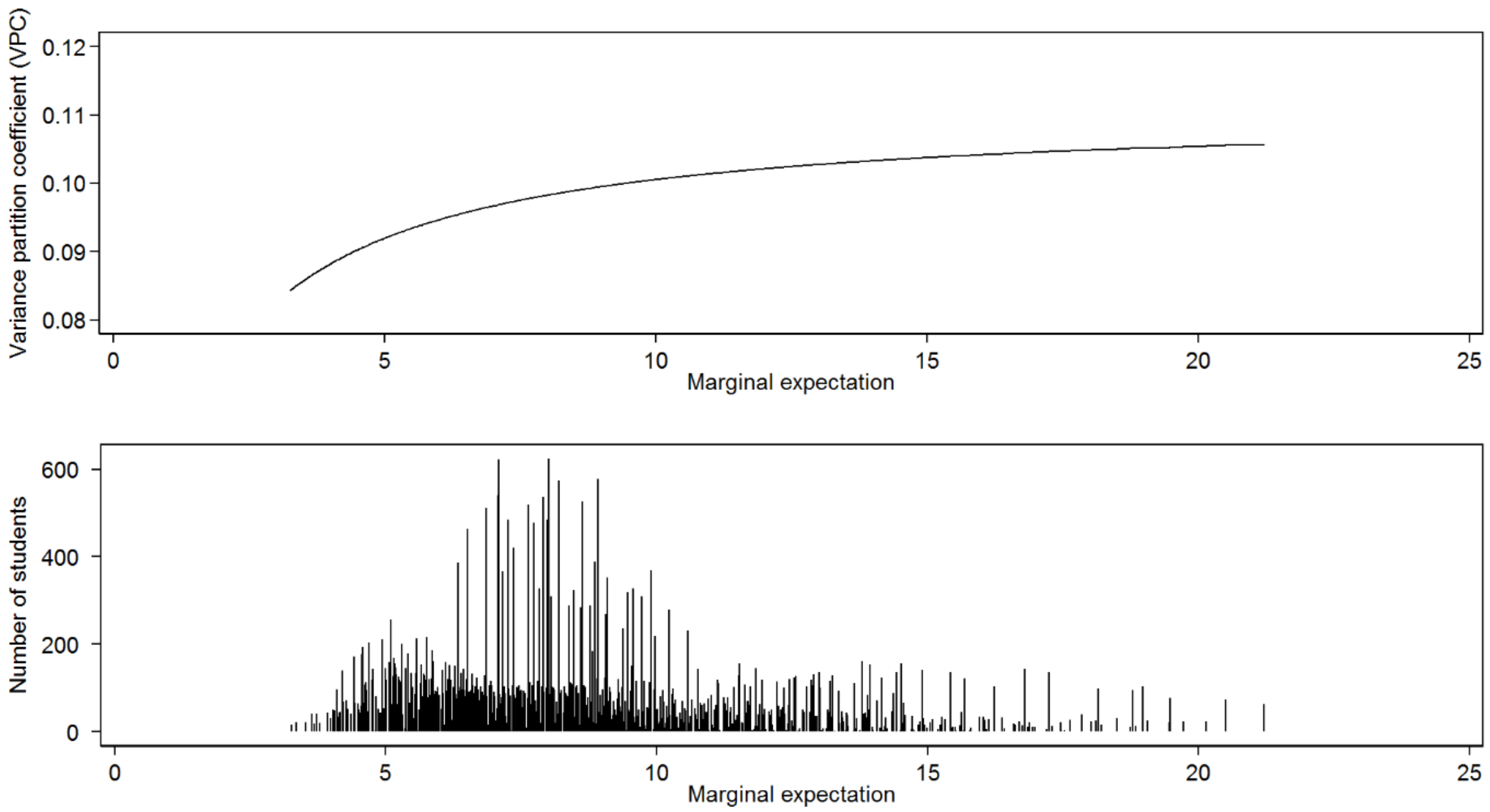

*Figure* 4. Scatterplots of predicted school effect values (left panel) and ranks (right panel) from unadjusted and covariate adjusted two-level random-intercept negative binomial models (Models 2 and 4).

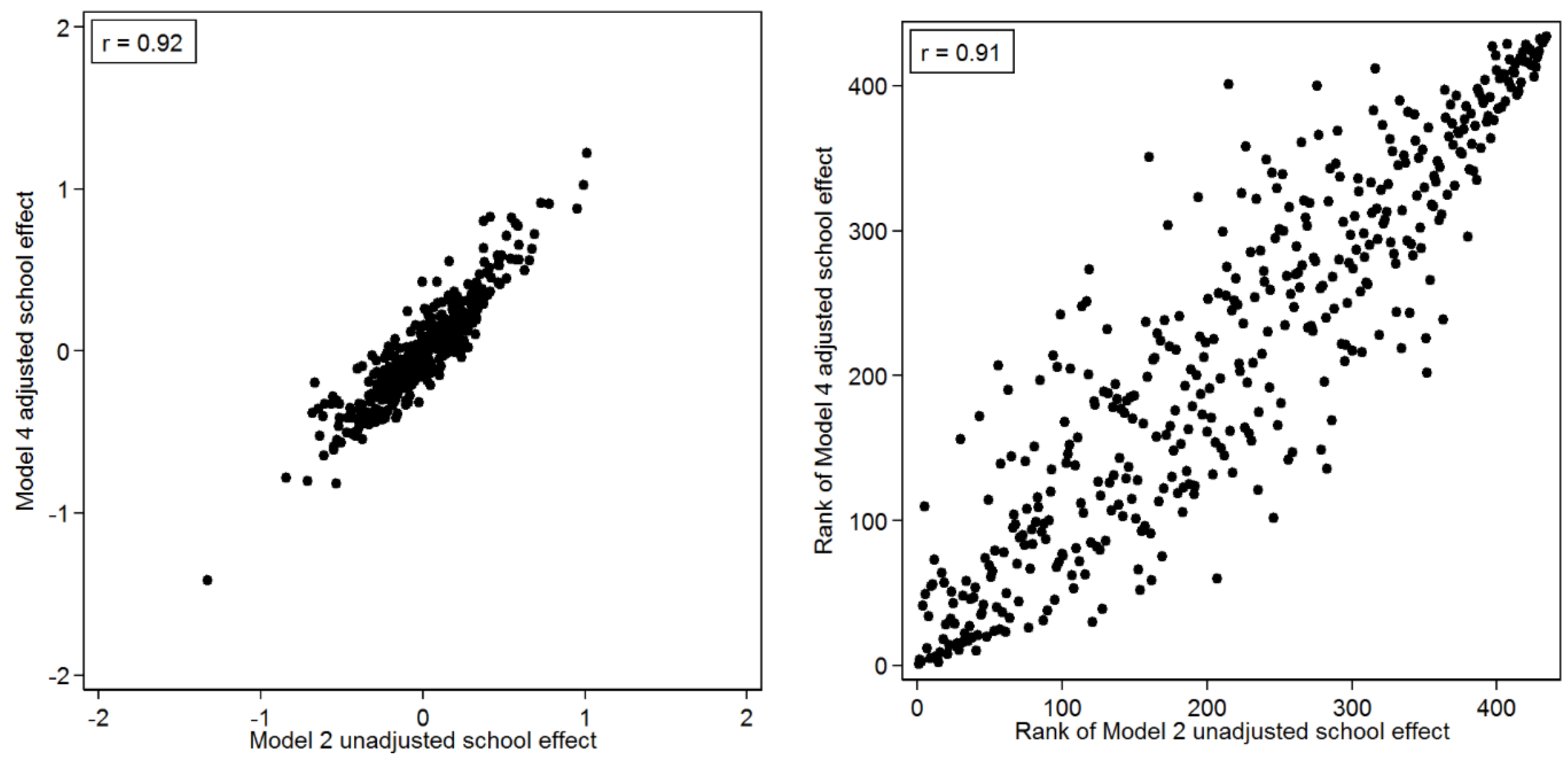

*Figure* 5. Relationship between the predicted school VPC and marginal expectation by student FSM status (top panel). Distribution of predicted student-level marginal expectation values by student FSM status (bottom panel). Plots are based on Model 5.

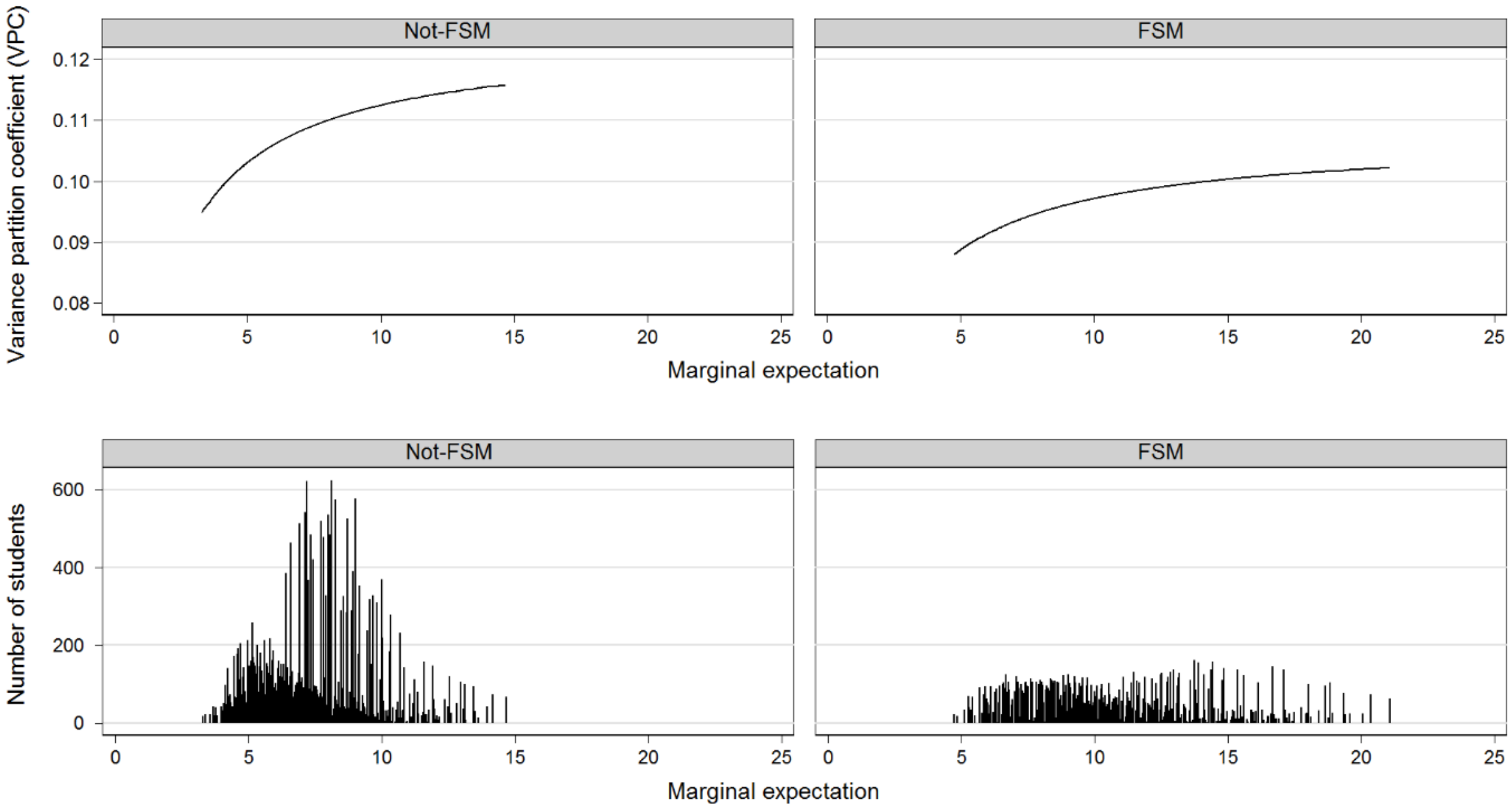

# Supplemental materials: Partitioning variation in multilevel models for count data

## S1. Review of two-level models for continuous responses

In this section we review two-level variance-components, random-intercept and random-coefficient models for continuous responses and their associated VPC and ICC expressions. The variance components model is simply the special case of the random-intercept model where we include no covariates (the "null" or "empty" model). To aid the discussion of each model, we present the conditional (given the covariates and random effects) and marginal (given the covariates but averaged over the random effects) expectations, variances, covariances and correlations of the continuous response. Reviewing this material will help readers better understand the more complex derivations we present in the article for multilevel count response models. When reading this section, readers may find it helpful to have an example application in mind such as the study of student math scores and student and school characteristics that we described in the Introduction.

### S1.1. Variance-components model for continuous responses

The two-level variance components model for continuous responses simply decomposes the total response variance into separate variance components at level-1 and level-2 in the data hierarchy.

#### Model

Let $y_{ij}$ denote the continuous response for unit $i$ ($i = 1, \dots, n_j$) in cluster $j$ ($j = 1, \dots, J$). In terms of the math score study, the units would be students, the clusters schools, and the continuous response would be the student math score. We can then write the model for $y_{ij}$ as follows

$$y_{ij} = \beta_0 + u_j + e_{ij} \tag{S1.1}$$

$$u_j \sim N(0, \sigma_u^2)$$

$$e_{ij} \sim N(0, \sigma_e^2)$$

where $\beta_0$ denotes the intercept, $u_j$ is the cluster random intercept effect, assumed normally distributed with zero mean and between-cluster variance $\sigma_u^2$, and $e_{ij}$ is the residual, assumed normally distributed with zero mean and within-cluster variance $\sigma_e^2$. The response variance (total, overall or marginal) is then given by the summation $\sigma_u^2 + \sigma_e^2$.

The VPC measures the proportion of response variance which lies between clusters and is calculated as the ratio of the between-cluster variance to the total response variance

$$\text{VPC} = \frac{\sigma_u^2}{\sigma_u^2 + \sigma_e^2} \tag{S1.2}$$

The ICC measures the expected correlation between two units from the same cluster and is again calculated as the ratio of the between-cluster variance to the total response variance

$$\text{ICC} = \frac{\sigma_u^2}{\sigma_u^2 + \sigma_e^2} \tag{S1.3}$$

The above expressions for the VPC and ICC are widely used by applied researchers. However, they are typically presented with little explanation.

### Conditional inference

The conditional expectation of $y_{ij}$ (given $u_j$) is given by

$$\mu_{ij}^C \equiv \mathrm{E}(y_{ij}|u_j) = \beta_0 + u_j \tag{S1.4}$$

and measures the mean response in each cluster $j$.

The conditional variance is given by

$$\omega_{ij}^C \equiv \mathrm{Var}(y_{ij}|u_j) = \sigma_e^2 \tag{S1.5}$$

and measures the variance of the responses around the mean response in each cluster.

The conditional covariance and correlation between the response measurements on two units $i$ and $i'$ for the same cluster $j$, $y_{ij}$ and $y_{i'j}$, are zero and so the responses are assumed independent in each cluster. This assumption is sometimes referred to as the conditional independence or local independence assumption. The conditional covariance and correlation between response measurements on two units from two different clusters are also assumed equal to zero.

### Marginal inference

The marginal expectation of $y_{ij}$ (now averaged over $u_j$) is given by

$$\mu_{ij}^M \equiv \mathrm{E}(y_{ij}) = \beta_0 \tag{S1.6}$$

and measures the overall mean response.

The marginal variance of $y_{ij}$ is given by

$$\omega_{ij}^{M} \equiv \mathrm{Var}(y_{ij}) = \sigma_u^2 + \sigma_e^2 \tag{S1.7}$$

and measures the total or overall response variance and is equal to the sum of the variance components.

The marginal covariance between $y_{ij}$ and $y_{i'j}$ is given by

$$\mathrm{Cov}(y_{ij}, y_{i'j}) = \sigma_u^2 \tag{S1.8}$$

The marginal correlation between $y_{ij}$ and $y_{i'j}$ can then be calculated as

$$\mathrm{ICC}_{ij,i'j} \equiv \mathrm{Corr}(y_{ij}, y_{i'j}) = \frac{\mathrm{Cov}(y_{ij}, y_{i'j})}{\sqrt{\mathrm{Var}(y_{ij})}\sqrt{\mathrm{Var}(y_{i'j})}} = \frac{\sigma_u^2}{\sigma_u^2 + \sigma_e^2} \tag{S1.9}$$

This marginal correlation can be interpreted as the ICC as it is the response correlation between two units in the same cluster. In contrast, the marginal covariance and correlation between response measurements on two units from two different clusters is equal to zero as the units do not share a cluster random effect.

The VPC is defined as the proportion of the marginal response variance which lies between clusters. Thus, we can only calculate the VPC after we have partitioned the marginal variance into level-specific components. The expression for the marginal variance (Equation S1.7) does just that. The expression for the VPC is therefore given by

$$\mathrm{VPC}_{ij} = \frac{\overbrace{\sigma_u^2}^{\text{level}-2\text{ variance}}}{\underbrace{\sigma_u^2}_{\text{level}-2\text{ variance}} + \underbrace{\sigma_e^2}_{\text{level}-1\text{ variance}}} \tag{S1.10}$$

and this expression is identical to that for the marginal correlation or ICC given in Equation S1.9. We see the usual result whereby the higher the between-cluster variance $\sigma_u^2$, the higher the VPC, but the higher the within-cluster variance $\sigma_e^2$, the lower the VPC. This VPC should strictly be referred to as the level-2 VPC. The level-1 VPC – the proportion of the marginal response variance which lies within clusters – is then equal to one minus the level-2 VPC.

### S1.2. Two-level random-intercept models with covariates for continuous responses

So far, we have reviewed the two-level variance-component model for continuous responses and we have shown how to calculate the VPC and ICC expressions for this model. We now extend this to the two-level random-intercept model with covariates and discuss the implications this has for calculating and interpreting the VPC and ICC.

#### Model

The model can be written as follows

$$y_{ij} = \mathbf{x}_{ij}'\boldsymbol{\beta} + u_j + e_{ij} \tag{S1.11}$$

$$u_j \sim N(0, \sigma_u^2)$$

$$e_{ij} \sim N(0, \sigma_e^2)$$

where $\mathbf{x}_{ij}$ denotes the vector of unit- and cluster-level covariates (including the intercept and any cross-level interactions), and $\boldsymbol{\beta}$ is the associated vector of regression coefficients. In terms of our math score study, the unit- and cluster-level covariates would be student and school characteristics believed to be relevant for explaining the variation in students' scores.

It may help the reader to consider the special case of a model with a random intercept and one covariate $x_{1ij}$. In terms of our math score study this might be student SES. Such a model can then be written as

$$y_{ij} = \beta_0 + \beta_1 x_{1ij} + u_j + e_{ij} \quad \text{(S1.12)}$$

$$u_j \sim N(0, \sigma_u^2)$$

$$e_{ij} \sim N(0, \sigma_e^2)$$

where $\beta_0$ denotes the intercept and $\beta_1$ denotes the slope coefficient on $x_{1ij}$.

The expressions for the VPC and ICC in the random-intercept model with covariates are the same as in the variance-components model. Thus, they again coincide to give

$$\text{VPC} \equiv \text{ICC} = \frac{\sigma_u^2}{\sigma_u^2 + \sigma_e^2} \quad \text{(S1.13)}$$

Here, the interpretation of these statistics is in terms of the response variation which remains after adjusting for the covariate. Thus, the VPC and ICC now answer the questions: What proportion of the residual math score variation lies between schools? And: What is the residual math score correlation between two students from the same school. As with the variance-components model, applied researchers widely report the VPC and ICC after fitting random-intercept models with covariates. However, they again typically present these statistics without derivation and so we review their derivation here.

The introduction of covariates mean that we must update the VPC and ICC expressions ( as well as the various other conditional and marginal statistics) given above for the variance-components model so that they now additionally condition on $\mathbf{x}_{ij}$. This then leads us to replace $\beta_0$ with $\mathbf{x}'_{ij}\boldsymbol{\beta}$ in any expression where $\beta_0$ was shown previously. As we shall see below, this affects only the expressions for the conditional and marginal expectations as $\beta_0$ does not appear in any of the other expressions. This is why the expressions for the VPC and ICC remain the same as those for the variance-components model (Equation S1.13 takes the same form as Equation S1.3). The interpretation of each statistic, however, does change upon the introduction of covariates as indicated above. Specifically, the interpretation of each statistic is now in terms of making statements about the variation in the response conditional on the covariates. Thus, these statistics now allow us to make statements about the response variation which remains once we have adjusted for the covariates (i.e., the unexplained or residual variation).

### Conditional statistics

The conditional expectation of $y_{ij}$ (now given $\mathbf{x}_{ij}$ and $u_j$) is given by

$$\mu^C_{ij} \equiv \mathrm{E}(y_{ij}|\mathbf{x}_{ij}, u_j) = \mathbf{x}'_{ij}\boldsymbol{\beta} + u_j \tag{S1.14}$$

and so is now a function of the covariates $\mathbf{x}_{ij}$. The conditional variance is given by

$$\omega_{ij}^{C} \equiv \mathrm{Var}\left(y_{ij} \middle| \mathbf{x}_{ij}, u_j\right) = \sigma_e^2 \tag{S1.15}$$

and so takes the same form as before. The conditional covariance and correlation between $y_{ij}$ and $y_{i'j}$ are again assumed to be zero.

## Marginal statistics

The marginal expectation of $y_{ij}$ (again given $\mathbf{x}_{ij}$ but now averaged over $u_j$) is given by

$$\mu_{ij}^{M} \equiv \mathrm{E}\left(y_{ij} \middle| \mathbf{x}_{ij}\right) = \mathbf{x}_{ij}'\boldsymbol{\beta} \tag{S1.16}$$

and so is now a function of the covariates $\mathbf{x}_{ij}$. The marginal variance of $y_{ij}$ is given by

$$\omega_{ij}^{M} \equiv \mathrm{Var}\left(y_{ij} \middle| \mathbf{x}_{ij}\right) = \sigma_u^2 + \sigma_e^2 \tag{S1.17}$$

and so is again the sum of the variance components. Note, however, these variance components no longer sum to give the response variance. Rather they sum to give the response variance having adjusted for the covariates. The marginal covariance between $y_{ij}$ and $y_{i'j}$ (given $\mathbf{x}_{ij}$ and $\mathbf{x}_{i'j}$ but averaged over $u_j$) is given by

$$\mathrm{Cov}\left(y_{ij}, y_{i'j} \middle| \mathbf{x}_{ij}, \mathbf{x}_{i'j}\right) = \sigma_u^2 \tag{S1.18}$$

and so also takes the same form as before. The marginal correlation between $y_{ij}$ and $y_{i'j}$ can then be calculated as

$$\text{ICC}_{ij,i'j} \equiv \text{Corr}\left(y_{ij}, y_{i'j} \middle| \mathbf{x}_{ij}, \mathbf{x}_{i'j}\right) = \frac{\text{Cov}\left(y_{ij}, y_{i'j} \middle| \mathbf{x}_{ij}, \mathbf{x}_{i'j}\right)}{\sqrt{\text{Var}\left(y_{ij} \middle| \mathbf{x}_{ij}\right)}\sqrt{\text{Var}\left(y_{i'j} \middle| \mathbf{x}_{i'j}\right)}} = \frac{\sigma_u^2}{\sigma_u^2 + \sigma_e^2} \qquad \text{(S1.19)}$$

and so takes the same form as before.

The VPC (given $\mathbf{x}_{ij}$) is derived in the usual way, as the ratio of the level-2 component of the marginal variance divided by the summation of the level-2 and -1 components. The expression for the marginal variance takes the same form as in the variance-component case and so the expression for the VPC also takes the same form and is given again by

$$\text{VPC}_{ij} = \frac{\overbrace{\sigma_u^2}^{\text{level-2 variance}}}{\underbrace{\sigma_u^2}_{\text{level-2 variance}} + \underbrace{\sigma_e^2}_{\text{level-1 variance}}} \qquad \text{(S1.20)}$$

Thus, the expressions for the VPC and the ICC are the same as one another and are also the same in both the variance-components model and in the random-intercept model with covariates.

### S1.3 Two-level random-coefficient models

Next, we consider the two-level random-coefficient model. We then discuss the implications this has for the calculation and interpretation of the VPC and ICC. The two-level random-coefficient model extends the two-level random-intercept model by allowing not just the intercept, but one or more of the slope coefficients to vary randomly across clusters.

## Model

The model can be written as follows

$$y_{ij} = \mathbf{x}'_{ij}\boldsymbol{\beta} + \mathbf{z}'_{ij}\mathbf{u}_j + e_{ij} \tag{S1.21}$$

$$\mathbf{u}_j \sim N(0, \boldsymbol{\Omega}_\mathbf{u})$$

$$e_{ij} \sim N(0, \sigma_e^2)$$

where $\mathbf{z}_{ij}$ denotes a vector of unit- and cluster-level covariates (typically an intercept and a subset of the unit-level covariates in $\mathbf{x}_{ij}$) and $\mathbf{u}_j$ is the associated vector of cluster random coefficient effects, assumed multivariate normally distributed with zero mean vector and covariance matrix $\boldsymbol{\Omega}_\mathbf{u}$.

It may help the reader to consider the special case of a model with a random-intercept and one covariate $x_{1ij}$ and where we enter that covariate with a random coefficient. Thus, we enter the covariate both in the fixed part of the model as $x_{1ij}$ and in the random part of the model as $z_{1ij} \equiv x_{1ij}$. In terms of our math score example application this might be student SES. The model can be written as

$$y_{ij} = \beta_0 + \beta_1 x_{1ij} + u_{0j} + u_{1j}x_{1ij} + e_{ij} \tag{S1.22}$$

$$\begin{pmatrix} u_{0j} \\ u_{1j} \end{pmatrix} \sim N\left\{\begin{pmatrix} 0 \\ 0 \end{pmatrix}, \begin{pmatrix} \sigma_{u0}^2 & \\ \sigma_{u01} & \sigma_{u1}^2 \end{pmatrix}\right\}$$

$$e_{ij} \sim N(0, \sigma_e^2)$$

where $\beta_0$ denotes the intercept and $\beta_1$ denotes the slope coefficient on $x_{1ij}$, $u_{0j}$ and $u_{1j}$ are the cluster random intercept and slope effects, assumed bivariate normally distributed with zero means and between-cluster intercept and slope variances and covariance $\sigma^2_{u0}$, $\sigma^2_{u1}$ and $\sigma_{u01}$, and $e_{ij}$ is the usual residual, assumed normally distributed with zero mean and within-cluster variance $\sigma^2_e$.

As we shall show below, the expressions for the VPC and ICC can also be derived for this more complex model.

## Conditional statistics

The conditional expectation of $y_{ij}$ (now given $\mathbf{x}_{ij}$, $\mathbf{z}_{ij}$ and $\mathbf{u}_j$) is given by

$$\mu^C_{ij} \equiv \mathrm{E}(y_{ij}|\mathbf{x}_{ij}, \mathbf{z}_{ij}, \mathbf{u}_j) = \mathbf{x}'_{ij}\boldsymbol{\beta} + \mathbf{z}'_{ij}\mathbf{u}_j \tag{S1.23}$$

The conditional variance is given by

$$\omega^C_{ij} \equiv \mathrm{Var}(y_{ij}|\mathbf{x}_{ij}, \mathbf{z}_{ij}, \mathbf{u}_j) = \sigma^2_e \tag{S1.24}$$

and so takes the same form as in the simple variance-components and random-intercept models. The conditional covariance and correlation between $y_{ij}$ and $y_{i'j}$ are again assumed zero.

## Marginal statistics

The marginal expectation of $y_{ij}$ (again given $\mathbf{x}_{ij}$ and $\mathbf{z}_{ij}$ but now averaged over $\mathbf{u}_j$) is given by

$$\mu_{ij}^{M} \equiv \mathrm{E}\left(y_{ij} \middle| \mathbf{x}_{ij}, \mathbf{z}_{ij}\right) = \mathbf{x}_{ij}'\boldsymbol{\beta} \tag{S1.25}$$

which is the same as in the random-intercept case. The marginal variance of $y_{ij}$ is given by

$$\omega_{ij}^{M} \equiv \mathrm{Var}\left(y_{ij} \middle| \mathbf{x}_{ij}, \mathbf{z}_{ij}\right) = \mathbf{z}_{ij}'\boldsymbol{\Omega}_{\mathbf{u}}\mathbf{z}_{ij} + \sigma_e^2 \tag{S1.26}$$

which now takes a more complex form than that seen previously. The first term $\mathbf{z}_{ij}'\boldsymbol{\Omega}_{\mathbf{u}}\mathbf{z}_{ij}$ is referred to as a cluster-level variance function as the response variance between clusters is now modelled as a function of the covariates $\mathbf{z}_{ij}$ (heteroskedasticity). It follows that the marginal variance is also a function of the covariates $\mathbf{z}_{ij}$ since to obtain this we simply add on the second term $\sigma_e^2$.

The marginal covariance between $y_{ij}$ and $y_{i'j}$ (given $\mathbf{x}_{ij}$, $\mathbf{x}_{i'j}$, $\mathbf{z}_{ij}$, $\mathbf{z}_{i'j}$ but averaged over $\mathbf{u}_j$) is given by

$$\mathrm{Cov}\left(y_{ij}, y_{i'j} \middle| \mathbf{x}_{ij}, \mathbf{x}_{i'j}, \mathbf{z}_{ij}, \mathbf{z}_{i'j}\right) = \mathbf{z}_{ij}'\boldsymbol{\Omega}_{\mathbf{u}}\mathbf{z}_{i'j} \tag{S1.27}$$

The marginal correlation between $y_{ij}$ and $y_{i'j}$ can then be calculated in the usual way to give

$$\begin{aligned}\mathrm{ICC}_{ij,i'j} \equiv \mathrm{Corr}\left(y_{ij}, y_{i'j} \middle| \mathbf{x}_{ij}, \mathbf{x}_{i'j}, \mathbf{z}_{ij}, \mathbf{z}_{i'j}\right) &= \frac{\mathrm{Cov}\left(y_{ij}, y_{i'j} \middle| \mathbf{x}_{ij}, \mathbf{x}_{i'j}, \mathbf{z}_{ij}, \mathbf{z}_{i'j}\right)}{\sqrt{\mathrm{Var}\left(y_{ij} \middle| \mathbf{x}_{ij}, \mathbf{z}_{ij}\right)}\sqrt{\mathrm{Var}\left(y_{i'j} \middle| \mathbf{x}_{i'j}, \mathbf{z}_{i'j}\right)}} \\ &= \frac{\mathbf{z}_{ij}'\boldsymbol{\Omega}_{\mathbf{u}}\mathbf{z}_{i'j}}{\sqrt{\mathbf{z}_{ij}'\boldsymbol{\Omega}_{\mathbf{u}}\mathbf{z}_{ij} + \sigma_e^2}\sqrt{\mathbf{z}_{i'j}'\boldsymbol{\Omega}_{\mathbf{u}}\mathbf{z}_{i'j} + \sigma_e^2}}\end{aligned} \tag{S1.28}$$

The ICC is therefore a function of $\mathbf{z}_{ij}$ and $\mathbf{z}_{i'j}$. To simplify matters, consider the correlation between two units $i$ and $i'$ within the same cluster $j$ who have the same covariate values $\mathbf{z}_{ij} = \mathbf{z}_{i'j}$. The expression for the ICC then simplifies to

$$\text{ICC}_{ij,i'j} \equiv \text{Corr}\left(y_{ij}, y_{i'j} \middle| \mathbf{x}_{ij}, \mathbf{x}_{i'j}, \mathbf{z}_{ij} = \mathbf{z}_{i'j}\right) = \frac{\mathbf{z}'_{ij}\boldsymbol{\Omega}_\mathbf{u}\mathbf{z}_{ij}}{\mathbf{z}'_{ij}\boldsymbol{\Omega}_\mathbf{u}\mathbf{z}_{ij}+\sigma_e^2} \quad \text{(S1.29)}$$

Now reconsider our example model with only one covariate $x_{1ij} \equiv z_{1ij}$. For this model, the ICC (Equation S1.28) becomes

$$\text{ICC}_{ij,i'j} \equiv \text{Corr}\left(y_{ij}, y_{i'j} \middle| x_{1ij}, x_{1i'j}\right) = \frac{\sigma_{u0}^2+\sigma_{u01}\left(x_{1ij}+x_{1i'j}\right)+\sigma_{u1}^2 x_{1ij}x_{1i'j}}{\sqrt{\sigma_{u0}^2+2\sigma_{u01}x_{1ij}+\sigma_{u1}^2x_{1ij}^2+\sigma_e^2}\sqrt{\sigma_{u0}^2+2\sigma_{u01}x_{1i'j}+\sigma_{u1}^2x_{1i'j}^2+\sigma_e^2}} \quad \text{(S1.30)}$$

The magnitude of the residual response correlation (i.e., the response correlation having adjusted for the covariate) between units $i$ and $i'$ for the same cluster $j$ therefore depends on the value of the covariate with the random coefficient for the two units being compared, $x_{1ij}$ and $x_{1i'j}$. Restricting our attention to the special case when $x_{1ij} \equiv x_{1i'j}$ leads this ICC expression (Equation S1.29) to simplify to

$$\text{ICC}_{ij,i'j} \equiv \text{Corr}\left(y_{ij}, y_{i'j} \middle| x_{1ij}\right) = \frac{\sigma_{u0}^2+2\sigma_{u01}x_{1ij}+\sigma_{u1}^2x_{1ij}^2}{\sigma_{u0}^2+2\sigma_{u01}x_{1ij}+\sigma_{u1}^2x_{1ij}^2+\sigma_e^2} \quad \text{(S1.31)}$$

The ICC is then a function of just one variable $x_{1ij}$ and this can be more easily plotted and evaluated.

The VPC (given $\mathbf{x}_{ij}$ and $\mathbf{z}_{ij}$) is derived again as the ratio of the level-2 component of the marginal variance divided by the summation of the level-2 and -1 components. The expression for the marginal variance is more complex than that for the variance-component and random-intercept models leading the expression for the VPC to also be more complex. The VPC is given by

$$\text{VPC}_{ij} = \frac{\overbrace{\mathbf{z}'_{ij}\boldsymbol{\Omega}_{\mathbf{u}}\mathbf{z}_{ij}}^{\text{level}-2\text{ variance}}}{\underbrace{\mathbf{z}'_{ij}\boldsymbol{\Omega}_{\mathbf{u}}\mathbf{z}_{ij}}_{\text{level}-2\text{ variance}} + \underbrace{\sigma_e^2}_{\text{level}-1\text{ variance}}} \tag{S1.32}$$

and this expression is identical to that for the simplified marginal correlation or ICC (where $\mathbf{z}_{ij} = \mathbf{z}_{i'j}$) given in Equation S1.29. This VPC is the level-2 VPC. The level-1 VPC – the proportion of the marginal response variance which lies within clusters – is simply equal to one minus the level-2 VPC.

Reconsider, once again, our example model with only one covariate $x_{1ij} \equiv z_{1ij}$. For this model, the VPC becomes

$$\text{VPC}_{ij} = \frac{\sigma_{u0}^2 + 2\sigma_{u01}x_{1ij} + \sigma_{u1}^2 x_{1ij}^2}{\sigma_{u0}^2 + 2\sigma_{u01}x_{1ij} + \sigma_{u1}^2 x_{1ij}^2 + \sigma_e^2} \tag{S1.33}$$

and this expression is identical to that for the simplified marginal correlation or ICC (where $x_{1ij} \equiv x_{1i'j}$) given in Equation S1.31.

## S2. Textbook and other coverage of conditional and marginal statistics for the two-level random-intercept Poisson model

In this section we review the presentation of the two-level random-intercept Poisson model in the standard multilevel modelling textbooks by Goldstein (2011b), Hox et al. (2017), Raudenbush and Bryk (2002), and Snijders and Bosker (2012). We focus on the extent to which these textbooks present and explain the conditional and marginal statistics implied by the model. We also include in our review the article by Austin et al. (2017) since this is the only article which presents the VPC and ICC for this model. One further resource we consider here as we have found it particularly useful is the advanced statistics text by McCulloch et al. (2008).

### S2.1 Review

Table S2.1 indicates the conditional and marginal statistics presented in each resource. We see that all resources present the conditional expectation and variance. The model is a conditional model and so this is expected. Where the resources differ is in their presentation of the marginal statistics: the marginal expectation, variance, covariance, correlation, including the ICC, level-2 and level-1 components of the marginal variance, and the VPC. Few of the resources present any of these statistics. Out of the standard textbooks on multilevel modelling, only Raudenbush and Bryk (2002) present any of these marginal statistics and even then they only present the marginal expectation.

Austin et al. (2017) do of course present the VPC. They also make clear that this expression coincides with that for the ICC (when we restrict comparisons to units with the same covariate values). They do not however present the four other marginal statistics reviewed here: the marginal expectation, variance, covariance, and correlation.

The best resource we have found is the advanced statistics text by McCulloch et al. (2008) which does presents the marginal expectation, variance, covariance and correlation for the two-level random-intercept Poisson model. However, this textbook does not additionally discuss the VPC and ICC. The ICC is simply the marginal correlation, but the authors do not make this connection clear to the reader. The expression for the VPC coincides with that for the ICC and therefore the marginal correlation, but neither do the authors make this link clear. We have shown that the VPC is derived by partitioning the marginal variance into variance components operating at the cluster and unit level. McCulloch et al. (2008) do in fact derive each of these level-specific terms prior to summing them to derive the marginal variance. The authors do not however make clear that these two terms can be interpreted as level-specific components of the marginal variance.

In sum, there is very little presentation or explanation in standard multilevel textbooks and other resources reader might turn to of the marginal statistics implied by the two-level random-intercept Poisson model. In the few cases where there is some coverage of the marginal statistics, this coverage ceases once we turn to the negative binomial, three-level and random-coefficient extensions to the two-level random-intercept Poisson model which are the focus of this article.

## S2.1 References not found in main reference list

McCulloch, C. E., Searle, S. R., & Neuhaus, J. M. (2008). *Generalized, Linear, and Mixed Models (2nd ed.)*. New York, USA: John Wiley & Since, Inc.

*Table S2.1*. Textbook and other coverage of conditional and marginal statistics for the two-level random-intercept Poisson model by resource

| Description | Notation | Goldstein (2011b) | Hox et al. (2017) | Raudennbush and Bryk (2002) | Snijders and Bosker (2012) | Austin et al. (2017) | McCulloch et al. (2008). |
|---|---|---|---|---|---|---|---|
| Conditional expectation | $\mu_{ij}^{C} \equiv \mathrm{E}(y_{ij} \mid \mathbf{x}_{ij}, u_j)$ | X | X | X | X | X | X |
| Conditional variance | $\omega_{ij}^{C} \equiv \mathrm{Var}(y_{ij} \mid \mathbf{x}_{ij}, u_j)$ | X | X | X | X | X | X |
| Marginal expectation | $\mu_{ij}^{M} \equiv \mathrm{E}(y_{ij} \mid \mathbf{x}_{ij})$ | | | X | | | X |
| Marginal variance | $\omega_{ij}^{M} \equiv \mathrm{Var}(y_{ij} \mid \mathbf{x}_{ij})$ | | | | | | X |
| Marginal covariance | $\mathrm{Cov}(y_{ij}, y_{i'j} \mid \mathbf{x}_{ij})$ | | | | | | X |
| Marginal correlation | $\mathrm{Corr}(y_{ij}, y_{i'j} \mid \mathbf{x}_{ij})$ | | | | | | X |
| Intraclass correlation coefficient (ICC) | | | | | | X | |
| Marginal variance: Level-2 component | $\mathrm{Var}(\mu_{ij}^{C} \mid \mathbf{x}_{ij}) \equiv \mathrm{Var}\{\mathrm{E}(y_{ij} \mid \mathbf{x}_{ij}, u_j) \mid \mathbf{x}_{ij}\}$ | | | | | | |
| Marginal variance: Level-1 component | $\mathrm{E}(\omega_{ij}^{C} \mid \mathbf{x}_{ij}) \equiv \mathrm{E}\{\mathrm{Var}(y_{ij} \mid \mathbf{x}_{ij}, u_j) \mid \mathbf{x}_{ij}\}$ | | | | | | |
| Variance partition coefficient (VPC) | | | | | | X | |

## S3. Review of two further two-level random-intercept models for count data

In the article, we review the two most widely applied multilevel models for count responses, the Poisson model and the negative binomial model (mean dispersion or NB2 version). In this section, we review two further count models, the Poisson model with an overdispersion random effect and the constant dispersion or NB1 version of the negative binomial model. Table S4.1 presents a summary table allowing readers to compare the conditional and marginal variances across all four models.

### S3.1 Poisson model with overdispersion random effect

The Poisson random-intercept model (Equation 1 where we substitute $\mathbf{x}'_{ij}\boldsymbol{\beta}$ for $\beta_0$) includes a normally distributed cluster random intercept effect to account for the clustering in the data. A natural way to deal with overdispersion in this model is to therefore add a normally distributed unit-level overdispersion random effect to represent omitted unit-level variables that are envisaged to be driving any overdispersion. In contrast to the conventional cluster random intercept effect, this does not induce any dependence among the units.

#### Model

The new model can be written as

$$y_{ij}|\mu_{ij} \sim \text{Poisson}(\mu_{ij})$$

$$\ln(\mu_{ij}) = \mathbf{x}'_{ij}\boldsymbol{\beta} + u_j + e_{ij} \quad \text{(S3.1)}$$

$$u_j \sim N(0, \sigma_u^2)$$

$$e_{ij} \sim N(0, \sigma_e^2)$$

where $e_{ij}$ denotes the overdispersion random effect. Thus, in this model, two units with the same covariate and random intercept effect value may nonetheless differ in their expected counts $\mu_{ij}$ with such differences attributed to the two units differing in terms of their values on omitted unit-level variables. This is also a feature of the two negative binomial models considered in this article (Equation 9 and Equation S3.9). The overdispersion random effect is assumed normally distributed with zero mean and variance or overdispersion parameter $\sigma_e^2$. The larger $\sigma_e^2$ is, the greater the overdispersion. When $\sigma_e^2 = 0$, the model simplifies to the Poisson model (Equation 1) permitting a likelihood-ratio test for whether any estimated overdispersion is statistically significant.

## Conditional statistics

In this model, we can again calculate the conditional expectation and variance of the response. However, here we must integrate out the overdispersion random effect since this is not typically of substantive interest. To do this, we exploit the fact that $\mu_{ij}^C$ is log normally distributed and so its expectation and variance have known forms. Thus, the conditional expectation of $y_{ij}$ (given $\mathbf{x}_{ij}$ and $u_j$ but averaged over $e_{ij}$) is now given by

$$\mu_{ij}^C \equiv \mathrm{E}(y_{ij} | \mathbf{x}_{ij}, u_j) = \exp(\mathbf{x}_{ij}'\boldsymbol{\beta} + u_j + \sigma_e^2/2) \quad \text{(S3.2)}$$

and we see that, in contrast to the Poisson model (Equation 2), $\mu_{ij}^C \neq \mu_{ij}$. The conditional variance is then given by

$$\omega_{ij}^{C} \equiv \mathrm{Var}\left(y_{ij}\middle|\mathbf{x}_{ij}, u_j\right) = \mu_{ij}^{C} + \left(\mu_{ij}^{C}\right)^2\{\exp(\sigma_e^2) - 1\} \tag{S3.3}$$

Thus, the conditional variance is now a quadratic function of the conditional expectation and is larger than the conditional expectation if $\sigma_e^2 > 0$. Therefore, the usual variance-mean relationship for the Poisson model (Equation 3) is relaxed, producing overdispersion with respect to the conditional expectation ($\omega_{ij}^{C} > \mu_{ij}^{C}$).

To help see the similarities between the mean dispersion negative binomial model (Equation 9) and the current model, recall that in the former we assume $\exp(e_{ij}) \sim \mathrm{Gamma}(1/\alpha\,, \alpha)$ while in the latter we assume $e_{ij} \sim N(0, \sigma_e^2)$, implying $\exp(e_{ij}) \sim \mathrm{logN}(0, \sigma_e^2)$. Thus, the two models differ only in the distribution they assume for the exponentiated overdispersion random effect (gamma vs. log-normal).

## Marginal statistics

The marginal expectation of $y_{ij}$ (given $\mathbf{x}_{ij}$ but now averaged over $u_j$ as well as $e_{ij}$) is given by

$$\mu_{ij}^{M} \equiv \mathrm{E}\left(y_{ij}\middle|\mathbf{x}_{ij}\right) = \exp(\beta_0 + \sigma_u^2/2 + \sigma_e^2/2) \tag{S3.4}$$

which differs from that for the Poisson and mean dispersion negative binomial models (Equations 4 and 12) via the inclusion of the additional term $\sigma_e^2/2$.

The marginal variance of $y_{ij}$ is given by

$$\omega_{ij}^{M} \equiv \mathrm{Var}\left(y_{ij}\middle|\mathbf{x}_{ij}\right) = \mu_{ij}^{M} + \left(\mu_{ij}^{M}\right)^2\{\exp(\sigma_u^2)\exp(\sigma_e^2) - 1\} \tag{S3.5}$$

which differs from that for the Poisson model (Equation 5) via the inclusion of the additional multiplicative term $\exp(\sigma_e^2)$. Thus, in this model, the marginal variance is larger than the marginal expectation if there is clustering $\sigma_u^2 > 0$ *or* overdispersion $\sigma_e^2 > 0$. Note that this expression takes the same form as that for the mean dispersion negative binomial model where $\exp(\sigma_e^2)$ takes the place of $(1 + \alpha)$.

The marginal covariance of $y_{ij}$ and $y_{i'j}$ (given $\mathbf{x}_{ij}$ and $\mathbf{x}_{i'j}$ but averaged over $u_j$, $e_{ij}$ and $e_{i'j}$) is given by

$$\text{Cov}\left(y_{ij}, y_{i'j} \middle| \mathbf{x}_{ij}, \mathbf{x}_{i'j}\right) = \left(\mu_{ij}^M\right)^2 \{\exp(\sigma_u^2) - 1\} \tag{S3.6}$$

and is the same as that for the Poisson and mean dispersion negative binomial models (Equations 6 and 14). The marginal correlation of $y_{ij}$ and $y_{i'j}$ is then given by

$$\text{ICC}_{ij,i'j} \equiv \text{Corr}\left(y_{ij}, y_{i'j} \middle| \mathbf{x}_{ij}, \mathbf{x}_{i'j}\right) = \frac{\left(\mu_{ij}^M\right)^2 \{\exp(\sigma_u^2) - 1\}}{\mu_{ij}^M + \left(\mu_{ij}^M\right)^2 \{\exp(\sigma_u^2)\exp(\sigma_e^2) - 1\}} \tag{S3.7}$$

and as with the Poisson and mean dispersion negative binomial models this statistic can be interpreted as the ICC. The expression differs from that for the Poisson model (Equation 7) only in the inclusion of the additional multiplicative term $\exp(\sigma_e^2)$ in the denominator. Here too, this expression takes the same form as that for the mean dispersion negative binomial model where $\exp(\sigma_e^2)$ takes the place of $(1 + \alpha)$.

As with the Poisson and mean dispersion negative binomial models, we can partition the marginal variance (Equation S3.5) into level-specific components which capture the within- and between-cluster variance in $y_{ij}$ (Supplemental materials S4.2). The resulting level-2 VPC is given by

$$\mathrm{VPC}_{ij} = \frac{\overbrace{\left(\mu_{ij}^{M}\right)^{2}\{\exp(\sigma_u^2)-1\}}^{\text{level}-2\ \text{variance}}}{\underbrace{\left(\mu_{ij}^{M}\right)^{2}\{\exp(\sigma_u^2)-1\}}_{\text{level}-2\ \text{variance}}+\underbrace{\mu_{ij}^{M}+\left(\mu_{ij}^{M}\right)^{2}\exp(\sigma_u^2)\{\exp(\sigma_e^2)-1\}}_{\text{level}-1\ \text{variance}}} \tag{S3.8}$$

and this expression is identical (after rearranging terms) to that for the marginal correlation or ICC given in Equation S3.7.

Studying Equation S3.8, we see that, as in the Poisson case (Equation 8), the VPC is an increasing function of both the marginal expectation $\mu_{ij}^{M}$ and the cluster variance $\sigma_u^2$. However, the VPC is now also a decreasing function of the overdispersion parameter $\sigma_e^2$. This makes sense. As the overdispersion increases, all else equal, the more unmodelled variation there is at level-1 and so the VPC decreases. Here too, this expression takes the same form as that for the mean dispersion negative binomial model where $\exp(\sigma_e^2) - 1$ takes the place of $\alpha$.

Comparing the three VPC expressions (Equations 8, 16 and S3.8), we see that the expression for the level-2 component of the marginal variance is the same and so it is only the expression for the level-1 component which varies across models. This makes sense as the models differ only in their treatment of overdispersion, which is viewed as a level-1 phenomenon. The overdispersion parameter in the current model (and the mean dispersion negative binomial model) leads the expression for the level-1 component of the marginal variance to exceed that of the Poisson model. The marginal variance is simply the summation of the level-2 and -1 variances and so is also expected to be higher in the current model (and mean dispersion negative binomial model) compared to that of the Poisson model.

## S3.2 Negative binomial model: Constant dispersion version

The constant dispersion or NB1 version of the negative binomial model cannot be derived from a version of the Poisson model with a unit-level overdispersion random effect.

### Model

The model is instead written as

$$y_{ij}|\mu_{ij} \sim \text{Poisson}(\mu_{ij})$$

$$\mu_{ij}|\mathbf{x}_{ij}, u_j \sim \text{Gamma}\left\{\frac{\exp(\mathbf{x}'_{ij}\boldsymbol{\beta} + u_j)}{\delta}, \delta\right\} \quad \text{(S3.9)}$$

$$u_j \sim N(0, \sigma_u^2)$$

where we now assume the expected count $\mu_{ij}$ has a conditional gamma distribution (given $\mathbf{x}_{ij}$ and $u_j$) with shape and scale parameters $\exp(\mathbf{x}'_{ij}\boldsymbol{\beta} + u_j)\,\delta^{-1}$ and $\delta$ and therefore expectation $\exp(\mathbf{x}'_{ij}\boldsymbol{\beta} + u_j)$ and variance $\exp(\mathbf{x}'_{ij}\boldsymbol{\beta} + u_j)\,\delta$ where $\delta$ is the overdispersion parameter. The larger $\delta$ is, the greater the overdispersion. When $\delta = 0$, the variance of this gamma distribution is equal to zero and the model simplifies to the Poisson model (Equation 1), once again permitting a likelihood-ratio test for whether the estimated overdispersion is statistically significant.

### Conditional statistics

It follows that the conditional expectation of $y_{ij}$ (given $\mathbf{x}_{ij}$ and $u_j$) is again the same as in the Poisson and mean dispersion negative binomial models (Equations 2 and 10), with

$$\mu_{ij}^{C} \equiv \mathrm{E}\left(y_{ij} \middle| \mathbf{x}_{ij}, u_j\right) = \exp\left(\mathbf{x}_{ij}' \boldsymbol{\beta} + u_j\right) \tag{S3.10}$$

However, the conditional variance of $y_{ij}$ is now

$$\omega_{ij}^{C} \equiv \mathrm{Var}\left(y_{ij} \middle| \mathbf{x}_{ij}, u_j\right) = \mu_{ij}^{C}(1 + \delta) \tag{S3.11}$$

Thus, in contrast to the quadratic form seen in the two unit-level random effect overdispersion models (Equation 11 and S3.3), the variance in the current model is a constant multiple of the conditional expectation.

To help see the similarities between the two versions of the negative binomial model, note that we can rewrite the mean dispersion model (Equation 9) as follows

$$\begin{gathered} y_{ij} | \mu_{ij} \sim \text{Poisson}\left(\mu_{ij}\right) \\ \mu_{ij} | \mathbf{x}_{ij}, u_j \sim \text{Gamma}\left\{\frac{1}{\alpha}, \alpha \exp\left(\mathbf{x}_{ij}' \boldsymbol{\beta} + u_j\right)\right\} \\ u_j \sim N(0, \sigma_u^2) \end{gathered} \tag{S3.12}$$

Thus, both versions of the negative binomial model assume the expected count $\mu_{ij}$ follows a conditional gamma distribution, but the models differ in terms of the shape and scale parameters of this distribution. In the mean dispersion model (Equation 9), the shape and scale parameters are chosen such that the conditional expectation of $\mu_{ij}$ is $\exp\left(\mathbf{x}'_{ij}\boldsymbol{\beta} + u_j\right)$ and conditional variance is $\exp\left(\mathbf{x}'_{ij}\boldsymbol{\beta} + u_j\right)^2 \alpha$ whereas in the constant dispersion model (Equation S3.12) the shape and scale parameters are chosen such that the conditional expectation of $\mu_{ij}$ is $\exp\left(\mathbf{x}'_{ij}\boldsymbol{\beta} + u_j\right)$ and conditional variance is $\exp\left(\mathbf{x}'_{ij}\boldsymbol{\beta} + u_j\right) \delta$. Thus, the difference between the two versions of the negative binomial model can be viewed as a difference in the relationship between the conditional variance and conditional expectation of $\mu_{ij}$. This difference then leads the models to differ in terms of the conditional variances of the response $\omega_{ij}^{C}$ (Equation 11 vs. S3.11).

## Marginal statistics

The marginal expectation of $y_{ij}$ (given $\mathbf{x}_{ij}$ but averaged over $u_j$) is given by

$$\mu_{ij}^{M} \equiv \mathrm{E}\left(y_{ij} \middle| \mathbf{x}_{ij}\right) = \exp(\beta_0 + \sigma_u^2/2) \qquad \text{(S3.13)}$$

which is the same as that for the Poisson and mean dispersion negative binomial models (Equation 4 and 12).

The marginal variance of $y_{ij}$ is given by

$$\omega_{ij}^{M} \equiv \mathrm{Var}\left(y_{ij} \middle| \mathbf{x}_{ij}\right) = \mu_{ij}^{M}(1+\delta) + \left(\mu_{ij}^{M}\right)^2\{\exp(\sigma_u^2) - 1\} \qquad \text{(S3.14)}$$

which differs from that for the Poisson model (Equation 5) in that the first term is now multiplied by $1 + \delta$. Thus, in this model, the marginal variance is larger than the marginal expectation if there is clustering $\sigma_u^2 > 0$ *or* overdispersion $\delta > 0$. Note that this expression takes a diferent form to that for the Poisson model with overdispersion random effect or the mean dispersion negative binomial model.

The marginal covariance of $y_{ij}$ and $y_{i'j}$ (given $\mathbf{x}_{ij}$ and $\mathbf{x}_{i'j}$ but averaged over $u_j$) is given by

$$\text{Cov}\left(y_{ij}, y_{i'j} \middle| \mathbf{x}_{ij}, \mathbf{x}_{i'j}\right) = \left(\mu_{ij}^{M}\right)^2 \{\exp(\sigma_u^2) - 1\} \tag{S3.15}$$

and is the same as that for the Poisson, mean dispersion negative binomial, and Poisson model with overdispersed random effect models (Equations 6, 14 and S3.6). The marginal correlation of $y_{ij}$ and $y_{i'j}$ is then given by

$$\text{ICC}_{ij,i'j} \equiv \text{Corr}\left(y_{ij}, y_{i'j} \middle| \mathbf{x}_{ij}, \mathbf{x}_{i'j}\right) = \frac{\left(\mu_{ij}^{M}\right)^2 \{\exp(\sigma_u^2) - 1\}}{\mu_{ij}^{M}(1+\delta) + \left(\mu_{ij}^{M}\right)^2 \{\exp(\sigma_u^2) - 1\}} \tag{S3.16}$$

and as with the other three models can be interpreted as the ICC. The expression differs from that for the Poisson model (Equation 7) only in the inclusion of the additional multiplicative term $1 + \delta$ in the denominator.

As with the other three models, we can partition the marginal variance (Equation S3.14) into level-specific components which capture the within- and between-cluster variance in $y_{ij}$ (Supplemental materials S4.2). The resulting level-2 VPC is given by

$$\text{VPC}_{ij} = \frac{\overbrace{\left(\mu_{ij}^{M}\right)^2 \{\exp(\sigma_u^2) - 1\}}^{\text{level-2 variance}}}{\underbrace{\left(\mu_{ij}^{M}\right)^2 \{\exp(\sigma_u^2) - 1\}}_{\text{level-2 variance}} + \underbrace{\mu_{ij}^{M}(1+\delta)}_{\text{level-1 variance}}} \tag{S3.17}$$

and this expression is identical (after rearranging terms) to that for the marginal correlation or ICC given in Equation S3.16.

Studying Equation S3.17, we see that, as with the other models which allow for overdispersion, the VPC is an increasing function of both the marginal expectation $\mu_{ij}^M$ and the cluster variance $\sigma_u^2$, but is a decreasing function of the overdispersion parameter, here $\delta$.

Comparing all four VPC expressions (Equations 8, 16, S3.8 and S3.17), we see that the expression for the level-2 component of the marginal variance is always the same and so it is only the expression for the level-1 component which varies across models. This makes sense as the models differ only in their treatment of overdispersion, which is viewed as a level-1 phenomenon. The overdispersion parameter in all three models which allow for overdispersion lead the expression for the level-1 component of the marginal variance to exceed that of the Poisson model. The marginal variance is simply the summation of the level-2 and -1 variances and so is also expected to be higher in all these models compared to that of the Poisson model.

## S4. Derivation of the marginal statistics in two-level random-intercept models: Marginal expectation, variance, covariance, correlation, VPCs and ICC

We now present general derivations for the marginal expectation, variance, covariance and correlation for two-level random-intercept models (McCulloch et al., 2008). These apply to the continuous response models reviewed in Supplemental materials S1 as well as all the count models explored in our work. We then show that the ICC is simply the marginal correlation. We then decompose the marginal variance into components of variance at each level and show that the VPC can be calculated as the ratio of the level-2 component to the sum of the level-1 and

level-2 components. All the expressions for the marginal statistics presented in the article have been obtained using these general derivations. Table S4.1 presents these expressions in tabular form to facilitate comparisons across models.

### S4.1 Marginal expectation

The marginal expectation of $y_{ij}$ (given $\mathbf{x}_{ij}$ but averaged over $u_j$) can be derived by exploiting the law of total expectations ($\mathrm{E}(A) = \mathrm{E}\{\mathrm{E}(A|B)\}$; law of iterated expectations; McCulloch et al., 2008)

$$\begin{aligned} \mu_{ij}^{M} \equiv \mathrm{E}(y_{ij}|\mathbf{x}_{ij}) &= \mathrm{E}\{\mathrm{E}(y_{ij}|\mathbf{x}_{ij}, u_j)|\mathbf{x}_{ij}\} \\ &= \mathrm{E}(\mu_{ij}^{C}|\mathbf{x}_{ij}) \end{aligned} \tag{S4.1}$$

### S4.2 Marginal variance

The marginal variance of $y_{ij}$ (given $\mathbf{x}_{ij}$ but averaged over $u_j$) can be derived by exploiting the law of total variance ($\mathrm{Var}(A) = \mathrm{Var}\{\mathrm{E}(A|B)\} + \mathrm{E}\{\mathrm{Var}(A|B)\}$; McCulloch et al., 2008)

$$\begin{aligned} \omega_{ij}^{M} \equiv \mathrm{Var}(y_{ij}|\mathbf{x}_{ij}) &= \mathrm{Var}\{\mathrm{E}(y_{ij}|\mathbf{x}_{ij}, u_j)|\mathbf{x}_{ij}\} + \mathrm{E}\{\mathrm{Var}(y_{ij}|\mathbf{x}_{ij}, u_j)|\mathbf{x}_{ij}\} \\ &= \underbrace{\mathrm{Var}(\mu_{ij}^{C}|\mathbf{x}_{ij})}_{\text{level}-2\text{ variance}} + \underbrace{\mathrm{E}(\omega_{ij}^{C}|\mathbf{x}_{ij})}_{\text{level}-1\text{ variance}} \end{aligned} \tag{S4.2}$$

It is instructive to realize that the law of total variance decomposes the marginal variance into separate level-2 and level-1 specific variance components. The level-2 variance component $\text{Var}(\mu_{ij}^C|\mathbf{x}_{ij})$ captures between cluster variation in the cluster specific expected counts $\mu_{ij}^C$ attributable to the cluster random intercept effect $u_j$ (i.e., given $\mathbf{x}_{ij}$). The level-1 variance component $\text{E}(\omega_{ij}^C|\mathbf{x}_{ij})$ captures within cluster variation in $y_{ij}$ around these cluster specific expected counts $\mu_{ij}^C$ (i.e., given $\mathbf{x}_{ij}$) averaged across all clusters (as $\omega_{ij}^C$ given $\mathbf{x}_{ij}$ still varies across clusters as function of $u_j$).

**S4.3 Marginal covariance**

To derive the marginal covariance of $y_{ij}$ and $y_{i'j}$ (given $\mathbf{x}_{ij}$ and $\mathbf{x}_{i'j}$ but averaged over $u_j$) , we exploit the law of total covariance ($\text{Cov}(A,B) = \text{Cov}\{\text{E}(A,B|C)\} + \text{E}\{\text{Cov}(A,B|C)\}$; McCulloch et al., 2008).

$$\begin{aligned}
&\text{Cov}(y_{ij}, y_{i'j}|\mathbf{x}_{ij}, \mathbf{x}_{i'j}) \\
&\qquad = \text{Cov}\{\text{E}(y_{ij}|\mathbf{x}_{ij}, u_j), \text{E}(y_{i'j}|\mathbf{x}_{i'j}, u_j)|\mathbf{x}_{ij}, \mathbf{x}_{i'j}\} \\
&\qquad + \text{E}\{\text{Cov}(y_{ij}, y_{i'j}|\mathbf{x}_{ij}, \mathbf{x}_{i'j}, u_j)|\mathbf{x}_{ij}, \mathbf{x}_{i'j}\} \\
&= \text{Cov}(\mu_{ij}^C, \mu_{i'j}^C|\mathbf{x}_{ij}, \mathbf{x}_{i'j}) + \text{E}(0|\mathbf{x}_{ij}, \mathbf{x}_{i'j})
\end{aligned} \tag{S4.3}$$

The second term equals zero due to the assumption of conditional independence among the responses for the same cluster given the covariates and random intercept effect. Namely, $y_{ij}$ and $y_{i'j}$ are assumed conditionally independent (given $\mathbf{x}_{ij}$, $\mathbf{x}_{i'j}$ and $u_j$) (i.e., assuming independent Poisson sampling variation). This is sometimes referred to as the local independence assumption.

**S4.4 Marginal correlation**

The marginal correlation or ICC can then be calculated in the usual way

$$\mathrm{ICC}_{ij,i'j} \equiv \mathrm{Corr}\left(y_{ij}, y_{i'j} \middle| \mathbf{x}_{ij}, \mathbf{x}_{i'j}\right) = \frac{\mathrm{Cov}\left(y_{ij}, y_{i'j} \middle| \mathbf{x}_{ij}, \mathbf{x}_{i'j}\right)}{\sqrt{\mathrm{Var}\left(y_{ij} \middle| \mathbf{x}_{ij}\right)}\sqrt{\mathrm{Var}\left(y_{i'j} \middle| \mathbf{x}_{i'j}\right)}}$$

$$= \frac{\mathrm{Cov}\left(\mu_{ij}^{C}, \mu_{i'j}^{C} \middle| \mathbf{x}_{ij}, \mathbf{x}_{i'j}\right)}{\sqrt{\mathrm{Var}\left(\mu_{ij}^{C} \middle| \mathbf{x}_{ij}\right) + \mathrm{E}\left(\omega_{ij}^{C} \middle| \mathbf{x}_{ij}\right)}\sqrt{\mathrm{Var}\left(\mu_{i'j}^{C} \middle| \mathbf{x}_{i'j}\right) + \mathrm{E}\left(\omega_{i'j}^{C} \middle| \mathbf{x}_{i'j}\right)}} \tag{S4.3}$$

**S4.5 Intraclass correlation coefficients (ICCs)**

We focus on the special case where the two units have the same covariate values $\mathbf{x}_{ij} = \mathbf{x}_{i'j}$ in which case $\mu_{ij}^{C} = \mu_{i'j}^{C}$ and $\omega_{ij}^{C} = \omega_{i'j}^{C}$ and so the ICC simplifies to

$$\mathrm{ICC}_{ij,i'j} \equiv \mathrm{Corr}\left(y_{ij}, y_{i'j} \middle| \mathbf{x}_{ij}\right) = \frac{\overbrace{\mathrm{Var}\left(\mu_{ij}^{C} \middle| \mathbf{x}_{ij}\right)}^{\text{level-2 variance}}}{\underbrace{\mathrm{Var}\left(\mu_{ij}^{C} \middle| \mathbf{x}_{ij}\right)}_{\text{level-2 variance}} + \underbrace{\mathrm{E}\left(\omega_{ij}^{C} \middle| \mathbf{x}_{ij}\right)}_{\text{level-1 variance}}} \tag{S4.4}$$

**S4.5 Variance partition coefficients (VPCs)**

Variance partition coefficients report the proportion of the total variation in the observed counts (given the covariates) which lies at each level of analysis. These can be calculated in the usual way, as ratios of each variance component to the marginal variance. The level-2 VPC is therefore calculated as

$$\text{VPC}(2)_{ijk} = \frac{\overbrace{\text{Var}\left(\mu_{ij}^{C}\middle|\mathbf{x}_{ij}\right)}^{\text{level-2 variance}}}{\underbrace{\text{Var}\left(\mu_{ij}^{C}\middle|\mathbf{x}_{ij}\right)}_{\text{level-2 variance}} + \underbrace{\text{E}\left(\omega_{ij}^{C}\middle|\mathbf{x}_{ij}\right)}_{\text{level-1 variance}}} \qquad \text{(S4.5)}$$

This is the same expression as that given above for the ICC (Equation S4.4).

The level-1 VPC is calculated as 1 minus the level-2 VPC

$$\text{VPC}(1)_{ijk} = \frac{\overbrace{\text{E}\left(\omega_{ij}^{C}\middle|\mathbf{x}_{ij}\right)}^{\text{level-1 variance}}}{\underbrace{\text{Var}\left(\mu_{ij}^{C}\middle|\mathbf{x}_{ij}\right)}_{\text{level-2 variance}} + \underbrace{\text{E}\left(\omega_{ij}^{C}\middle|\mathbf{x}_{ij}\right)}_{\text{level-1 variance}}} \qquad \text{(S4.5)}$$

**S4.8 References not found in main reference list**

McCulloch, C. E., Searle, S. R., & Neuhaus, J. M. (2008). *Generalized, Linear, and Mixed Models (2nd ed.)*. New York, USA: John Wiley & Since, Inc.

Table S4.1.

Two-level count response models: Expressions for the conditional and marginal expectations and variances of the response and the level-1 and -2 components of the marginal variance used in the calculation of the VPCs.

| Description | Notation | Poisson model | Poisson model with overdispersion random effect | Negative binomial model: Mean dispersion or NB2 version | Negative binomial model: Constant dispersion or NB1 version |
|---|---|---|---|---|---|
| Conditional expectation | $\mu_{ij}^C \equiv \mathrm{E}(y_{ij} \mid \mathbf{x}_{ij}, u_j)$ | $\exp(\mathbf{x}'_{ij}\boldsymbol{\beta} + u_j)$ | $\exp(\mathbf{x}'_{ij}\boldsymbol{\beta} + u_j + \sigma_e^2/2)$ | $\exp(\mathbf{x}'_{ij}\boldsymbol{\beta} + u_j)$ | $\exp(\mathbf{x}'_{ij}\boldsymbol{\beta} + u_j)$ |
| Conditional variance | $\omega_{ij}^C \equiv \mathrm{Var}(y_{ij} \mid \mathbf{x}_{ij}, u_j)$ | $\mu_{ij}^C$ | $\mu_{ij}^C + (\mu_{ij}^C)^2\{\exp(\sigma_e^2) - 1\}$ | $\mu_{ij}^C + (\mu_{ij}^C)^2\alpha$ | $\mu_{ij}^C(1+\delta)$ |
| Marginal expectation | $\mu_{ij}^M \equiv \mathrm{E}(y_{ij} \mid \mathbf{x}_{ij})$ | $\exp(\mathbf{x}'_{ij}\boldsymbol{\beta} + \sigma_u^2/2)$ | $\exp(\mathbf{x}'_{ij}\boldsymbol{\beta} + \sigma_u^2/2 + \sigma_e^2/2)$ | $\exp(\mathbf{x}'_{ij}\boldsymbol{\beta} + \sigma_u^2/2)$ | $\exp(\mathbf{x}'_{ij}\boldsymbol{\beta} + \sigma_u^2/2)$ |
| Marginal variance | $\omega_{ij}^M \equiv \mathrm{Var}(y_{ij} \mid \mathbf{x}_{ij})$ | $\mu_{ij}^M + (\mu_{ij}^M)^2\{\exp(\sigma_u^2) - 1\}$ | $\mu_{ij}^M + (\mu_{ij}^M)^2\{\exp(\sigma_u^2)\exp(\sigma_e^2) - 1\}$ | $\mu_{ij}^M + (\mu_{ij}^M)^2\{\exp(\sigma_u^2)(1+\alpha) - 1\}$ | $\mu_{ij}^M(1+\delta) + (\mu_{ij}^M)^2\{\exp(\sigma_u^2) - 1\}$ |
| Level-2 component | $\mathrm{Var}(\mu_{ij}^C \mid \mathbf{x}_{ij}) \equiv \mathrm{Var}\{\mathrm{E}(y_{ij} \mid \mathbf{x}_{ij}, u_j) \mid \mathbf{x}_{ij}\}$ | $(\mu_{ij}^M)^2\{\exp(\sigma_u^2) - 1\}$ | $(\mu_{ij}^M)^2\{\exp(\sigma_u^2) - 1\}$ | $(\mu_{ij}^M)^2\{\exp(\sigma_u^2) - 1\}$ | $(\mu_{ij}^M)^2\{\exp(\sigma_u^2) - 1\}$ |
| Level-1 component | $\mathrm{E}(\omega_{ij}^C \mid \mathbf{x}_{ij}) \equiv \mathrm{E}\{\mathrm{Var}(y_{ij} \mid \mathbf{x}_{ij}, u_j) \mid \mathbf{x}_{ij}\}$ | $\mu_{ij}^M$ | $\mu_{ij}^M + (\mu_{ij}^M)^2\exp(\sigma_u^2)\{\exp(\sigma_e^2) - 1\}$ | $\mu_{ij}^M + (\mu_{ij}^M)^2\exp(\sigma_u^2)\,\alpha$ | $\mu_{ij}^M(1+\delta)$ |

Note.

The above expressions are for two-level random-intercept models. The corresponding expressions for the two-level models with random coefficients are obtained by replacing $u_j$ and $\sigma_u^2$ in all expressions with $\mathbf{z}'_{ij}\mathbf{u}_j$ and $\mathbf{z}'_{ij}\boldsymbol{\Omega}_{\mathbf{u}}\mathbf{z}_{ij}$.

## S5. Derivation of the marginal statistics in three-level random-intercept models: Marginal expectation, variance, covariance, correlation, VPCs and ICCs

We now present general derivations for the marginal expectation, variance, covariance and correlation for three-level random-intercept models. We then decompose the marginal variance into components of variance at each level and show that the different VPCs can be calculated as ratios of the level-specific variance components. All the expressions for the marginal statistics presented in the article have been obtained using these general derivations. Table S5.1 presents these expressions in tabular form to facilitate comparisons across models.

### S5.1 Marginal expectation

The marginal expectation of $y_{ijk}$ (given $\mathbf{x}_{ijk}$ but averaged over $v_k$ and $u_{jk}$) can be derived by exploiting the law of total expectations ($\mathrm{E}(A) = \mathrm{E}\{\mathrm{E}(A|B)\}$; law of iterated expectations)

$$\mu_{ijk}^{M} \equiv \mathrm{E}\left(y_{ijk}|\mathbf{x}_{ij}\right) = \mathrm{E}\left\{\mathrm{E}\left(y_{ijk}|\mathbf{x}_{ijk}, v_k, u_{jk}\right)|\mathbf{x}_{ijk}\right\}$$
$$= \mathrm{E}\left(\mu_{ijk}^{C2}|\mathbf{x}_{ij}\right) \quad \text{(S5.1)}$$

### S5.2 Marginal variance

The marginal variance of $y_{ijk}$ (given $\mathbf{x}_{ijk}$ but averaged over $v_k$ and $u_{jk}$) can be derived by repetitively exploiting the law of total variance ($\mathrm{Var}(A) = \mathrm{Var}\{\mathrm{E}(A|B)\} + \mathrm{E}\{\mathrm{Var}(A|B)\}$

First decompose the marginal variance of $y_{ijk}$ (given $\mathbf{x}_{ijk}$ but averaged over $v_k$ and $u_{jk}$) into a between supercluster and within supercluster components

$$\underbrace{\text{Var}(y_{ijk}|\mathbf{x}_{ijk})}_{\text{Total variance}} = \underbrace{\text{Var}\{\text{E}(y_{ijk}|\mathbf{x}_{ijk}, v_k)|\mathbf{x}_{ijk}\}}_{\text{level}-3\text{ variance}} + \underbrace{\text{E}\{\text{Var}(y_{ijk}|\mathbf{x}_{ijk}, v_k)|\mathbf{x}_{ijk}\}}_{\text{level}-2\text{ and}-1\text{ combined variance}} \quad \text{(S5.2)}$$

Next decompose the conditional variance of $y_{ijk}$ (given $\mathbf{x}_{ijk}$ and $v_k$ but averaged over $u_{jk}$) into a between cluster and within cluster component (i.e., decompose the contents of the expectation in the second term of the above expressions).

$$\underbrace{\text{Var}(y_{ijk}|\mathbf{x}_{ijk}, v_k)}_{\text{level}-2\text{ and}-1\text{ combined variance in supercluster }k} = \underbrace{\text{Var}\{\text{E}(y_{ijk}|\mathbf{x}_{ijk}, v_k, u_{jk})|\mathbf{x}_{ijk}\}}_{\text{level}-2\text{ variance in supercluster }k}$$
$$+ \underbrace{\text{E}\{\text{Var}(y_{ijk}|\mathbf{x}_{ijk}, v_k, u_{jk})|\mathbf{x}_{ijk}\}}_{\text{level}-1\text{ variance in supercluster }k} \quad \text{(S5.3)}$$

Substitute Equation S5.3 into Equation S5.2 to give

$$\underbrace{\text{Var}(y_{ijk}|\mathbf{x}_{ijk})}_{\text{Total variance}} = \underbrace{\text{Var}\{\text{E}(y_{ijk}|\mathbf{x}_{ijk}, v_k)|\mathbf{x}_{ijk}\}}_{\text{level}-3\text{ variance}} + \underbrace{\text{E}\{\text{Var}\{\text{E}(y_{ijk}|\mathbf{x}_{ijk}, v_k, u_{jk})|\mathbf{x}_{ijk}\}|\mathbf{x}_{ijk}\}}_{\text{level}-2\text{ variance}}$$
$$+ \underbrace{\text{E}\{\text{Var}(y_{ijk}|\mathbf{x}_{ijk}, v_k, u_{jk})|\mathbf{x}_{ijk}\}}_{\text{level}-1\text{ variance}} \quad \text{(S5.4)}$$

This expression can be written more concisely as

$$\underbrace{\text{Var}(y_{ijk}|\mathbf{x}_{ijk})}_{\text{Total variance}} = \underbrace{\text{Var}(\mu_{ijk}^{C3}|\mathbf{x}_{ijk})}_{\text{level}-3\text{ variance}} + \underbrace{\text{E}\{\text{Var}\{\mu_{ijk}^{C2}|\mathbf{x}_{ijk}\}|\mathbf{x}_{ijk}\}}_{\text{level}-2\text{ variance}} + \underbrace{\text{E}(\omega_{ijk}^{C2}|\mathbf{x}_{ijk})}_{\text{level}-1\text{ variance}} \quad \text{(S5.5)}$$

The level-3 variance component $\text{Var}(\mu_{ijk}^{C3}|\mathbf{x}_{ij})$ captures between supercluster variation in the supercluster specific expected counts $\mu_{ijk}^{C3}$ attributable to the supercluster random intercept effect $v_k$ (i.e., given $\mathbf{x}_{ij}$). The level-2 variance component $\mathrm{E}\{\mathrm{Var}\{\mu_{ijk}^{C2}|\mathbf{x}_{ijk}\}|\mathbf{x}_{ijk}\}$ captures the within supercluster, between cluster variation in the cluster specific expected counts $\mu_{ijk}^{C2}$ attributable to the cluster random intercept effect $u_{jk}$ (i.e., given $\mathbf{x}_{ij}$) and averaged across all superclusters. The level-1 variance component $\mathrm{E}(\omega_{ijk}^{C2}|\mathbf{x}_{ijk})$ captures within cluster variation in $y_{ij}$ around these cluster specific expected counts $\mu_{ijk}^{C2}$ (i.e., given $\mathbf{x}_{ij}$) averaged across all clusters (as $\omega_{ijk}^{C2}$ given $\mathbf{x}_{ij}$ still varies across clusters as function of $v_k$ and $u_{jk}$).

### S5.3 Variance partition coefficients (VPCs)

Variance partition coefficients report the proportion of the total variation in the observed counts (given the covariates) which lies at each level of analysis. These can be calculated in the usual way, as ratios of each variance component to the marginal variance. The level-3 VPC is therefore calculated as

$$\mathrm{VPC}(3)_{ijk} = \frac{\overbrace{\mathrm{Var}(\mu_{ijk}^{C3}|\mathbf{x}_{ijk})}^{\text{level-3 variance}}}{\underbrace{\mathrm{Var}(\mu_{ijk}^{C3}|\mathbf{x}_{ijk})}_{\text{level-3 variance}} + \underbrace{\mathrm{E}\{\mathrm{Var}\{\mu_{ijk}^{C2}|\mathbf{x}_{ijk}\}|\mathbf{x}_{ijk}\}}_{\text{level-2 variance}} + \underbrace{\mathrm{E}(\omega_{ijk}^{C2}|\mathbf{x}_{ijk})}_{\text{level-1 variance}}} \tag{S5.5}$$

The level-2 VPC is calculated as

$$\mathrm{VPC}(3)_{ijk} = \frac{\overbrace{\mathrm{E}\{\mathrm{Var}\{\mu_{ijk}^{C2}|\mathbf{x}_{ijk}\}|\mathbf{x}_{ijk}\}}^{\text{level-2 variance}}}{\underbrace{\mathrm{Var}(\mu_{ijk}^{C3}|\mathbf{x}_{ijk})}_{\text{level-3 variance}} + \underbrace{\mathrm{E}\{\mathrm{Var}\{\mu_{ijk}^{C2}|\mathbf{x}_{ijk}\}|\mathbf{x}_{ijk}\}}_{\text{level-2 variance}} + \underbrace{\mathrm{E}(\omega_{ijk}^{C2}|\mathbf{x}_{ijk})}_{\text{level-1 variance}}} \tag{S5.6}$$

The level-1 VPC is calculated as

$$\text{VPC}(3)_{ijk} = \frac{\overbrace{\text{E}\left(\omega_{ijk}^{C2} \middle| \mathbf{x}_{ijk}\right)}^{\text{level−1 variance}}}{\underbrace{\text{Var}\left(\mu_{ijk}^{C3} \middle| \mathbf{x}_{ijk}\right)}_{\text{level−3 variance}} + \underbrace{\text{E}\left\{\text{Var}\left\{\mu_{ijk}^{C2} \middle| \mathbf{x}_{ijk}\right\} \middle| \mathbf{x}_{ijk}\right\}}_{\text{level−2 variance}} + \underbrace{\text{E}\left(\omega_{ijk}^{C2} \middle| \mathbf{x}_{ijk}\right)}_{\text{level−1 variance}}} \tag{S5.7}$$

Table S5.1.

Three-level count response models: Expressions for the conditional and marginal expectations and variances of the response and the level-1, -2 and -3 components of the marginal variance used in the calculation of the VPCs.

| Description | Notation | Poisson model | Poisson model with overdispersion random effect | Negative binomial model: Mean dispersion or NB2 version | Negative binomial model: Constant dispersion or NB1 version |
|---|---|---|---|---|---|
| Conditional expectation (level-2) | $\mu_{ijk}^{C2} \equiv \mathrm{E}(y_{ijk} \mid \mathbf{x}_{ijk}, v_k, u_{jk})$ | $\exp(\mathbf{x}'_{ijk}\boldsymbol{\beta} + v_k + u_{jk})$ | $\exp(\mathbf{x}'_{ijk}\boldsymbol{\beta} + v_k + u_{jk} + \sigma_e^2/2)$ | $\exp(\mathbf{x}'_{ijk}\boldsymbol{\beta} + v_k + u_{jk})$ | $\exp(\mathbf{x}'_{ijk}\boldsymbol{\beta} + v_k + u_{jk})$ |
| Conditional variance (level-2) | $\omega_{ijk}^{C2} \equiv \mathrm{Var}(y_{ijk} \mid \mathbf{x}_{ijk}, v_k, u_{jk})$ | $\mu_{ijk}^{C2}$ | $\mu_{ijk}^{C2} + (\mu_{ijk}^{C2})^2\{\exp(\sigma_e^2) - 1\}$ | $\mu_{ijk}^{C2} + (\mu_{ijk}^{C2})^2\alpha$ | $\mu_{ijk}^{C2}(1+\delta)$ |
| Conditional expectation (level-3) | $\mu_{ijk}^{C3} \equiv \mathrm{E}(y_{ijk} \mid \mathbf{x}_{ijk}, v_k)$ | $\exp(\mathbf{x}'_{ijk}\boldsymbol{\beta} + v_k + \sigma_u^2/2)$ | $\exp(\mathbf{x}'_{ijk}\boldsymbol{\beta} + v_k + \sigma_u^2/2 + \sigma_e^2/2)$ | $\exp(\mathbf{x}'_{ijk}\boldsymbol{\beta} + v_k + \sigma_u^2/2)$ | $\exp(\mathbf{x}'_{ijk}\boldsymbol{\beta} + v_k + \sigma_u^2/2)$ |
| Conditional variance (level-3) | $\omega_{ijk}^{C3} \equiv \mathrm{Var}(y_{ijk} \mid \mathbf{x}_{ijk}, v_k)$ | $\mu_{ijk}^{C3} + (\mu_{ijk}^{C3})^2\{\exp(\sigma_u^2) - 1\}$ | $\mu_{ijk}^{C3} + (\mu_{ijk}^{C3})^2\{\exp(\sigma_u^2 + \sigma_e^2) - 1\}$ | $\mu_{ijk}^{C3} + (\mu_{ijk}^{C3})^2\{\exp(\sigma_u^2)(1+\alpha) - 1\}$ | $\mu_{ijk}^{C3}(1+\delta) + (\mu_{ijk}^{C3})^2\{\exp(\sigma_u^2) - 1\}$ |
| Marginal expectation | $\mu_{ijk}^{M} \equiv \mathrm{E}(y_{ijk} \mid \mathbf{x}_{ijk})$ | $\exp(\mathbf{x}'_{ijk}\boldsymbol{\beta} + \sigma_v^2/2 + \sigma_u^2/2)$ | $\exp(\mathbf{x}'_{ijk}\boldsymbol{\beta} + \sigma_v^2/2 + \sigma_u^2/2 + \sigma_e^2/2)$ | $\exp(\mathbf{x}'_{ijk}\boldsymbol{\beta} + \sigma_v^2/2 + \sigma_u^2/2)$ | $\exp(\mathbf{x}'_{ijk}\boldsymbol{\beta} + \sigma_v^2/2 + \sigma_u^2/2)$ |
| Marginal variance | $\omega_{ijk}^{M} \equiv \mathrm{Var}(y_{ijk} \mid \mathbf{x}_{ijk})$ | $\mu_{ijk}^{M} + (\mu_{ijk}^{M})^2\{\exp(\sigma_v^2 + \sigma_u^2) - 1\}$ | $\mu_{ijk}^{M} + (\mu_{ijk}^{M})^2\{\exp(\sigma_v^2 + \sigma_u^2 + \sigma_e^2) - 1\}$ | $\mu_{ijk}^{M} + (\mu_{ijk}^{M})^2\{\exp(\sigma_v^2 + \sigma_u^2)(1+\alpha) - 1\}$ | $(1+\delta)\mu_{ijk}^{M} + (\mu_{ijk}^{M})^2\{\exp(\sigma_v^2 + \sigma_u^2) - 1\}$ |
| Level-3 component | $\mathrm{Var}\{\mu_{ijk}^{C3} \equiv \mid \mathbf{x}_{ijk}\}$ | $(\mu_{ijk}^{M})^2\{\exp(\sigma_v^2) - 1\}$ | $(\mu_{ijk}^{M})^2\{\exp(\sigma_v^2) - 1\}$ | $(\mu_{ijk}^{M})^2\{\exp(\sigma_v^2) - 1\}$ | $(\mu_{ijk}^{M})^2\{\exp(\sigma_v^2) - 1\}$ |
| Level-2 component | $\mathrm{E}\{\mathrm{Var}(\mu_{ijk}^{C23} \mid \mathbf{x}_{ijk}, v_k) \mid \mathbf{x}_{ijk}\}$ | $(\mu_{ijk}^{M})^2 \exp(\sigma_v^2)\{\exp(\sigma_u^2) - 1\}$ | $(\mu_{ijk}^{M})^2 \exp(\sigma_v^2)\{\exp(\sigma_u^2) - 1\}$ | $(\mu_{ijk}^{M})^2 \exp(\sigma_v^2)\{\exp(\sigma_u^2) - 1\}$ | $(\mu_{ijk}^{M})^2 \exp(\sigma_v^2)\{\exp(\sigma_u^2) - 1\}$ |
| Level-1 component | $\mathrm{E}\{\omega_{ijk}^{C23} \mid \mathbf{x}_{ijk}\}$ | $\mu_{ijk}^{M}$ | $\mu_{ijk}^{M} + (\mu_{ijk}^{M})^2 \exp(\sigma_v^2 + \sigma_u^2)\{\exp(\sigma_e^2) - 1\}$ | $\mu_{ijk}^{M} + (\mu_{ijk}^{M})^2 \exp(\sigma_v^2 + \sigma_u^2)\,\alpha$ | $\mu_{ijk}^{M}(1+\delta)$ |

Note.

The above expressions are for three-level random-intercept models. The corresponding expressions for the three-level models with random coefficients are obtained by replacing $u_j$ and $\sigma_u^2$ in all expressions with $\mathbf{z}'_{\mathbf{u}ijk}\mathbf{u}_{jk}$ and $\mathbf{z}'_{\mathbf{u}ijk}\boldsymbol{\Omega}_{\mathbf{u}}\mathbf{z}_{\mathbf{u}ijk}$ and by replacing $v_k$ and $\sigma_v^2$ in all expressions with $\mathbf{z}'_{\mathbf{u}ijk}\mathbf{v}_k$ and $\mathbf{z}'_{\mathbf{v}ijk}\boldsymbol{\Omega}_{\mathbf{v}}\mathbf{z}_{ijk}$.

**S6. Derivation of the marginal expectation, variance, covariance and correlation in two- and three-level random-coefficient models**

All expressions derived in Supplemental materials S4 were for two-level random-intercept models. The corresponding expressions for two-level models with random coefficients are obtained by replacing $u_j$ and $\sigma_u^2$ in all expressions with $\mathbf{z}'_{ij}\mathbf{u}_j$ and $\mathbf{z}'_{ij}\mathbf{\Omega}_\mathbf{u}\mathbf{z}_{ij}$.

All expressions derived in Supplemental materials S5 were for three-level random-intercept models. The corresponding expressions for three-level models with random coefficients are obtained by replacing $u_j$ and $\sigma_u^2$ in all expressions with $\mathbf{z}'_{\mathbf{u}ijk}\mathbf{u}_{jk}$ and $\mathbf{z}'_{\mathbf{u}ijk}\mathbf{\Omega}_\mathbf{u}\mathbf{z}_{\mathbf{u}ijk}$ and by replacing $v_k$ and $\sigma_v^2$ in all expressions with $\mathbf{z}'_{\mathbf{u}ijk}\mathbf{v}_k$ and $\mathbf{z}'_{\mathbf{v}ijk}\mathbf{\Omega}_\mathbf{v}\mathbf{z}_{ijk}$.

## S7. Additional analyses for student absenteeism application

Table S7.1 presents variable definitions and summary statistics for the student covariates used in the models.

*Table S7.1*. Covariate distributions, including mean days absent.

| Student characteristic | N | Percent | Mean days absent |
|---|---|---|---|
| Prior attainment (quintile) | | | |
| 1 (lowest prior attainment) | 14366 | 21% | 10.1 |
| 2 | 13288 | 20% | 9.0 |
| 3 | 12703 | 19% | 8.3 |
| 4 | 16000 | 24% | 7.4 |
| 5 (highest prior attainment) | 10598 | 16% | 7.0 |
| Age | | | |
| Summer (youngest in year) | 17032 | 25% | 8.1 |
| Spring | 16495 | 25% | 8.2 |
| Winter | 16551 | 25% | 8.5 |
| Autumn (oldest in year) | 16877 | 25% | 8.8 |
| Gender | | | |
| Male | 33628 | 50% | 8.0 |
| Female | 33327 | 50% | 8.8 |
| Ethnicity | | | |
| White | 27803 | 41% | 9.6 |
| Mixed | 6181 | 09% | 9.6 |
| Asian | 13914 | 21% | 7.1 |
| Black | 14409 | 22% | 7.0 |
| Other | 4648 | 7% | 7.9 |
| Language | | | |
| English | 40529 | 61% | 9.2 |
| Not English | 26426 | 39% | 7.2 |
| SEN | | | |
| Not SEN | 57157 | 85% | 7.9 |
| SEN | 9798 | 15% | 11.4 |
| FSM | | | |
| Not FSM | 41606 | 62% | 7.3 |
| FSM | 25349 | 38% | 10.2 |

*Note*. Number of school districts: $K = 32$; number of schools: $J = 434$; number of students: $N = 66{,}955$. Prior attainment quintiles are based on an average test score across separate tests in English and maths taken five years earlier at the end of primary schooling, just before the start of secondary schooling.

## S8. Simulation method for calculating the VPC

Until now the only way to calculate the VPC and ICC (and other marginal statistics) after fitting multilevel count models was to use the simulation method. The simulation method can be used to approximate the unknown expressions for calculating the VPC and ICC. Now that we have derived these expressions, the simulation method is redundant. In this section, we confirm that our VPC and ICC derivations are correct by showing that had we used the simulation method rather than our expressions, we would have calculated the same values for the VPC and ICC in each model in our application.

The simulation method works by treating the fitted model as a data generating process. We then then simulate a single very large dataset. The marginal statistics are then estimated by forming different summaries of the simulated data. For example, in a model with no covariates, the marginal mean and variance are simple the mean and variance of the simulated counts.

**S8.1 Model 1: Two-level variance-components Poisson model for count responses**

Model 1 is a two-level variance-components Poisson model for count responses (Equation 1). The model, written out again for convenience, is as follows.

$$y_{ij}|\mu_{ij} \sim \text{Poisson}(\mu_{ij})$$
$$\ln(\mu_{ij}) = \beta_0 + u_j \qquad \text{(S8.1)}$$
$$u_j \sim N(0, \sigma_u^2)$$

The full results are presented in Table 1.

The parameter estimates are $\hat{\beta}_0 = 2.085$, $\hat{\sigma}_u^2 = 0.100$ and the VPCs and other marginal statistics are presented in Table S8.1 under the column "exact method". The simulation method for calculating these statistics consists of the following steps:

1. Create a new two-level dataset with $J$ schools and $n$ students per school
2. Assign each school a random effect value from the estimated distribution $u_j \sim N(0, \hat{\sigma}_u^2)$
3. Calculate the expected count $\mu_{ij} = \exp(\hat{\beta}_0 + u_j)$
4. Simulate the observed counts $y_{ij}|\mu_{ij} \sim \text{Poisson}(\mu_{ij})$
5. Calculate the marginal expectation as the sample mean of $y_{ij}$
6. Calculate the marginal variance as the sample variance of $y_{ij}$
7. Calculate the school (level-2) component of the marginal variance as the variance of the school means of $y_{ij}$
8. Calculate the student (level-1) component of the marginal variance as the variance of the student observed count $y_{ij}$ deviated from their school means

9. Calculate the school (level-2) VPC as the ratio of the school component of the marginal variance to the overall marginal variance
10. Calculate the student (level-1) VPC as the ratio of the student component of the marginal variance to the overall marginal variance

where $J$ and $n$ should be set to large values to minimise Monte Carlo error in the resulting marginal statistics.

Applying this method gives the VPCs (and other marginal statistics) presented under the column "Simulation method" in Table S8.1. As expected, these statistics effectively identical to those based on the exact method confirming that our derivations are correct.

*Table S8.1*. Comparison of the estimated marginal statistics for the two-level variance-components Poisson model (Model 1) based on the exact method and the simulation method.

| | Marginal statistics | |
|---|---|---|
| | Exact method | Simulation method |
| Marginal expectation | 8.46 | 8.42 |
| Marginal variance | 15.98 | 16.02 |
| School (level-2) component | 7.52 | 7.61 |
| Student (level-1) component | 8.46 | 8.84 |
| School (level-2) VPC | 0.47 | 0.48 |
| Student (level-1) VPC | 0.53 | 0.52 |

*Note*. Number of simulated schools $J = 10000$. Number of simulated students per school $n = 1000$. The estimates presented in the "Exact method" column replicate those presented in Table 1 (Model 1).

**S8.2 Model 2: Two-level variance-components negative binomial model for count responses**

Model 2 is a two-level variance-components negative binomial model for count responses (Equation 9). The model, written out again for convenience, is as follows.

$$y_{ij}|\mu_{ij} \sim \text{Poisson}(\mu_{ij})$$
$$\ln(\mu_{ij}) = \beta_0 + u_j + e_{ij} \tag{S8.2}$$
$$u_j \sim N(0, \sigma_u^2)$$
$$\exp(e_{ij}) \sim \text{Gamma}\left(\frac{1}{\alpha}, \alpha\right)$$

The full results are presented in Table 1.

The parameter estimates are $\hat{\beta}_0 = 2.088$, $\hat{\sigma}_u^2 = 0.093$, $\hat{\alpha} = 0.877$ and the VPCs and other marginal statistics are presented in Table S8.2 under the column "exact method". The simulation method for calculating these statistics consists of the following steps:

1. Create a new two-level dataset with $J$ schools and $n$ students per school
2. Assign each school a random effect value from the estimated distribution $u_j \sim N(0, \hat{\sigma}_u^2)$
3. Assign each student an exponentiated overdispersion random effect value from the estimated distribution $\exp(e_{ij}) \sim \text{Gamma}\left(\frac{1}{\hat{\alpha}}, \hat{\alpha}\right)$
4. Calculate the expected count $\mu_{ij} = \exp(\hat{\beta}_0 + u_j + e_{ij})$
5. Simulate the observed counts $y_{ij}|\mu_{ij} \sim \text{Poisson}(\mu_{ij})$
6. Calculate the marginal expectation as the sample mean of $y_{ij}$
7. Calculate the marginal variance as the sample variance of $y_{ij}$

8. Calculate the school (level-2) component of the marginal variance as the variance of the school means of $y_{ij}$
9. Calculate the student (level-1) component of the marginal variance as the variance of the student observed count $y_{ij}$ deviated from their school means
10. Calculate the school (level-2) VPC as the ratio of the school component of the marginal variance to the overall marginal variance
11. Calculate the student (level-1) VPC as the ratio of the student component of the marginal variance to the overall marginal variance

Applying this method gives the VPCs (and other marginal statistics) presented under the column “Simulation method” in Table S8.2. As expected, these statistics are effectively identical to those based on the exact method confirming that our derivations are correct.

*Table S8.2*. Comparison of the estimated marginal statistics for the two-level variance-components negative binomial model (Model 2) based on the exact method and the simulation method.

| | Marginal statistics | |
|---|---|---|
| | Exact method | Simulation method |
| Marginal expectation | 8.45 | 8.41 |
| Marginal variance | 84.10 | 88.25 |
| School (level-2) component | 6.95 | 6.92 |
| Student (level-1) component | 77.15 | 76.33 |
| School (level-2) VPC | 0.08 | 0.08 |
| Student (level-1) VPC | 0.92 | 0.92 |

*Note*. Number of simulated schools $J = 10000$. Number of simulated students per school $n = 1000$. The estimates presented in the "Exact method" column replicate those presented in Table 1 (Model 2).

**S8.3 Model 3: Three-level variance-components negative binomial model for count responses**

Model 3 is a three-level variance-components negative binomial model for count responses (Equation 17). The model, written out again for convenience, is as follows.

$$
\begin{gathered}
y_{ijk}|\mu_{ijk} \sim \text{Poisson}(\mu_{ijk}) \\
\ln(\mu_{ijk}) = \beta_0 + v_k + u_{jk} + e_{ijk} \\
v_k \sim N(0, \sigma_v^2) \\
u_{jk} \sim N(0, \sigma_u^2) \\
\exp(e_{ijk}) \sim \text{Gamma}\left(\frac{1}{\alpha}, \alpha\right)
\end{gathered} \tag{S8.3}
$$

The full results are presented in Table 1.

The parameter estimates are $\hat{\beta}_0 = 2.088$, $\hat{\sigma}_v^2 = 0.006$, $\hat{\sigma}_u^2 = 0.087$, $\hat{\alpha} = 0.877$ and the VPCs and other marginal statistics are presented in Table S8.3 under the column “exact method”. The simulation method for calculating these statistics consists of the following steps:

1. Create a new three-level dataset with $K$ districts, $J$ schools per district, and $n$ students per school
2. Assign each district a random effect value from the estimated distribution $v_k \sim N(0, \hat{\sigma}_v^2)$
3. Assign each school a random effect value from the estimated distribution $u_{jk} \sim N(0, \hat{\sigma}_u^2)$
4. Assign each student an exponentiated overdispersion random effect value from the estimated distribution $\exp(e_{ijk}) \sim \text{Gamma}\left(\frac{1}{\hat{\alpha}}, \hat{\alpha}\right)$

5. Calculate the expected count $\mu_{ijk} = \exp(\hat{\beta}_0 + v_k + u_{jk} + e_{ijk})$
6. Simulate the observed counts $y_{ijk}|\mu_{ijk} \sim \text{Poisson}(\mu_{ijk})$
7. Calculate the marginal expectation as the sample mean of $y_{ijk}$
8. Calculate the marginal variance as the sample variance of $y_{ijk}$
9. Calculate the district (level-3) component of the marginal variance as the variance of the district means of $y_{ijk}$
10. Calculate the school (level-2) component of the marginal variance as the variance of the school means of $y_{ijk}$ deviated from their district means
11. Calculate the student (level-1) component of the marginal variance as the variance of the of the student observed count $y_{ijk}$ deviated from their school means
12. Calculate the district (level-3) VPC as the ratio of the district component of the marginal variance to the overall marginal variance
13. Calculate the school (level-2) VPC as the ratio of the school component of the marginal variance to the overall marginal variance
14. Calculate the student (level-1) VPC as the ratio of the student component of the marginal variance to the overall marginal variance

Applying this method gives the VPCs (and other marginal statistics) presented under the column "Simulation method" in Table S8.3. As expected, these statistics are effectively identical to those based on the exact method confirming that our derivations are correct.

*Table S8.3*. Comparison of the estimated marginal statistics for the three-level variance-components negative binomial model (Model 3) based on the exact method and the simulation method.

| | Marginal statistics | |
|---|---|---|
| | Exact method | Simulation method |
| Marginal expectation | 8.44 | 8.53 |
| Marginal variance | 83.79 | 85.32 |
| District (level-3) component | 0.42 | 0.42 |
| School (level-2) component | 6.50 | 6.58 |
| Student (level-1) component | 76.87 | 78.32 |
| District (level-3) VPC | 0.005 | 0.005 |
| School (level-2) VPC | 0.08 | 0.08 |
| Student (level-1) VPC | 0.92 | 0.92 |

*Note*. Number of simulated districts $K = 100$. Number of simulated schools per district $J = 100$. Number of simulated students per school $n = 1000$. The estimates presented in the “Exact method” column replicate those presented in Table 1 (Model 3).

**S8.4 Model 4: Two-level random-intercept negative binomial model for count responses**

Model 4 is a two-level random-intercept negative binomial model for count responses (Equation 17 where we have substituted $\mathbf{x}'_{ijk}\boldsymbol{\beta}$ for $\beta_0$). The model, written out again for convenience, is as follows.

$$
\begin{gathered}
y_{ij}|\mu_{ij} \sim \text{Poisson}(\mu_{ij}) \\
\ln(\mu_{ij}) = \mathbf{x}'_{ijk}\boldsymbol{\beta} + u_j + e_{ij} \\
u_j \sim N(0, \sigma_u^2) \\
\exp(e_{ij}) \sim \text{Gamma}\left(\frac{1}{\alpha}, \alpha\right)
\end{gathered} \tag{S8.2}
$$

The full results are presented in Table 2.

In models with covariates, the VPCs and marginal statistics vary as a function of these covariates. For simplicity, we focus first on the reference student (a student who takes a zero value for each covariate). The only relevant parameters estimates for calculating the VPC for this student are $\hat{\beta}_0 = 2.126$, $\hat{\sigma}_u^2 = 0.103$, $\hat{\alpha} = 0.782$. The marginal statistics calculated using our expressions are presented in Table S8.2 under the column "exact method". The simulation method for calculating these statistics consists of the following steps:

1. Create a new two-level dataset with $J$ schools and $n$ students per school
2. Assign each school a random effect value from the estimated distribution $u_j \sim N(0, \hat{\sigma}_u^2)$
3. Assign each student an exponentiated overdispersion random effect value from the estimated distribution $\exp(e_{ij}) \sim \text{Gamma}\left(\frac{1}{\hat{\alpha}}, \hat{\alpha}\right)$

4. Calculate the expected count $\mu_{ij} = \exp(\hat{\beta}_0 + u_j + e_{ij})$
5. Simulate the observed counts $y_{ij}|\mu_{ij} \sim \text{Poisson}(\mu_{ij})$
6. Calculate the marginal expectation as the sample mean of $y_{ij}$
7. Calculate the marginal variance as the sample variance of $y_{ij}$
8. Calculate the school (level-2) component of the marginal variance as the variance of the school means of $y_{ij}$
9. Calculate the student (level-1) component of the marginal variance as the variance of the student observed count $y_{ij}$ deviated from their school means
10. Calculate the school (level-2) VPC as the ratio of the school component of the marginal variance to the overall marginal variance
11. Calculate the student (level-1) VPC as the ratio of the student component of the marginal variance to the overall marginal variance

Next, we contrast this reference student to an otherwise equivalent student who is eligible for free school meals ($x_{15ij}$ now takes the value 1, all other covariates continue to be held at 0). Now we must additionally consider $\hat{\beta}_{15} = 0.377$. Thus we repeat the above steps, but where we now replace Step 4 above with:

4. Calculate the expected count $\mu_{ij} = \exp(\hat{\beta}_0 + \hat{\beta}_{15} + u_j + e_{ij})$

Applying this method for these two hypothetical students gives the VPCs (and other marginal statistics) presented under the column “Simulation method” in Table S8.4. As expected, these statistics are effectively identical to those based on the exact method confirming that our derivations are correct.

*Table S8.4*. Comparison of the estimated marginal statistics for the two-level random-intercept negative binomial model (Model 4) based on the exact method and the simulation method.

| | Marginal statistics | |
|---|---|---|
| | Exact method | Simulation method |
| | Non-FSM student | |
| Marginal expectation | 8.82 | 8.80 |
| Marginal variance | 84.77 | 84.24 |
| School (level-2) component | 8.45 | 8.45 |
| Student (level-1) component | 76.32 | 75.79 |
| School (level-2) VPC | 0.10 | 0.10 |
| Student (level-1) VPC | 0.90 | 0.90 |
| | FSM student | |
| Marginal expectation | 12.86 | 12.83 |
| Marginal variance | 174.29 | 173.20 |
| School (level-2) component | 17.96 | 17.94 |
| Student (level-1) component | 156.33 | 155.26 |
| School (level-2) VPC | 0.10 | 0.10 |
| Student (level-1) VPC | 0.90 | 0.90 |

*Note*. Number of simulated schools $J = 10000$. Number of simulated students per school $n = 1000$.

**S8.5 Model 5: Two-level random-coefficient negative binomial model for count responses**

Model 5 is a two-level random-coefficient negative binomial model for count responses (Equation 21 where we have substituted $\mathbf{x}'_{ijk}\boldsymbol{\beta}$ for $\beta_0$). Recall that the model simply extends the previous random-intercept model by allowing the coefficient on the binary indicator of free school meal (FSM) eligibility to vary randomly across schools. The model, written out again for convenience, is as follows.

$$y_{ij}|\mu_{ij} \sim \text{Poisson}(\mu_{ij})$$

$$\ln(\mu_{ij}) = \mathbf{x}'_{ijk}\boldsymbol{\beta} + \mathbf{z}'_{ijk}\mathbf{u}_j + e_{ij} \quad \text{(S8.2)}$$

$$u_j \sim N(0, \sigma_u^2)$$

$$\exp(e_{ij}) \sim \text{Gamma}\left(\frac{1}{\alpha}, \alpha\right)$$

where $\mathbf{z}_{ijk} = x_{15ij}$ (the binary indicator of FSM eligibility). The full results are presented in Table 2.

As in the previous random-intercept model, we first focus on the reference student (a student who takes a zero value for each covariate). The only relevant parameters estimates for calculating the VPC for this student are $\hat{\beta}_0 = 2.126$, $\hat{\sigma}_{u0}^2 = 0.116$, $\hat{\alpha} = 0.775$. The marginal statistics calculated using our expressions are presented in Table S8.2 under the column "exact method". The simulation method for calculating these statistics consists of the following steps:

1. Create a new two-level dataset with $J$ schools and $n$ students per school
2. Assign each school a random effect value from the estimated distribution $u_{0j} \sim N(0, \hat{\sigma}_{u0}^2)$

3. Assign each student an exponentiated overdispersion random effect value from the estimated distribution $\exp(e_{ij}) \sim \text{Gamma}\left(\frac{1}{\hat{\alpha}}, \hat{\alpha}\right)$
4. Calculate the expected count $\mu_{ij} = \exp(\hat{\beta}_0 + u_{0j} + e_{ij})$
5. Simulate the observed counts $y_{ij}|\mu_{ij} \sim \text{Poisson}(\mu_{ij})$
6. Calculate the marginal expectation as the sample mean of $y_{ij}$
7. Calculate the marginal variance as the sample variance of $y_{ij}$
8. Calculate the school (level-2) component of the marginal variance as the variance of the school means of $y_{ij}$
9. Calculate the student (level-1) component of the marginal variance as the variance of the student observed count $y_{ij}$ deviated from their school means
10. Calculate the school (level-2) VPC as the ratio of the school component of the marginal variance to the overall marginal variance
11. Calculate the student (level-1) VPC as the ratio of the student component of the marginal variance to the overall marginal variance

Next, we contrast this reference student to an otherwise equivalent student who is eligible for free school meals ($x_{15ij}$ now takes the value 1, all other covariates continue to be held at 0). Now we must additionally consider $\hat{\beta}_{15} = 0.372$, $\hat{\sigma}^2_{u15} = 0.035$, $\hat{\sigma}_{u015} = -0.027$. Thus we repeat the above steps, but where we now replace Steps 2 and 4 above with:

2. Assign each school a random effect value from the estimated distribution

$$\begin{pmatrix} u_{0j} \\ u_{15j} \end{pmatrix} \sim N \left\{ \begin{pmatrix} 0 \\ 0 \end{pmatrix}, \begin{pmatrix} \hat{\sigma}^2_{u0} & \\ \hat{\sigma}_{u015} & \hat{\sigma}^2_{u15} \end{pmatrix} \right\}$$

and

4. Calculate the expected count $\mu_{ij} = \exp(\hat{\beta}_0 + \hat{\beta}_{15} + u_{0j} + u_{15j} + e_{ij})$

Applying this method for these two hypothetical students gives the VPCs (and other marginal statistics) presented under the column “Simulation method” in Table S8.5. As expected, these statistics are effectively identical to those based on the exact method confirming that our derivations are correct.

*Table S8.5*. Comparison of the estimated marginal statistics for the two-level random-coefficient negative binomial model (Model 4) based on the exact method and the simulation method.

| | Marginal statistics | |
|---|---|---|
| | Exact method | Simulation method |
| | Non-FSM student | |
| Marginal expectation | 8.88 | 8.85 |
| Marginal variance | 87.24 | 86.68 |
| School (level-2) component | 9.70 | 9.71 |
| Student (level-1) component | 77.54 | 76.97 |
| School (level-2) VPC | 0.11 | 0.11 |
| Student (level-1) VPC | 0.89 | 0.89 |
| | FSM student | |
| Marginal expectation | 12.76 | 12.77 |
| Marginal variance | 168.44 | 168.84 |
| School (level-2) component | 16.59 | 16.79 |
| Student (level-1) component | 154.85 | 152.05 |
| School (level-2) VPC | 0.10 | 0.10 |
| Student (level-1) VPC | 0.90 | 0.90 |

*Note*. Number of simulated schools $J = 10000$. Number of simulated students per school $n = 1000$.

## S9. Stata syntax and output for student absenteeism application

In this section we present Stata syntax and output to replicate the results for models 1-5 presented in Tables 1 and 2 of the paper. The syntax is annotated to help the reader.

### S9.1 Stata syntax

```
********************************************************************************
* Variance partitioning in multilevel count models
********************************************************************************
* Stata do-file to replicate results presented in the article
********************************************************************************

********************************************************************************
* Basics
********************************************************************************

* Set model results display format to three decimal places
set cformat %9.3f

* Load the data
use "absence.dta", clear

* Figure 1: Spikeplot of days absent
spikeplot y

* Figure 2: Bar chart of district means
graph bar (mean) y, over(district, sort(y) label(nolabel))

* Figure 2: Bar chart of school means
graph bar (mean) y, over(school, sort(y) label(nolabel))

* Table S7.1 Covariate distributions, including mean days absent
tabstat y, statistics(count mean) by(quintile) nototal
tabstat y, statistics(count mean) by(season) nototal
tabstat y, statistics(count mean) by(female) nototal
tabstat y, statistics(count mean) by(ethnicity) nototal
tabstat y, statistics(count mean) by(notenglish) nototal
tabstat y, statistics(count mean) by(sen) nototal
tabstat y, statistics(count mean) by(fsm) nototal



********************************************************************************
* Table 1: Model 1: Two-level variance-components Poisson model
********************************************************************************

* Clear memory
clear all

* Load the data
use "absence.dta", clear

* Fit model
mepoisson y || school:, startvalues(iv)

* Deviance
```

```
scalar define deviance = -2*e(ll)
scalar list deviance

* Intercept
scalar define beta0 = _b[_cons]
scalar list beta0

* Cluster variance
scalar define sigma2u = _b[/var(_cons[school])]
scalar list sigma2u

* Marginal expectation
scalar define expectation = exp(beta0 + sigma2u/2)
scalar list expectation

* Marginal variance
scalar define variance = expectation + expectation^2*(exp(sigma2u) - 1)
scalar list variance

* Marginal variance: Level-2 component
scalar define variance2 = expectation^2*(exp(sigma2u) - 1)
scalar list variance2

* Marginal variance: Level-1 component
scalar define variance1 = expectation
scalar list variance1

* Level-2 VPC
scalar define vpc2 = variance2/(variance2 + variance1)
scalar list vpc2

* Level-1 VPC
scalar define vpc1 = variance1/(variance2 + variance1)
scalar list vpc1


********************************************************************************
* Table 1: Model 2: Two-level variance-components negative binomial model
********************************************************************************

* Clear memory
clear all

* Load the data
use "absence.dta", clear

* Fit model
menbreg y || school:, startvalues(iv)

* Deviance
scalar define deviance = -2*e(ll)
scalar list deviance

* Intercept
scalar define beta0 = _b[_cons]
scalar list beta0

* Cluster variance
scalar define sigma2u = _b[/var(_cons[school])]
scalar list sigma2u

* Overdispersion parameter
```

```
scalar define alpha = exp(_b[/lnalpha])
scalar list alpha

* Marginal expectation
scalar define expectation = exp(beta0 + sigma2u/2)
scalar list expectation

* Marginal variance
scalar define variance = expectation ///
  + expectation^2*(exp(sigma2u)*(1 + alpha) - 1)
scalar list variance

* Marginal variance: Level-2 component
scalar define variance2 = expectation^2*(exp(sigma2u) - 1)
scalar list variance2

* Marginal variance: Level-1 component
scalar define variance1 = expectation + expectation^2*exp(sigma2u)*alpha
scalar list variance1

* Level-2 VPC
scalar define vpc2 = variance2/(variance2 + variance1)
scalar list vpc2

* Level-1 VPC
scalar define vpc1 = variance1/(variance2 + variance1)
scalar list vpc1

* Predict cluster random intercept effects
predict u, reffects

* Keep the cluster identifier and the predicted cluster effect
keep school u

* Collapse the data down to one observation per cluster
duplicates drop

* Save the predicted cluster effects
save "model2.dta", replace



********************************************************************************
* Table 1: Model 3: Three-level variance-components negative binomial model
********************************************************************************

* Clear memory
clear all

* Load the data
use "absence.dta", clear

* Fit model
menbreg y || district: || school:, startvalues(iv)

* Deviance
scalar define deviance = -2*e(ll)
scalar list deviance

* Intercept
scalar define beta0 = _b[_cons]
scalar list beta0
```

```
* Supercluster variance
scalar define sigma2v = _b[/var(_cons[district])]
scalar list sigma2v

* Cluster variance
scalar define sigma2u = _b[/var(_cons[school<district])]
scalar list sigma2u

* Overdispersion parameter
scalar define alpha = exp(_b[/lnalpha])
scalar list alpha

* Marginal expectation
scalar define expectation = exp(beta0 + sigma2v/2 + sigma2u/2)
scalar list expectation

* Marginal variance
scalar define variance = expectation ///
  + expectation^2*(exp(sigma2v + sigma2u)*(1 + alpha) - 1)
scalar list variance

* Marginal variance: Level-3 component
scalar define variance3 = expectation^2*(exp(sigma2v) - 1)
scalar list variance3

* Marginal variance: Level-2 component
scalar define variance2 = expectation^2*exp(sigma2v)*(exp(sigma2u) - 1)
scalar list variance2

* Marginal variance: Level-1 component
scalar define variance1 = expectation ///
  + expectation^2*exp(sigma2v + sigma2u)*alpha
scalar list variance1

* Level-3 VPC
scalar define vpc3 = variance3/(variance3 + variance2 + variance1)
scalar list vpc3

* Level-2 VPC
scalar define vpc2 = variance2/(variance3 + variance2 + variance1)
scalar list vpc2

* Level-1 VPC
scalar define vpc1 = variance1/(variance3 + variance2 + variance1)
scalar list vpc1



********************************************************************************
* Table 2: Model 4: Two-level random-intercept negative binomial model
********************************************************************************

* Clear memory
clear all

* Load the data
use "absence.dta", clear

* Fit model
menbreg y x || school:, startvalues(iv)

* Deviance
scalar define deviance = -2*e(ll)
```

```
scalar list deviance

* Linear predictor
predict xb, xb

* Cluster variance
scalar define sigma2u = _b[/var(_cons[school])]
scalar list sigma2u

* Overdispersion parameter
scalar define alpha = exp(_b[/lnalpha])
scalar list alpha

* Marginal expectation
generate expectation = exp(xb + sigma2u/2)

* Marginal variance
generate variance = expectation + expectation^2*(exp(sigma2u)*(1 + alpha) - 1)

* Marginal variance: Level-2 component
generate variance2 = expectation^2*(exp(sigma2u) - 1)

* Marginal variance: Level-1 component
generate variance1 = expectation + expectation^2*exp(sigma2u)*alpha

* Level-2 VPC
generate vpc2 = variance2/(variance2 + variance1)

* Level-1 VPC
generate vpc1 = variance1/(variance2 + variance1)

* Summarize marginal statistics
summarize expectation variance variance2 variance1 vpc2 vpc1

* Figure 3: Line plot of Level-2 VPC against the marginal expectation
line vpc2 expectation, sort

* Figure 3: Spikeplot of marginal expectation
spikeplot expectation

* Predict cluster random intercept effects
predict u, reffects

* Keep the cluster identifier and the predicted cluster effect
keep school u

* Collapse the data down to one observation per cluster
duplicates drop

* Save the predicted cluster effects
save "model4.dta", replace

* Load model 2 predicted cluster random effects
use "model2.dta", clear
rename u model2u

* Merge in model 4 predicted cluster random effects
merge 1:1 school using "model4"
rename u model4u

* Figure 4: Scatterplot of model 4 vs. model 2 predicted cluster random effects
scatter model4u model2u
correlate model4u model2u
```

```
* Rank the model 2 predicted cluster random effects
sort model2u
generate model2urank = _n

* Rank the model 4 predicted cluster random effects
sort model4u
generate model4urank = _n

* Figure 4: Scatterplot of ranks of model 4 vs. model 2 predicted effects
scatter model4urank model2urank
correlate model4urank model2urank



********************************************************************************
* Table 2: Model 5: Two-level random-coefficient negative binomial model
********************************************************************************

* Clear memory
clear all

* Load the data
use "absence.dta", clear

* Fit model
menbreg y x  ///
  || school: x, covariance(unstructured) ///
  startvalues(iv)

* Deviance
scalar define deviance = -2*e(ll)
scalar list deviance

* Linear predictor
predict xb, xb

* Cluster intercept variance
scalar define sigma2u0 = _b[/var(_cons[school])]
scalar list sigma2u0

* Cluster slope variance
scalar define sigma2u15 = _b[/var(fsm[school])]
scalar list sigma2u15

* Cluster intercept-slope covariance
scalar define sigmau015 = _b[/cov(fsm[school],_cons[school])]
scalar list sigmau015

* Overdispersion parameter
scalar define alpha = exp(_b[/lnalpha])
scalar list alpha

* Cluster-level variance function
generate zomegauz = sigma2u0 + 2*sigmau015*fsm + sigma2u15*fsm^2

* Marginal expectation
generate expectation = exp(xb + zomegauz/2)

* Marginal variance
generate variance = expectation + expectation^2*(exp(zomegauz)*(1 + alpha) - 1)

* Marginal variance: Level-2 component
```

```
generate variance2 = expectation^2*(exp(zomegauz) - 1)

* Marginal variance: Level-1 component
generate variance1 = expectation + expectation^2*exp(zomegauz)*alpha

* Level-2 VPC
generate vpc2 = variance2/(variance2 + variance1)

* Level-1 VPC
generate vpc1 = variance1/(variance2 + variance1)

* Summarize marginal statistics
summarize expectation variance variance2 variance1 vpc2 vpc1

* Figure 5: Line plot of Level-2 VPC against the marginal expectation by FSM status
line vpc2 expectation, sort by(fsm)

* Figure 5: Spikeplot of marginal expectation by FSM status
spikeplot expectation,  by(fsm)


********************************************************************************
exit
```

## S9.2 Stata output

```
. ********************************************************************************
. * Variance partitioning in multilevel count models
. ********************************************************************************
. * Stata do-file to replicate results presented in the article
. ********************************************************************************
.
. ********************************************************************************
. * Basics
. ********************************************************************************
.
. * Set model results display format to three decimal places
. set cformat %9.3f

.
. * Load the data
. use "absence.dta", clear

.
. * Figure 1: Spikeplot of days absent
. spikeplot y

.
. * Figure 2: Bar chart of district means
. graph bar (mean) y, over(district, sort(y) label(nolabel))

.
. * Figure 2: Bar chart of school means
. graph bar (mean) y, over(school, sort(y) label(nolabel))

.
. * Table S7.1 Covariate distributions, including mean days absent
. tabstat y, statistics(count mean) by(quintile) nototal

Summary for variables: y
```

```
     by categories of: quintile (5 quantiles of x)

quintile |         N      mean
---------+--------------------
       1 |     14366  10.05972
       2 |     13288  8.992023
       3 |     12703  8.334567
       4 |     16000  7.421375
       5 |     10598  7.029251
------------------------------

. tabstat y, statistics(count mean) by(season) nototal

Summary for variables: y
     by categories of: season

season |         N      mean
-------+--------------------
Summer |     17032  8.120538
Spring |     16495  8.225523
Winter |     16551  8.540995
Autumn |     16877  8.755407
----------------------------

. tabstat y, statistics(count mean) by(female) nototal

Summary for variables: y
     by categories of: female (Female)

  female |         N      mean
---------+--------------------
       0 |     33628  8.025009
       1 |     33327  8.799202
------------------------------

. tabstat y, statistics(count mean) by(ethnicity) nototal

Summary for variables: y
     by categories of: ethnicity (Ethnicity)

ethnicity |         N      mean
----------+--------------------
    White |     27803  9.608711
    Mixed |      6181  9.639541
    Asian |     13914  7.072229
    Black |     14409  7.013672
    Other |      4648  7.943201
-------------------------------

. tabstat y, statistics(count mean) by(notenglish) nototal

Summary for variables: y
     by categories of: notenglish (English as an additional language)

notenglish |         N      mean
-----------+--------------------
         0 |     40529  9.231439
         1 |     26426  7.151101
--------------------------------

. tabstat y, statistics(count mean) by(sen) nototal

Summary for variables: y
```

```
     by categories of: sen (SEN)

     sen |         N      mean
---------+--------------------
       0 |     57157  7.892612
       1 |      9798   11.4307
------------------------------

. tabstat y, statistics(count mean) by(fsm) nototal

Summary for variables: y
     by categories of: fsm (FSM)

     fsm |         N      mean
---------+--------------------
       0 |     41606  7.311253
       1 |     25349  10.21437
------------------------------

.
.
. ******************************************************************************
. * Table 1: Model 1: Two-level variance-components Poisson model
. ******************************************************************************
.
. * Clear memory
. clear all

.
. * Load the data
. use "absence.dta", clear

.
. * Fit model
. mepoisson y || school:, startvalues(iv)

Fitting fixed-effects model:

Iteration 0:   log likelihood = -761235.06
Iteration 1:   log likelihood =  -421985.3
Iteration 2:   log likelihood = -419174.27
Iteration 3:   log likelihood = -419168.08
Iteration 4:   log likelihood = -419168.08

Refining starting values:

Grid node 0:   log likelihood = -393452.09

Refining starting values (unscaled likelihoods):

Grid node 0:   log likelihood = -393459.83

Fitting full model:

Iteration 0:   log likelihood = -393459.83  (not concave)
Iteration 1:   log likelihood = -393002.29
Iteration 2:   log likelihood = -392866.51
Iteration 3:   log likelihood = -392768.39
Iteration 4:   log likelihood = -392698.35
Iteration 5:   log likelihood = -392649.56
Iteration 6:   log likelihood = -392616.78
Iteration 7:   log likelihood = -392595.86
Iteration 8:   log likelihood = -392583.38
```

```
Iteration 9:   log likelihood = -392572.82
Iteration 10:  log likelihood = -392571.14
Iteration 11:  log likelihood = -392571.07
Iteration 12:  log likelihood = -392571.07

Mixed-effects Poisson regression               Number of obs     =     66,955
Group variable:          school                Number of groups  =        434

                                               Obs per group:
                                                             min =         31
                                                             avg =      154.3
                                                             max =        315

Integration method: mvaghermite                Integration pts.  =          7

                                               Wald chi2(0)      =          .
Log likelihood = -392571.07                    Prob > chi2       =          .
------------------------------------------------------------------------------
           y |      Coef.   Std. Err.      z    P>|z|     [95% Conf. Interval]
-------------+----------------------------------------------------------------
       _cons |      2.085      0.015   136.72   0.000        2.055       2.115
-------------+----------------------------------------------------------------
school       |
   var(_cons)|      0.100      0.007                         0.087       0.114
------------------------------------------------------------------------------
LR test vs. Poisson model: chibar2(01) = 53194.02     Prob >= chibar2 = 0.0000

.
. * Deviance
. scalar define deviance = -2*e(ll)

. scalar list deviance
  deviance =  785142.14

.
. * Intercept
. scalar define beta0 = _b[_cons]

. scalar list beta0
     beta0 =  2.0852543

.
. * Cluster variance
. scalar define sigma2u = _b[/var(_cons[school])]

. scalar list sigma2u
   sigma2u =  .09998112

.
. * Marginal expectation
. scalar define expectation = exp(beta0 + sigma2u/2)

. scalar list expectation
expectation =  8.4591173

.
. * Marginal variance
. scalar define variance = expectation + expectation^2*(exp(sigma2u) - 1)

. scalar list variance
  variance =  15.983304

.
```

```
. * Marginal variance: Level-2 component
. scalar define variance2 = expectation^2*(exp(sigma2u) - 1)

. scalar list variance2
 variance2 =  7.5241871

.
. * Marginal variance: Level-1 component
. scalar define variance1 = expectation

. scalar list variance1
 variance1 =  8.4591173

.
. * Level-2 VPC
. scalar define vpc2 = variance2/(variance2 + variance1)

. scalar list vpc2
      vpc2 =  .47075291

.
. * Level-1 VPC
. scalar define vpc1 = variance1/(variance2 + variance1)

. scalar list vpc1
      vpc1 =  .52924709

.
.
.
. ********************************************************************************
. * Table 1: Model 2: Two-level variance-components negative binomial model
. ********************************************************************************
.
. * Clear memory
. clear all

.
. * Load the data
. use "absence.dta", clear

.
. * Fit model
. menbreg y || school:, startvalues(iv)

Fitting fixed-effects model:

Iteration 0:   log likelihood =  -216299.9
Iteration 1:   log likelihood = -213455.56
Iteration 2:   log likelihood = -213346.63
Iteration 3:   log likelihood = -213346.62

Refining starting values:

Grid node 0:   log likelihood = -211608.67

Fitting full model:

Iteration 0:   log likelihood = -211608.67
Iteration 1:   log likelihood = -211226.32
Iteration 2:   log likelihood = -211072.56
Iteration 3:   log likelihood = -211029.19
Iteration 4:   log likelihood = -211023.25
```

```
Iteration 5:   log likelihood = -211023.09
Iteration 6:   log likelihood = -211023.09

Mixed-effects nbinomial regression             Number of obs     =     66,955
Overdispersion:             mean
Group variable:           school               Number of groups  =        434

                                               Obs per group:
                                                             min =         31
                                                             avg =      154.3
                                                             max =        315

Integration method: mvaghermite                Integration pts.  =          7

                                               Wald chi2(0)      =          .
Log likelihood = -211023.09                    Prob > chi2       =          .
------------------------------------------------------------------------------
           y |      Coef.   Std. Err.      z    P>|z|     [95% Conf. Interval]
-------------+----------------------------------------------------------------
       _cons |      2.088      0.015   137.30   0.000        2.058       2.118
-------------+----------------------------------------------------------------
    /lnalpha |     -0.132      0.006   -21.24   0.000       -0.144      -0.120
-------------+----------------------------------------------------------------
school       |
   var(_cons)|      0.093      0.007                         0.080       0.107
------------------------------------------------------------------------------
LR test vs. nbinomial model: chibar2(01) = 4647.07    Prob >= chibar2 = 0.0000

.
. * Deviance
. scalar define deviance = -2*e(ll)

. scalar list deviance
  deviance =  422046.17

.
. * Intercept
. scalar define beta0 = _b[_cons]

. scalar list beta0
     beta0 =  2.0878598

.
. * Cluster variance
. scalar define sigma2u = _b[/var(_cons[school])]

. scalar list sigma2u
   sigma2u =  .09284542

.
. * Overdispersion parameter
. scalar define alpha = exp(_b[/lnalpha])

. scalar list alpha
     alpha =    .876623

.
. * Marginal expectation
. scalar define expectation = exp(beta0 + sigma2u/2)

. scalar list expectation
expectation =  8.4509804
```

```
.
. * Marginal variance
. scalar define variance = expectation ///
>   + expectation^2*(exp(sigma2u)*(1 + alpha) - 1)

. scalar list variance
  variance =  84.098316

.
. * Marginal variance: Level-2 component
. scalar define variance2 = expectation^2*(exp(sigma2u) - 1)

. scalar list variance2
 variance2 =  6.9485112

.
. * Marginal variance: Level-1 component
. scalar define variance1 = expectation + expectation^2*exp(sigma2u)*alpha

. scalar list variance1
 variance1 =  77.149805

.
. * Level-2 VPC
. scalar define vpc2 = variance2/(variance2 + variance1)

. scalar list vpc2
      vpc2 =  .08262367

.
. * Level-1 VPC
. scalar define vpc1 = variance1/(variance2 + variance1)

. scalar list vpc1
      vpc1 =  .91737633

.
. * Predict cluster random intercept effects
. predict u, reffects
(calculating posterior means of random effects)
(using 7 quadrature points)

.
. * Keep the cluster identifier and the predicted cluster effect
. keep school u

.
. * Collapse the data down to one observation per cluster
. duplicates drop

Duplicates in terms of all variables

(66,521 observations deleted)

.
. * Save the predicted cluster effects
. save "model2.dta", replace
file model2.dta saved

.
.
.
. ********************************************************************************
```

```
. * Table 1: Model 3: Three-level variance-components negative binomial model
. ***************************************************************************
.
. * Clear memory
. clear all

.
. * Load the data
. use "absence.dta", clear

.
. * Fit model
. menbreg y || district: || school:, startvalues(iv)

Fitting fixed-effects model:

Iteration 0:   log likelihood =  -216299.9
Iteration 1:   log likelihood = -213455.56
Iteration 2:   log likelihood = -213346.63
Iteration 3:   log likelihood = -213346.62

Refining starting values:

Grid node 0:   log likelihood = -211551.43

Fitting full model:

Iteration 0:   log likelihood = -211551.43
Iteration 1:   log likelihood = -211198.25
Iteration 2:   log likelihood = -211061.47
Iteration 3:   log likelihood = -211024.08
Iteration 4:   log likelihood = -211019.52
Iteration 5:   log likelihood = -211019.44
Iteration 6:   log likelihood = -211019.44

Mixed-effects nbinomial regression              Number of obs     =     66,955
Overdispersion:            mean

-------------------------------------------------------------
                |     No. of       Observations per Group
 Group Variable |     Groups    Minimum    Average    Maximum
----------------+--------------------------------------------
       district |         32        665    2,092.3      3,182
         school |        434         31      154.3        315
-------------------------------------------------------------

Integration method: mvaghermite                 Integration pts.  =          7

                                                Wald chi2(0)      =          .
Log likelihood = -211019.44                     Prob > chi2       =          .
--------------------------------------------------------------------------------
              y |      Coef.   Std. Err.      z    P>|z|     [95% Conf. Interval]
----------------+---------------------------------------------------------------
          _cons |      2.086      0.020   103.17   0.000        2.046       2.126
----------------+---------------------------------------------------------------
       /lnalpha |     -0.132      0.006   -21.24   0.000       -0.144      -0.120
----------------+---------------------------------------------------------------
district        |
      var(_cons)|      0.006      0.003                         0.002       0.017
----------------+---------------------------------------------------------------
district>school |
      var(_cons)|      0.087      0.007                         0.075       0.101
--------------------------------------------------------------------------------
```

```
LR test vs. nbinomial model: chi2(2) = 4654.36           Prob > chi2 = 0.0000

Note: LR test is conservative and provided only for reference.

.
. * Deviance
. scalar define deviance = -2*e(ll)

. scalar list deviance
  deviance =  422038.88

.
. * Intercept
. scalar define beta0 = _b[_cons]

. scalar list beta0
     beta0 =  2.0860497

.
. * Supercluster variance
. scalar define sigma2v = _b[/var(_cons[district])]

. scalar list sigma2v
   sigma2v =  .00582819

.
. * Cluster variance
. scalar define sigma2u = _b[/var(_cons[school<district])]

. scalar list sigma2u
   sigma2u =  .08692447

.
. * Overdispersion parameter
. scalar define alpha = exp(_b[/lnalpha])

. scalar list alpha
     alpha =   .8766216

.
. * Marginal expectation
. scalar define expectation = exp(beta0 + sigma2v/2 + sigma2u/2)

. scalar list expectation
expectation =  8.4353062

.
. * Marginal variance
. scalar define variance = expectation ///
>   + expectation^2*(exp(sigma2v + sigma2u)*(1 + alpha) - 1)

. scalar list variance
  variance =  83.788592

.
. * Marginal variance: Level-3 component
. scalar define variance3 = expectation^2*(exp(sigma2v) - 1)

. scalar list variance3
 variance3 =  .41591198

.
. * Marginal variance: Level-2 component
```

```
. scalar define variance2 = expectation^2*exp(sigma2v)*(exp(sigma2u) - 1)

. scalar list variance2
 variance2 =  6.4996057

.
. * Marginal variance: Level-1 component
. scalar define variance1 = expectation ///
>   + expectation^2*exp(sigma2v + sigma2u)*alpha

. scalar list variance1
 variance1 =  76.873075

.
. * Level-3 VPC
. scalar define vpc3 = variance3/(variance3 + variance2 + variance1)

. scalar list vpc3
      vpc3 =  .00496383

.
. * Level-2 VPC
. scalar define vpc2 = variance2/(variance3 + variance2 + variance1)

. scalar list vpc2
      vpc2 =  .07757149

.
. * Level-1 VPC
. scalar define vpc1 = variance1/(variance3 + variance2 + variance1)

. scalar list vpc1
      vpc1 =  .91746469

.
.
.
. ********************************************************************************
. * Table 2: Model 4: Two-level random-intercept negative binomial model
. ********************************************************************************
.
. * Clear memory
. clear all

.
. * Load the data
. use "absence.dta", clear

.
. * Fit model
. menbreg y quintile2 quintile3 quintile4 quintile5 spring winter autumn ///
>   female mixed asian black other notenglish sen fsm || school:, startvalues(iv)

Fitting fixed-effects model:

Iteration 0:   log likelihood = -213128.96
Iteration 1:   log likelihood = -210651.83
Iteration 2:   log likelihood = -210545.23
Iteration 3:   log likelihood = -210545.17
Iteration 4:   log likelihood = -210545.17

Refining starting values:
```

```
Grid node 0:   log likelihood = -208325.29

Fitting full model:

Iteration 0:   log likelihood = -208325.29
Iteration 1:   log likelihood = -207903.92
Iteration 2:   log likelihood = -207750.23
Iteration 3:   log likelihood = -207720.21
Iteration 4:   log likelihood = -207718.81
Iteration 5:   log likelihood =  -207718.8

Mixed-effects nbinomial regression              Number of obs     =     66,955
Overdispersion:             mean
Group variable:           school                Number of groups  =        434

                                                Obs per group:
                                                              min =         31
                                                              avg =      154.3
                                                              max =        315

Integration method: mvaghermite                 Integration pts.  =          7

                                                Wald chi2(15)     =    6724.95
Log likelihood =  -207718.8                     Prob > chi2       =     0.0000
------------------------------------------------------------------------------
           y |      Coef.   Std. Err.      z    P>|z|     [95% Conf. Interval]
-------------+----------------------------------------------------------------
   quintile2 |     -0.051      0.012    -4.34   0.000       -0.074      -0.028
   quintile3 |     -0.118      0.012    -9.71   0.000       -0.142      -0.094
   quintile4 |     -0.222      0.012   -18.77   0.000       -0.245      -0.199
   quintile5 |     -0.330      0.014   -23.60   0.000       -0.358      -0.303
      spring |      0.026      0.011     2.50   0.013        0.006       0.047
      winter |      0.077      0.011     7.33   0.000        0.057       0.098
      autumn |      0.112      0.011    10.59   0.000        0.091       0.132
      female |      0.122      0.009    13.83   0.000        0.104       0.139
       mixed |     -0.073      0.014    -5.28   0.000       -0.100      -0.046
       asian |     -0.194      0.013   -15.25   0.000       -0.219      -0.169
       black |     -0.422      0.011   -38.32   0.000       -0.444      -0.400
       other |     -0.194      0.017   -11.49   0.000       -0.227      -0.161
  notenglish |     -0.244      0.009   -26.01   0.000       -0.262      -0.226
         sen |      0.267      0.011    23.42   0.000        0.245       0.290
         fsm |      0.377      0.008    44.91   0.000        0.360       0.393
       _cons |      2.126      0.021   103.07   0.000        2.086       2.167
-------------+----------------------------------------------------------------
    /lnalpha |     -0.246      0.006   -38.41   0.000       -0.259      -0.234
-------------+----------------------------------------------------------------
school       |
   var(_cons)|      0.103      0.007                         0.089       0.118
------------------------------------------------------------------------------
LR test vs. nbinomial model: chibar2(01) = 5652.76    Prob >= chibar2 = 0.0000

.
. * Deviance
. scalar define deviance = -2*e(ll)

. scalar list deviance
  deviance =  415437.59

.
. * Linear predictor
. predict xb, xb

.
```

```
. * Cluster variance
. scalar define sigma2u = _b[/var(_cons[school])]

. scalar list sigma2u
   sigma2u =  .10259547

.
. * Overdispersion parameter
. scalar define alpha = exp(_b[/lnalpha])

. scalar list alpha
     alpha =  .78186163

.
. * Marginal expectation
. generate expectation = exp(xb + sigma2u/2)

.
. * Marginal variance
. generate variance = expectation + expectation^2*(exp(sigma2u)*(1 + alpha) - 1)

.
. * Marginal variance: Level-2 component
. generate variance2 = expectation^2*(exp(sigma2u) - 1)

.
. * Marginal variance: Level-1 component
. generate variance1 = expectation + expectation^2*exp(sigma2u)*alpha

.
. * Level-2 VPC
. generate vpc2 = variance2/(variance2 + variance1)

.
. * Level-1 VPC
. generate vpc1 = variance1/(variance2 + variance1)

.
. * Summarize marginal statistics
. summarize expectation variance variance2 variance1 vpc2 vpc1

    Variable |        Obs        Mean    Std. Dev.       Min        Max
-------------+---------------------------------------------------------
 expectation |     66,955    8.502207    2.885131   3.258426   21.21049
    variance |     66,955    87.04828    61.42828   13.60374   459.5689
   variance2 |     66,955    8.709502    6.498338    1.14713   48.60693
   variance1 |     66,955    78.33878     54.9303   12.45661   410.9619
        vpc2 |     66,955    .0979414    .0035452   .0843246   .1057664
-------------+---------------------------------------------------------
        vpc1 |     66,955    .9020586    .0035452   .8942336   .9156754

.
. * Figure 3: Line plot of Level-2 VPC against the marginal expectation
. line vpc2 expectation, sort

.
. * Figure 3: Spikeplot of marginal expectation
. spikeplot expectation

.
. * Predict cluster random intercept effects
. predict u, reffects
(calculating posterior means of random effects)
```

```
(using 7 quadrature points)

.
. * Keep the cluster identifier and the predicted cluster effect
. keep school u

.
. * Collapse the data down to one observation per cluster
. duplicates drop

Duplicates in terms of all variables

(66,521 observations deleted)

.
. * Save the predicted cluster effects
. save "model4.dta", replace
file model4.dta saved

.
. * Load model 2 predicted cluster random effects
. use "model2.dta", clear

. rename u model2u

.
. * Merge in model 4 predicted cluster random effects
. merge 1:1 school using "model4"
(label urn already defined)

    Result                           # of obs.
    -----------------------------------------
    not matched                             0
    matched                               434  (_merge==3)
    -----------------------------------------

. rename u model4u

.
. * Figure 4: Scatterplot of model 4 vs. model 2 predicted cluster random effects
. scatter model4u model2u

. correlate model4u model2u
(obs=434)

             |  model4u  model2u
-------------+------------------
     model4u |   1.0000
     model2u |   0.9184   1.0000


.
. * Rank the model 2 predicted cluster random effects
. sort model2u

. generate model2urank = _n

.
. * Rank the model 4 predicted cluster random effects
. sort model4u

. generate model4urank = _n
```

```
.
. * Figure 4: Scatterplot of ranks of model 4 vs. model 2 predicted effects
. scatter model4urank model2urank

. correlate model4urank model2urank
(obs=434)

             | model4~k model2~k
-------------+------------------
 model4urank |   1.0000
 model2urank |   0.9091   1.0000


.
.
. ******************************************************************************
. * Table 2: Model 5: Two-level random-coefficient negative binomial model
. ******************************************************************************
.
. * Clear memory
. clear all

.
. * Load the data
. use "absence.dta", clear

.
. * Fit model
. menbreg y quintile2 quintile3 quintile4 quintile5 spring winter autumn ///
>   female mixed asian black other notenglish sen fsm  ///
>   || school: fsm, covariance(unstructured) ///
>   startvalues(iv)

Fitting fixed-effects model:

Iteration 0:   log likelihood = -213128.96
Iteration 1:   log likelihood = -210651.83
Iteration 2:   log likelihood = -210545.23
Iteration 3:   log likelihood = -210545.17
Iteration 4:   log likelihood = -210545.17

Refining starting values:

Grid node 0:   log likelihood = -208637.83

Fitting full model:

Iteration 0:   log likelihood = -208637.83  (not concave)
Iteration 1:   log likelihood = -208461.37  (not concave)
Iteration 2:   log likelihood =  -208287.8  (not concave)
Iteration 3:   log likelihood = -208213.99
Iteration 4:   log likelihood = -208042.22
Iteration 5:   log likelihood = -207760.47
Iteration 6:   log likelihood =  -207661.3
Iteration 7:   log likelihood = -207635.69
Iteration 8:   log likelihood = -207633.84
Iteration 9:   log likelihood = -207633.82
Iteration 10:  log likelihood = -207633.82

Mixed-effects nbinomial regression              Number of obs     =     66,955
Overdispersion:           mean
Group variable:          school                 Number of groups  =        434
```

```
                                                Obs per group:
                                                              min =         31
                                                              avg =      154.3
                                                              max =        315

Integration method: mvaghermite                 Integration pts.  =          7

                                                Wald chi2(15)     =    5401.93
Log likelihood = -207633.82                     Prob > chi2       =     0.0000
------------------------------------------------------------------------------
           y |      Coef.   Std. Err.      z    P>|z|     [95% Conf. Interval]
-------------+----------------------------------------------------------------
   quintile2 |     -0.048      0.012    -4.09   0.000       -0.072      -0.025
   quintile3 |     -0.116      0.012    -9.52   0.000       -0.140      -0.092
   quintile4 |     -0.219      0.012   -18.49   0.000       -0.242      -0.196
   quintile5 |     -0.326      0.014   -23.25   0.000       -0.353      -0.299
      spring |      0.026      0.011     2.47   0.014        0.005       0.047
      winter |      0.078      0.011     7.37   0.000        0.057       0.099
      autumn |      0.112      0.011    10.64   0.000        0.091       0.133
      female |      0.122      0.009    13.86   0.000        0.105       0.139
       mixed |     -0.074      0.014    -5.36   0.000       -0.101      -0.047
       asian |     -0.198      0.013   -15.49   0.000       -0.223      -0.173
       black |     -0.421      0.011   -38.14   0.000       -0.443      -0.400
       other |     -0.195      0.017   -11.56   0.000       -0.228      -0.162
  notenglish |     -0.242      0.009   -25.81   0.000       -0.261      -0.224
         sen |      0.267      0.011    23.35   0.000        0.244       0.289
         fsm |      0.372      0.013    29.56   0.000        0.347       0.397
       _cons |      2.126      0.021    99.26   0.000        2.084       2.168
-------------+----------------------------------------------------------------
    /lnalpha |     -0.255      0.006   -39.51   0.000       -0.267      -0.242
-------------+----------------------------------------------------------------
school       |
      var(fsm)|     0.035      0.005                         0.027       0.046
    var(_cons)|     0.116      0.009                         0.100       0.135
-------------+----------------------------------------------------------------
school       |
cov(_cons,fsm)|    -0.027      0.005    -5.23   0.000       -0.037      -0.017
------------------------------------------------------------------------------
LR test vs. nbinomial model: chi2(3) = 5822.71           Prob > chi2 = 0.0000

Note: LR test is conservative and provided only for reference.

.
. * Deviance
. scalar define deviance = -2*e(ll)

. scalar list deviance
  deviance =  415267.64

.
. * Linear predictor
. predict xb, xb

.
. * Cluster intercept variance
. scalar define sigma2u0 = _b[/var(_cons[school])]

. scalar list sigma2u0
  sigma2u0 =  .11603906

.
. * Cluster slope variance
. scalar define sigma2u15 = _b[/var(fsm[school])]
```

```
. scalar list sigma2u15
 sigma2u15 =  .03503611

.
. * Cluster intercept-slope covariance
. scalar define sigmau015 = _b[/cov(fsm[school],_cons[school])]

. scalar list sigmau015
 sigmau015 = -.02662019

.
. * Overdispersion parameter
. scalar define alpha = exp(_b[/lnalpha])

. scalar list alpha
     alpha =  .77526043

.
. * Cluster-level variance function
. generate zomegauz = sigma2u0 + 2*sigmau015*fsm + sigma2u15*fsm^2

.
. * Marginal expectation
. generate expectation = exp(xb + zomegauz/2)

.
. * Marginal variance
. generate variance = expectation + expectation^2*(exp(zomegauz)*(1 + alpha) - 1)

.
. * Marginal variance: Level-2 component
. generate variance2 = expectation^2*(exp(zomegauz) - 1)

.
. * Marginal variance: Level-1 component
. generate variance1 = expectation + expectation^2*exp(zomegauz)*alpha

.
. * Level-2 VPC
. generate vpc2 = variance2/(variance2 + variance1)

.
. * Level-1 VPC
. generate vpc1 = variance1/(variance2 + variance1)

.
. * Summarize marginal statistics
. summarize expectation variance variance2 variance1 vpc2 vpc1

    Variable |        Obs        Mean    Std. Dev.       Min        Max
-------------+---------------------------------------------------------
 expectation |     66,955    8.524475    2.850676   3.302331   21.06587
    variance |     66,955    87.20277     59.4723   14.13889   446.0752
   variance2 |     66,955    9.036841    6.074755   1.341797   45.61101
   variance1 |     66,955    78.16592    53.43194   12.79709   400.4642
        vpc2 |     66,955    .1038891    .0063813   .0879661   .1158658
-------------+---------------------------------------------------------
        vpc1 |     66,955    .8961109    .0063813   .8841342   .9120339

.
. * Figure 5: Line plot of Level-2 VPC against the marginal expectation by FSM status
. line vpc2 expectation, sort by(fsm)
```

```
.
. * Figure 5: Spikeplot of marginal expectation by FSM status
. spikeplot expectation,  by(fsm)

.
.
.
. ********************************************************************************
. exit
```

## S10. R code for student absenteeism application

In this section we present R code to replicate the results for models 1-5 presented in Tables 1 and 2 of the paper. The code is annotated to help the reader.

### 10.1 R code

```
################################################################################
# Variance partitioning in multilevel count models
################################################################################
# R script file to replicate results presented in the article
################################################################################

################################################################################
# Basics
################################################################################

# Load library
#library(lme4)
library(haven)
library(ggplot2)
library(glmmTMB)

# Load the data
absence <- read_dta("absence.dta")
head(absence)

# Figure 1: Spikeplot of days absent
ggplot(data = absence, mapping = aes(x = y)) + geom_bar()

# Figure 2: Bar chart of district means
ggplot(data = absence, mapping = aes(x = reorder(district, y), y = y)) +
  geom_bar(stat = "summary", fun.y = "mean")

# Figure 2: Bar chart of school means
ggplot(data = absence, mapping = aes(x = reorder(school, y), y = y)) +
  geom_bar(stat = "summary", fun.y = "mean")

# Table S7.1 Covariate distributions, including mean days absent
by(absence$y, absence$quintile, function(x) c(N = length(x), mean = mean(x)))
by(absence$y, absence$season, function(x) c(N = length(x), mean = mean(x)))
by(absence$y, absence$female, function(x) c(N = length(x), mean = mean(x)))
by(absence$y, absence$ethnicity, function(x) c(N = length(x), mean = mean(x)))
by(absence$y, absence$notenglish, function(x) c(N = length(x), mean = mean(x)))
by(absence$y, absence$sen, function(x) c(N = length(x), mean = mean(x)))
by(absence$y, absence$fsm, function(x) c(N = length(x), mean = mean(x)))



################################################################################
# Table 1: Model 1: Two-level variance-components Poisson model
################################################################################

# Load the data
```

```
absence <- read_dta("absence.dta")
head(absence)

# Fit model
fm1 <- glmmTMB(y ~ 1 + (1|school), data = absence, family = poisson)
summary(fm1)

# Intercept
str(summary(fm1))
beta0 <- summary(fm1)$coefficients$cond[1,1]
beta0

# Cluster variance
str(summary(fm1))
sigma2u <- summary(fm1)$varcor$cond$school[1,1]
sigma2u

# Marginal expectation
expectation <- exp(beta0 + sigma2u/2)
expectation

# Marginal variance
variance <- expectation + expectation^2*(exp(sigma2u) - 1)
variance

# Marginal variance: Level-2 component
variance2 <- expectation^2*(exp(sigma2u) - 1)
variance2

# Marginal variance: Level-1 component
variance1 <- expectation
variance1

# Level-2 VPC
vpc2 <- variance2/(variance2 + variance1)
vpc2

# Level-1 VPC
vpc1 <- variance1/(variance2 + variance1)
vpc1



################################################################################
# Table 1: Model 2: Two-level variance-components negative binomial model
################################################################################

# Load the data
absence <- read_dta("absence.dta")
head(absence)

# Fit model
fm2 <- glmmTMB(y ~ 1 + (1|school), data = absence, family = nbinom2)
summary(fm2)

# Intercept
str(summary(fm2))
beta0 <- summary(fm2)$coefficients$cond[1,1]
beta0

# Cluster variance
str(summary(fm2))
sigma2u <- summary(fm2)$varcor$cond$school[1,1]
```

```
sigma2u

# Overdispersion parameter
str(summary(fm2))
alpha <- 1/(summary(fm2)$sigma)
alpha

# Marginal expectation
expectation <- exp(beta0 + sigma2u/2)
expectation

# Marginal variance
variance <- expectation + expectation^2*(exp(sigma2u)*(1 + alpha) - 1)
variance

# Marginal variance: Level-2 component
variance2 <- expectation^2*(exp(sigma2u) - 1)
variance2

# Marginal variance: Level-1 component
variance1 <- expectation + expectation^2*exp(sigma2u)*alpha
variance1

# Level-2 VPC
vpc2 <- variance2/(variance2 + variance1)
vpc2

# Level-1 VPC
vpc1 <- variance1/(variance2 + variance1)
vpc1

# Predict cluster random intercept effects
fm2u <- ranef(fm2)
fm2u



################################################################################
# Table 1: Model 3: Three-level variance-components negative binomial model
################################################################################

# Load the data
absence <- read_dta("absence.dta")
head(absence)

# Fit model
fm3 <- glmmTMB(y ~ 1 + (1|district) + (1|school),
               data = absence, family = nbinom2)
summary(fm3)

# Intercept
str(summary(fm3))
beta0 <- summary(fm3)$coefficients$cond[1,1]
beta0

# Supercluster variance
str(summary(fm3))
sigma2v <- summary(fm3)$varcor$cond$district[1,1]
sigma2v

# Cluster variance
str(summary(fm3))
sigma2u <- summary(fm3)$varcor$cond$school[1,1]
```

```
sigma2u

# Overdispersion parameter
str(summary(fm3))
alpha <- 1/(summary(fm3)$sigma)
alpha

# Marginal expectation
expectation <- exp(beta0 + sigma2v/2 + sigma2u/2)
expectation

# Marginal variance
variance <- expectation +
  expectation^2*(exp(sigma2v + sigma2u)*(1 + alpha) - 1)
variance

# Marginal variance: Level-3 component
variance3 < (expectation^2*(exp(sigma2v) - 1)
variance3

# Marginal variance: Level-2 component
variance2 <- expectation^2*exp(sigma2v)*(exp(sigma2u) - 1)
variance2

# Marginal variance: Level-1 component
variance1 <- expectation + expectation^2*exp(sigma2v + sigma2u)*alpha
variance1

# Level-3 VPC
vpc3 <- variance3/(variance3 + variance2 + variance1)
vpc3

# Level-2 VPC
vpc2 <- variance2/(variance3 + variance2 + variance1)
vpc2

# Level-1 VPC
vpc1 <- variance1/(variance3 + variance2 + variance1)
vpc1



################################################################################
# Table 2: Model 4: Two-level random-intercept negative binomial model
################################################################################

# Load the data
absence <- read_dta("absence.dta")
head(absence)

# Fit model
fm4 <- glmmTMB(y ~ 1 + fsm + (1|school), data = absence, family = nbinom2)
summary(fm4)

# Linear predictor
absence$xb <- predict(fm4)
head(absence)

# Cluster variance
str(summary(fm4))
sigma2u <- summary(fm4)$varcor$cond$school[1,1]
sigma2u
```

```
# Overdispersion parameter
str(summary(fm4))
alpha <- 1/(summary(fm4)$sigma)
alpha

# Marginal expectation
absence$expectation <- exp(absence$xb + sigma2u/2)
head(absence)

# Marginal variance
absence$variance <- absence$expectation +
  absence$expectation^2*(exp(sigma2u)*(1 + alpha) - 1)
head(absence)

# Marginal variance: Level-2 component
absence$variance2 <- absence$expectation^2*(exp(sigma2u) - 1)
head(absence)

# Marginal variance: Level-1 component
absence$variance1 <- absence$expectation +
  absence$expectation^2*exp(sigma2u)*alpha
head(absence)

# Level-2 VPC
absence$vpc2 <- absence$variance2/(absence$variance2 + absence$variance1)
head(absence)

# Level-1 VPC
absence$vpc1 <- absence$variance1/(absence$variance2 + absence$variance1)
head(absence)

# Summarize marginal statistics
colnames(absence)
sapply(absence[7:12], mean)

# Figure 3: Line plot of Level-2 VPC against the marginal expectation
ggplot(data = absence, mapping = aes(x = expectation, y = vpc2)) + geom_line()

# Figure 3: Spikeplot of marginal expectation
ggplot(data = absence, mapping = aes(x = expectation)) +
  geom_histogram(binwidth=1)

# Predict cluster random intercept effects
fm4u <- ranef(fm4)
str(fm4u)
head(fm4u$cond$school)

# Figure 4: Scatterplot of model 4 vs. model 2 predicted cluster random effects
fm4vsfm2 <- cbind(fm2u$cond$school,fm4u$cond$school)
colnames(fm4vsfm2)
colnames(fm4vsfm2) <- c("fm2u", "fm4u")
colnames(fm4vsfm2)
head(fm4vsfm2)
ggplot(data = fm4vsfm2, mapping = aes(x = fm2u, y = fm4u)) + geom_point()
cor(fm4vsfm2)

# Rank the model 2 predicted cluster random effects
fm4vsfm2$fm2urank <- rank(fm4vsfm2$fm2u)
head(fm4vsfm2)

# Rank the model 4 predicted cluster random effects
fm4vsfm2$fm4urank <- rank(fm4vsfm2$fm4u)
head(fm4vsfm2)
```

```
# Figure 4: Scatterplot of ranks of model 4 vs. model 2 predicted effects
ggplot(data = fm4vsfm2, mapping = aes(x = fm2urank, y = fm4urank)) +
  geom_point()
colnames(fm4vsfm2)
cor(fm4vsfm2[,3:4])



################################################################################
# Table 2: Model 5: Two-level random-coefficient negative binomial model
################################################################################

# Load the data
absence <- read_dta("absence.dta")
head(absence)

# Fit model
fm5 <- glmmTMB(y ~ 1 + fsm + (1 + fsm |school), data = absence, family = nbinom2)
summary(fm5)

# Linear predictor
absence$xb <- predict(fm5)
head(absence)

# Cluster intercept variance
str(summary(fm5))
sigma2u0 <- summary(fm5)$varcor$cond$school[1,1]
sigma2u0

# Cluster slope variance
str(summary(fm5))
sigma2u15 <- summary(fm5)$varcor$cond$school[2,2]
sigma2u15

# Cluster intercept-slope covariance
str(summary(fm5))
sigmau015 <- summary(fm5)$varcor$cond$school[1,2]
sigmau015

# Overdispersion parameter
str(summary(fm5))
alpha <- 1/(summary(fm5)$sigma)
alpha

# Cluster-level variance function
absence$zomegauz = sigma2u0 + 2*sigmau015*absence$fsm + sigma2u15*absence$fsm^2
head(absence)

# Marginal expectation
absence$expectation = exp(absence$xb + absence$zomegauz/2)
head(absence)

# Marginal variance
absence$variance = absence$expectation +
  absence$expectation^2*(exp(absence$zomegauz)*(1 + alpha) - 1)
head(absence)

# Marginal variance: Level-2 component
absence$variance2 = absence$expectation^2*(exp(absence$zomegauz) - 1)
head(absence)

# Marginal variance: Level-1 component
```

```
absence$variance1 = absence$expectation +
  absence$expectation^2*exp(absence$zomegauz)*alpha
head(absence)

# Level-2 VPC
absence$vpc2 = absence$variance2/(absence$variance2 + absence$variance1)
head(absence)

# Level-1 VPC
absence$vpc1 = absence$variance1/(absence$variance2 + absence$variance1)
head(absence)

# Summarize marginal statistics
colnames(absence)
sapply(absence[8:13], mean)

# Figure 5: Line plot of Level-2 VPC against the marginal expectation by FSM status
ggplot(data = absence, mapping = aes(x = expectation, y = vpc2)) +
  geom_line(aes(group = fsm))

# Figure 5: Spikeplot of marginal expectation by FSM status
ggplot(data = absence, mapping = aes(x = expectation)) +
  geom_histogram(binwidth=1)


################################################################################
```